\documentclass[12pt]{article}
\usepackage{amscd,amssymb,amsmath,latexsym,enumerate}
\usepackage{amsfonts,amsthm,hyperref,mathrsfs,ulem,tikz}
\usepackage[mathscr]{euscript}
\usepackage{graphicx}
\usepackage{mathrsfs}
\usepackage{epsfig}
\usepackage{fancybox}
\usepackage{verbatim}
\usepackage{tikz}
\usepackage{tikz-cd}
\usetikzlibrary{calc}
\usepackage{float}
\usepackage{todonotes}
\usepackage{multicol}
\usepackage{mathtools}
\usepackage[final]{pdfpages}

\usepackage{xcolor}
\definecolor{MyBlue}{cmyk}{1,0.13,0,0.63}
\definecolor{MyGreen}{cmyk}{0.91,0,0.88,0.52}
\newcommand{\mylinkcolor}{MyBlue}
\newcommand{\mycitecolor}{MyGreen}
\newcommand{\myurlcolor}{black}

\usepackage{color}

\newcommand{\CM}{{\mathbb C}}
\newcommand{\NM}{{\mathbb N}}

\newcommand{\ZM}{{\mathbb Z}}

\newcommand{\EM}{{\mathbb E}}

\newcommand{\Oo}{{\cal O}}
\newcommand{\Tr}{\mbox{\rm Tr}}

\newcommand{\Nn}{{\cal N}}

\def\essinf{\mathop{\rm ess\,inf}}
\def\esssup{\mathop{\rm ess\,sup}}

\newcommand{\one}{{\bf 1}}

\newcommand{\sgn}{\mbox{\rm sgn}}

\newcommand{\diag}{{\mbox{\rm diag}}}

\newcommand{\arccot}{{\mbox{\rm arccot}}}

\newcommand{\ParaEps}{\delta}
\newcommand{\ChiVar}{w}

\newcommand{\tick}[1]{\filldraw[black!100,line width=0.5mm,fill=none] {#1} ellipse (0.0 and 0.2)}

\newcommand{\vtick}[1]{\filldraw[black!100,line width=1pt,fill=none] {#1} ellipse (0.0 and 0.2)}

\def\centerarc[#1](#2)(#3:#4:#5){\draw[#1] ($(#2)+({#5*cos(#3)},{#5*sin(#3)})$) arc (#3:#4:#5)} 
\newcommand{\shortrightarrow}{\parbox{5pt}{\tikz{\draw[-{>[scale=0.7]}](0,0)--(5pt,0);}}}

\newcommand{\scriptshortrightarrow}{\parbox{4pt}{\tikz{\draw[-{>[scale=0.6]}](0,0)--(4pt,0);}}}

\usepackage{hyperref}
\hypersetup{%
  bookmarksnumbered=true,bookmarksopen=false,%
  plainpages=false,
  linktocpage=true,%
  colorlinks=true,breaklinks=true,%
  linkcolor=\mylinkcolor,citecolor=\mycitecolor,urlcolor=\myurlcolor,%
  pdfpagelayout=OneColumn,%
  pageanchor=true,%
}

\usepackage{etoolbox}
\makeatletter
\patchcmd{\math@cr@@@align}{\cr}{\global\let\df@label\@empty\cr}{}{}
\makeatother

\title{Scaling of the Lyapunov exponent \\ at a balanced hyperbolic critical point}

\author{Joris De Moor$^1$, Christian Sadel$^2$, Hermann Schulz-Baldes$^1$
\\
\\
{\small $^1$Friedrich-Alexander-Universit\"at Erlangen-N\"urnberg}
\\
{\small Department Mathematik, Cauerstr.~11, D-91058 Erlangen, Germany}
\\
{\small $^2$Pontifica Universidad Cath\'olica de Chile}
\\
{\small Facultad de Matem\'aticas, Av. Vicu\~na Mackaenna 4860, Santiago 7820436, Chile}
}

\date{ }

\newtheorem{theorem}{Theorem}

\newtheorem{lemma}[theorem]{Lemma}

\newtheorem{definition}[theorem]{Definition}
\newtheorem{remark}[theorem]{Remark}

\def\essinf{\mathop{\rm ess\,inf}}
\def\esssup{\mathop{\rm ess\,sup}}

\begin{document}

\maketitle

\begin{abstract}
In both the random hopping model and at topological phase transitions in one-dimensional chiral systems, the Lyapunov exponent vanishes at zero energy, but is here shown to have an inverse logarithmic increase with a coefficient that is computed explicitly. This is the counterpart of the Dyson spike in the density of states. The argument also transposes to the free energy density of the random field Ising model,  and more generally to many so-called balanced hyperbolic critical points. It is based on the fact that the Furstenberg measure in rescaled logarithmic Dyson-Schmidt variables can be well-approximated by an absolutely continuous measure with trapezoidal density.

\vspace{.1cm}

\end{abstract}



\section{Main results and outline}
\label{sec-Intro}

This work is about the Lyapunov exponent of i.i.d. products of smooth one-parameter random families  $\epsilon\mapsto T^\epsilon$ of real random $2\times 2$ matrices of the form 
\begin{equation}
\label{eq:expansion}
T^{\epsilon}
\;=\;
\pm
\left[\mathbf{1}\,+\,\epsilon a
\begin{pmatrix}
0 & -1 \\
1 & 0
\end{pmatrix}
\,+\,\epsilon b
\begin{pmatrix}
0 & 1 \\
1 & 0
\end{pmatrix}
\,+\,\epsilon c
\begin{pmatrix}
1 & 0 \\
0 & -1
\end{pmatrix}
\,+\,
\Oo(\epsilon^2)
\right]
\begin{pmatrix}
\kappa & 0 \\ 0 &\frac{1}{\kappa}
\end{pmatrix}
\,,
\end{equation}
where $a,b,c$ are compactly supported random variables satisfying the conditions $a>| b|$,  the term $\Oo(\epsilon^2)$ contains a compactly supported random $2\times 2$ matrix and $\kappa>0$ is a compactly supported non-constant random variable satisfying $\EM(\log(\kappa))=0$. Here and below, $\EM$ denotes the expectation value. Because the dominating term $a$ is in front of the generator of rotations, the set-up~\eqref{eq:expansion} is referred to as a balanced hyperbolic critical point of rotating type. The Lyapunov exponent  associated to an i.i.d. sequence $(T^\epsilon_n)_{n\geq 1}$ is defined as usual~\cite{BL,BQ} by
\begin{equation}
\label{eq-LyapDef}
\gamma^\epsilon
\;=\;
\lim_{N \to \infty} \,\frac{1}{N}\,
\mathbb{E}\big(\log(\|T^\epsilon_{\vphantom{\widehat{k}}N}\cdots T^\epsilon_{1}\|)\big)
\;.
\end{equation}

\begin{theorem}
\label{theo-Lyapunov}
The Lyapunov exponent of a balanced hyperbolic critical point of rotating type satisfies
\begin{equation}
\gamma^\epsilon
\;=\;
\frac{\EM\big(\log(\kappa)^2\big)}{\big|\log(|\epsilon|)\big|}
\;+\;
\Oo\Big(\frac{\log(|\log(|\epsilon|)|)^3}{\log(|\epsilon|)^2}\Big)
\;.
\label{eq-Lyap}
\end{equation}
\end{theorem}

Note that the overall sign in~\eqref{eq:expansion} is irrelevant for the Lyapunov exponent. Moreover, for $\epsilon=0$, the matrices $T^\epsilon$ are always diagonal and hence they all commute. The upper diagonal entry of $T^0_{\vphantom{\widehat{k}}N}\cdots T^0_{1}$ is $\prod_{n=1}^N\kappa_n$, the lower one the inverse of this value. Due to the balancing equation $\EM(\log(\kappa))=0$, one thus has $\gamma^0=0$. Theorem~\ref{theo-Lyapunov} hence shows that the growth of $\epsilon\mapsto\gamma^\epsilon$ near $\epsilon=0$ is merely log-H\"older continuous which is the minimal type of regularity of the Lyapunov exponent associated to a smooth family of random matrix products \cite{TV}.  In fact, if $\mathbb{E}(\log(\kappa))\neq 0$, the two Lyapunov exponents $\pm \mathbb{E}(\log(\kappa))$ associated to the random matrix products for $\epsilon=0$ are distinct and $\epsilon\mapsto \gamma^\epsilon$ would be H\"older continuous at $\epsilon=0$. Let us note that the example of a not H\"older continuous family provided in Section 5 of \cite{TV} is precisely of the form as studied here. In \cite{TV}, the cocycle is constructed from a Schr\"odinger operator and one uses connections between the so-called
$(\gamma,\beta)\,$log H\"older continuities of the Lyapunov exponent and the integrated density of states together with the Dyson spike, {\it cf.}~\eqref{eq-IDS} below, to deduce the non-H\"older behavior of the Lyapunov exponent. But a perturbative formula for the Lyapunov exponent as in \eqref{eq-Lyap} is not developed in \cite{TV}. One further implication of \cite{TV} is that hyperbolic critical energies give the worst possible form of regularity of the Lyapunov exponent.
 
 \vspace{.2cm}
 
On first sight, it may seem that Theorem~\ref{theo-Lyapunov} is already obtained in the recent works by Giacomin and Greenblatt~\cite{GG} as well as Collin~\cite{Col} (see also \cite{CGGH} which appeared later than the present work) that rigorously confirm a formula by Derrida and Hilhorst~\cite{DH}, but as explained towards the end of this introduction, the hypothesis $a>|b|$ is not covered in these works and leads to crucial differences and, in particular, a completely different Furstenberg measure. Finally let us note that the compact support assumption on the random variable $T^\epsilon$ can, in principle and with further technicalities, be relaxed to a finite moment condition as in \cite{DS} and \cite{Col}.

\vspace{.2cm}

\begin{figure}[h]
\begin{center}
\includegraphics[width=0.325\textwidth]{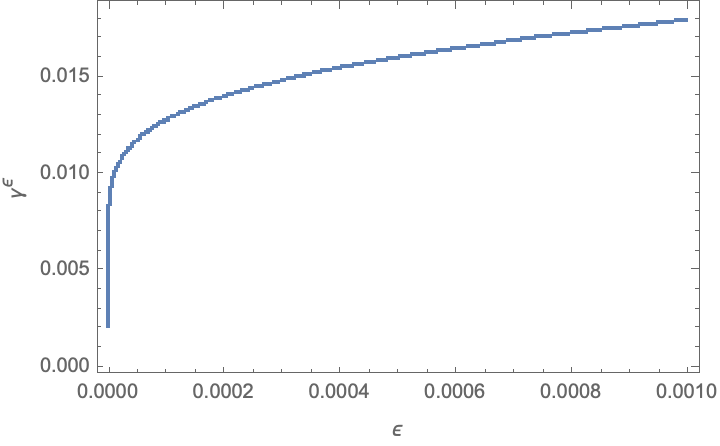}
\includegraphics[width=0.325\textwidth]{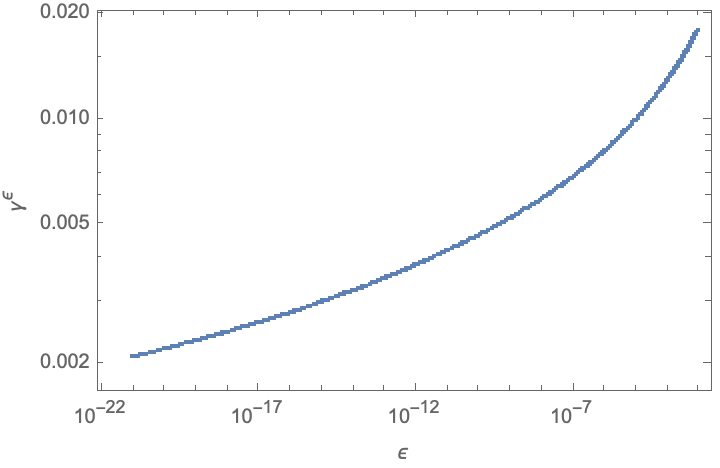}
\includegraphics[width=0.325\textwidth]{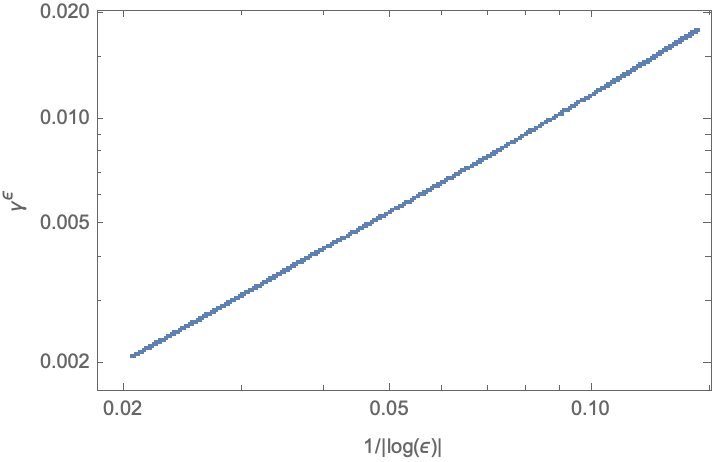}
\caption{\it Numerical computation of the Lyapunov exponent for the random hopping model with $t - 1.1$ being uniformly distributed on $[-0.4,0.4]$.  The data for $\gamma^\epsilon$ is obtained from Birkhoff sums over orbits of length $10^6$ and represented in three ways: normal plot, log-log plot and as a function of $1/\log(\epsilon^{-1})$.}
\label{fig-Lyap}
\end{center}
\end{figure}

Balanced hyperbolic critical points of rotating type appear in numerous applications in which the parameter $\epsilon$ is an energy difference from a so-called critical energy. As a first example, let us consider the random hopping model which was already studied by Dyson~\cite{Dys}. It is a bounded selfadjoint Hamiltonian on the one-dimensional tight-binding Hilbert space $\ell^2(\mathbb{Z})$, given by the purely off-diagonal random Jacobi matrix
$$
(H\psi)_n
\;=\;
-\,{t}_{n+1} \psi_{n+1}\,-\,{t}_{n}\psi_{n-1}
\,,
\qquad
\psi=(\psi_n)_{n\in\mathbb{Z}}\in\ell^2(\mathbb{Z})
\;,
$$
with independent and identically distributed (i.i.d.) positive random variables $({t}_{n})_{n\in\ZM}$ which are called the hopping parameters. It is well-known that the formal solutions of the Schr\"odinger equation $H\psi=\epsilon\psi$ at energy $\epsilon$ can be computed by $2\times 2$ transfer matrices. Due to the sublattice symmetry $\mathcal{J}H\mathcal{J}=-H$ for $(\mathcal{J}\psi)_n=(-1)^n\psi_n$, it is natural to rather work with the transfer matrix over two sites:
\begin{align}
\begin{aligned}
\label{eq-TransferHopping}
T^\epsilon_{n}
&
\;=\;
\begin{pmatrix}
-\epsilon\,t_{2n}^{-1} & - t_{2n} \\ t_{2n}^{-1} & 0
\end{pmatrix}
\begin{pmatrix}
-\epsilon\,t_{2n-1}^{-1} & - t_{2n-1} \\ t_{2n-1}^{-1} & 0
\end{pmatrix}
\\
&\;=\;
-\,
\left[
\one
\;+\;
\epsilon
\begin{pmatrix}
0 & -1 \\ \tfrac{1}{t_{2n}^2} & 0
\end{pmatrix}
\;+\;
\epsilon^2
\begin{pmatrix}
-\tfrac{1}{t_{2n}^2} & 0 \\ 0 & 0
\end{pmatrix}
\right]
\begin{pmatrix}
\kappa_n & 0 \\ 0 & \tfrac{1}{\kappa_n}
\end{pmatrix}
\;,
\end{aligned}
\end{align}
where $\kappa_n=\frac{t_{2n}}{t_{2n-1}}$.  This is indeed of the form~\eqref{eq:expansion} with $a_n=(t_{2n})^{-2}+1$, $b_n=(t_{2n})^{-2}-1$ and $c_n=0$ satisfying the conditions stated above, so that Theorem~\ref{theo-Lyapunov} applies. Let us stress that while the definition~\eqref{eq-LyapDef} of the Lyapunov exponent is the standard one in the theory of products of random matrices, it is often modified by a factor $\frac{1}{2}$ in this context because each $T^\epsilon_n$ is the transfer matrices over $2$ sites. In particular, the localization length at energy $\epsilon$ is $\frac{2}{\gamma^\epsilon}$ rather than $\frac{1}{\gamma^\epsilon}$. Implementing the random hopping model numerically, Figure~\ref{fig-Lyap} illustrates Theorem~\ref{theo-Lyapunov}.

\vspace{.2cm}

For the random hopping model, the lowest order term in~\eqref{eq-Lyap} has been well-known in the physics community for decades~\cite{ER,Zim,HJ,McK,BF,BCR}, and these references also stress its relevance for quantum phase transitions in random field quantum Ising chains because the random hopping model is their Jordan-Wigner transformation. As argued in~\cite{DSS}, variations of the random hopping model describe topological phase transitions in chiral one-dimensional topological insulators, such as the disordered Su-Schrieffer-Heeger model~\cite{SSH,MSHP}, and hence Theorem~\ref{theo-Lyapunov} also applies at such topological phase transitions. Furthermore certain random Dirac operators also lead to a balanced hyperbolic critical energy of rotating type, see Section~\ref{sec-Examples}.

\vspace{.2cm}

The singular behavior of the Lyapunov exponent in the random hopping model (and, more generally, in random polymer models  with a hyperbolic critical energy and chiral one-channel Hamiltonians) is accompanied by its integrated density of states of the form
\begin{equation}
\Nn^\epsilon
\;=\;
\frac{\EM\big(\log(\kappa)^2\big)}{4\,\log(\epsilon)^2}
\;+\;
\Oo\Big(\frac{\log(|\log(\epsilon)|)^3}{|\log(\epsilon)|^3}\Big)
\;,
\label{eq-IDS}
\end{equation}
namely a so-called Dyson spike for the density of states. While this was already found by Dyson~\cite{Dys}, a rigorous proof was only given more recently by Kotowski and Vir\'ag with a weaker error bound~\cite{KV} and as stated in~\eqref{eq-IDS} in a previous work~\cite{DSS}. Actually, the techniques of the latter work and~\cite{DS} (renewal theory, optional stopping theorem) are here extended and complemented for the proof of Theorem~\ref{theo-Lyapunov}. As density of states and Lyapunov exponent are real and imaginary part of the disorder averaged Green function on the real axis, it is possible to obtain both \eqref{eq-Lyap} and~\eqref{eq-IDS} using complex Dyson-Schmidt variables \cite{Zim}. A rigorous treatment seems conceivable by a control of the random dynamics of complex Pr\"ufer variables, similar to~\cite{DS1} which dealt with an elliptic critical energy where the commuting matrices $(T^0_n)_{n\geq 1}$ have spectrum on the unit circle.  Let us also mention here that the log-H\"older regularity of the integrated density of states as in \eqref{eq-IDS} is known in very general terms \cite{BoK}.

\vspace{.2cm}

Critical energies of such a different type (elliptic or parabolic) appear in the random dimer and random polymer models~\cite{DWP,JSS}, at the band edges of the Anderson model~\cite{SaS} and in certain random Kronig-Penney models~\cite{DKS}. All these models have a characteristic vanishing of the Lyapunov exponent (described in Section~\ref{sec-Dyn}) and may exhibit a non-trivial  quantum transport (anomalous diffusion). For the random hopping model, supersymmetric analysis suggests that the growth of the second moment of the position operator is logarithmic in time, a behavior called quantum Sinai diffusion due to a classical counterpart in certain Langevin equations~\cite{BAK}. This transport only results from low-energy states, as all others are known to be localized~\cite{Sh}. A rigorous confirmation of Sinai diffusion based on~\eqref{eq-Lyap} and~\eqref{eq-IDS} seems to be within reach and is currently under investigation. Non-quantitative rigorous results merely state that there is no Anderson localization in the sense of the fractional moment method~\cite{PS,SS}. 

\vspace{.2cm}

The proof of Theorem~\ref{theo-Lyapunov} and other results in the paper use the same formalism as in \cite{DSS}, with minor  technical improvements though which are explained throughout the text. Let us now describe the new elements of the arguments.  The first ingredient is a statement about the Furstenberg measure which is of considerable independent interest. Recall that the one-dimensional real projective space
$$
\mathbb{R} {\rm P}(1)
\;=\;
\big\{e_\theta\,:\;\theta\in[0,\pi)
\big\}
\;\cong\;[0,\pi)
\;,
\qquad
e_\theta\,=\,
\begin{pmatrix} \cos(\theta) \\ \sin(\theta) \end{pmatrix}
\;,
$$
can naturally be identified with an interval of so-called Pr\"ufer angles $\theta$, and that real invertible $2\times 2$ matrices naturally act on $\mathbb{R}$P$(1)$. If such matrices are drawn i.i.d., this generates a random dynamical system on $\mathbb{R}$P$(1)$ which is a Markov process with state space $\mathbb{R}$P$(1)$ and is, under mild non-triviality conditions (irreducibility, see~\cite{BL,BQ}), known to have a unique invariant probability measure on $\mathbb{R}$P$(1)$ called the Furstenberg measure. If the i.i.d. matrices are given by~\eqref{eq:expansion}, then this measure $\mu^\epsilon$ is characterized by
\begin{equation}
\label{eq-FurstenbergDef}
\int \mu^\epsilon(d\theta)\,f(e_\theta)
\;=\;
\EM\;
\int \mu^\epsilon(d\theta)\,f\Big(\pm\frac{T^\epsilon e_\theta}{\|T^\epsilon e_\theta\|}\Big)
\;,
\qquad
f\in C(\mathbb{R} {\rm P}(1))
\;.
\end{equation}
The Furstenberg measure is always H\"older continuous and absolutely continuous if the distribution of the $T^\epsilon$ is absolutely continuous. Actually the former is known to hold under much milder assumptions \cite{BL,Gui,AG}. More detailed information on $\mu^\epsilon$, such as a formula for its density, can only be attained in particular situations, often in a perturbative manner: At elliptic critical energies and away from so-called anomalies, it is known to be the Lebesgue measure up to errors~\cite{PF,JSS}. At anomalies it is absolutely continuous with a density that can be computed as the ground state of a certain Fokker-Planck operator~\cite{BK,SB2}. At band edges of Anderson-type models, a similar result can be obtained after appropriate rescaling~\cite{DG,SaS}. Finally, for a balanced hyperbolic critical energy of rotating type (and thus, in particular, for the random hopping model), it was shown in~\cite{DS,DSS} that $\mu^\epsilon$ converges weakly  in the limit $\epsilon\to 0$ to a linear combination of two Dirac peaks on $\theta=0$ and $\theta=\frac{\pi}{2}$ which are the two fixed points of the random dynamics at $\epsilon=0$. The novel insight provided in Theorem~\ref{theo-InvMeasure} below is that in suitable coordinates for $\mathbb{R} {\rm P}(1)$ this measure is, up to controllable errors, given by an explicitly computed absolutely continuous measure. The needed orientation-preserving change of variables are
\begin{align}
\theta\in[0,\pi)\cong\mathbb{R} {\rm P}(1)
&
\;\mapsto\;x\;=\;-\cot(\theta)\in\dot{\mathbb{R}}
\label{eq-DysonSchmidt}
\\
&
\;\mapsto\;
(y,\nu)=\Big(\frac{\sgn(x)}{2C_0}\log(|x|),\sgn(x)\Big)\in\dot{\mathbb{R}}\times\{-,+\}
\label{eq-VariableTransforms}
\\
&
\;\mapsto\;
(z,\nu)
=\Big(\frac{2C_0}{\log(|\epsilon|^{-1})}\,y ,\nu\Big)\in\dot{\mathbb{R}}\times\{-,+\}
\;,
\label{eq-normalizedLogDysonSchmidt}
\end{align}
where $\dot{\mathbb{R}}=\mathbb{R}\cup\{\infty\}$. Here $x$ is called the Dyson-Schmidt variable, $y$ the logarithmic and $z$ the rescaled logarithmic Dyson-Schmidt variable which satisfies $x=\nu \epsilon^{-\nu z}$. We will also refer to these representations as the $x$-picture, $y$-picture and $z$-picture. The constant $C_0$ in~\eqref{eq-VariableTransforms} is introduced for technical convenience in the $y$-picture and is the one appearing in the Main Hypothesis stated in Section~\ref{sec-Dyn}. In~\eqref{eq-DysonSchmidt}, $\dot{\mathbb{R}}$ is equipped with the topology of one-point compactification, while in~\eqref{eq-VariableTransforms} and~\eqref{eq-normalizedLogDysonSchmidt} the two copies are rather connected to form a circle as well. As this is not of importance in the present context, it will not be notationally distinguished.  The push-forward of $\mu^\epsilon(d\theta)$ under the concatenation of these maps will be denoted by $(\mu_+^\epsilon(dz),\mu_-^\epsilon(dz))$. Then the sum $\mu_s^\epsilon(dz)= \mu_+^\epsilon(dz)+\mu_-^\epsilon(dz)$ is again a probability measure. The following is then the main result of Section~\ref{sec:mu}.

\begin{theorem}
\label{theo-InvMeasure}
Near a balanced hyperbolic critical point of rotating type, one has for every closed interval $I\subset (-1,1)$
\begin{equation}
\label{eq-FurstenbergApprox}
\mu^\epsilon_\pm(I)
\;=\;
\frac{1}{2}\int_I\frac{1-z}{2}\,dz
\;+\;
\Oo\Big(\frac{\log(\log(\epsilon^{-1}))}{\log(\epsilon^{-1})}\Big)
\;.
\end{equation}
\end{theorem}

\begin{figure}[h]
\begin{center}
\includegraphics[width=0.32\textwidth]{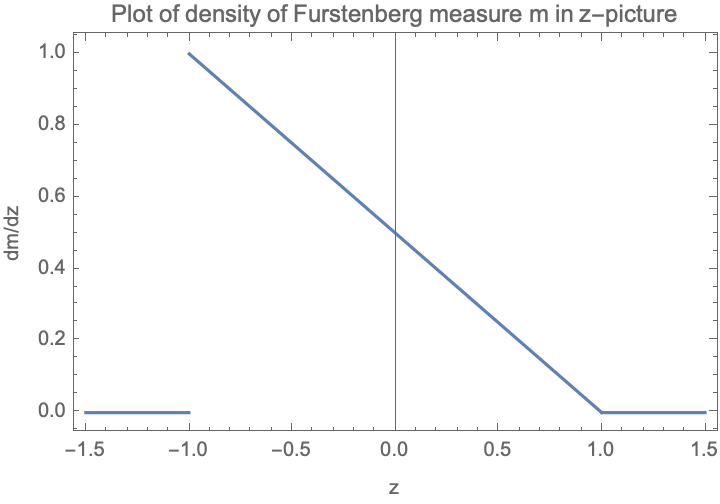}
\includegraphics[width=0.32\textwidth]{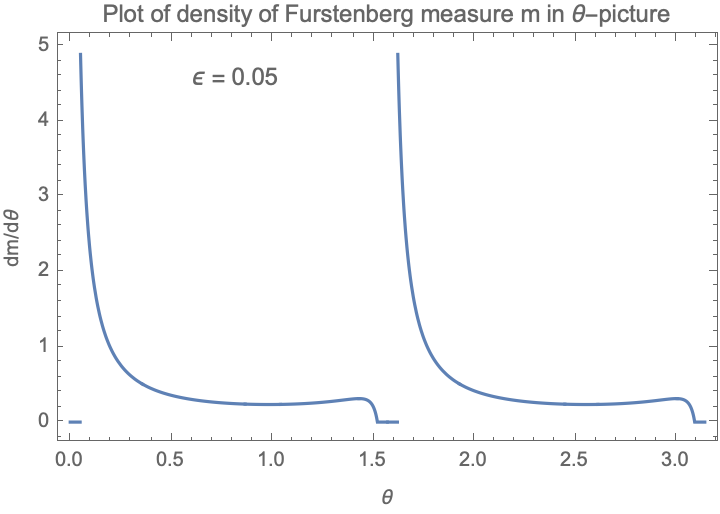}
\includegraphics[width=0.32\textwidth]{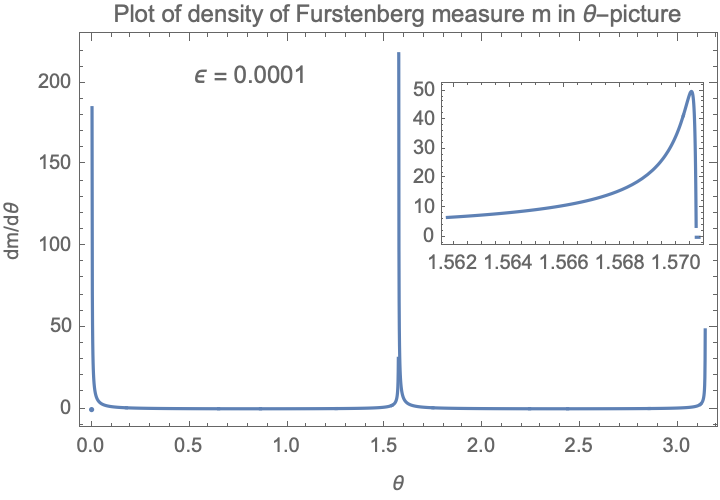}
\caption{\it The first plot shows the lowest order approximation to $\mu_s^\epsilon(dz)$ as given by the sum of the r.h.s. of~\eqref{eq-FurstenbergApprox}. The second plot shows its pull-back under~\eqref{eq-DysonSchmidt} to~\eqref{eq-normalizedLogDysonSchmidt}, namely the approximation to the Furstenberg measure $\mu^\epsilon(d\theta)$ on the Pr\"ufer phases for $\epsilon=0.05$. Note that there are two intervals of size $\epsilon$ around the $\theta=0$ and $\theta=\frac{\pi}{2}$ on which the density of the approximate Furstenberg measure vanishes. The third plot is as the second one with $\epsilon=0.0001$. Its inlay shows that for $\epsilon$ small enough, a peak to the left of $\frac{\pi}{2}$ develops {\rm (}to the left of $\pi$ one has the same{\rm )}. 
}
\label{fig-Furstenberg}
\end{center}
\end{figure}

\begin{figure}[h]
\begin{center}
\includegraphics[width=0.31\textwidth]{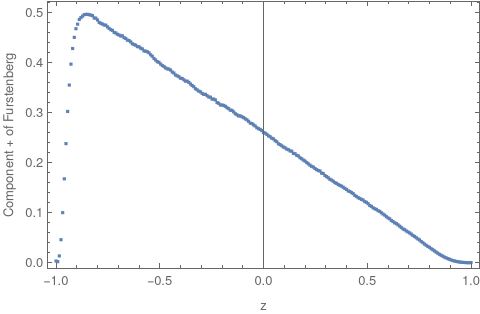}
\hspace{.1cm}
\includegraphics[width=0.31\textwidth]{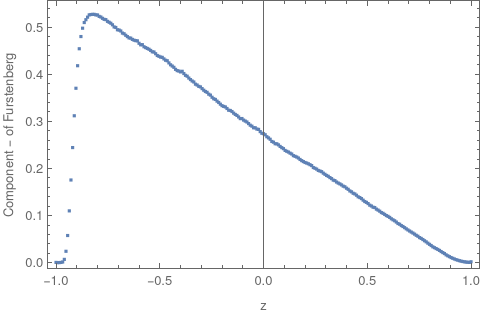}
\hspace{.1cm}
\includegraphics[width=0.31\textwidth]{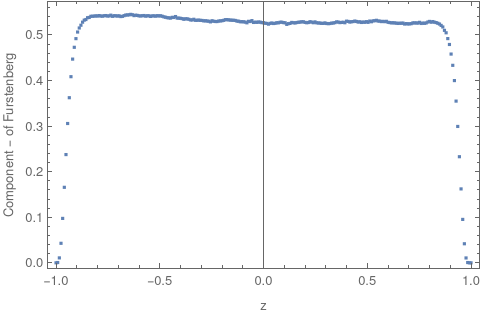}
\caption{\it Plots approximating $\mu^\epsilon_+$ and $\mu^\epsilon_-$ in the $z$-picture computed from random orbits $(z^\epsilon_n,\nu^\epsilon_n)_{n=1,\ldots,N}$ of length $N=8\cdot 10^7$ for the random hopping model with $\epsilon = 10^{-19}$ and i.i.d. hopping parameters $t$ distributed such that $t-0.5$ follows the uniform distribution on $[-0.15,0.15]$. The last picture shows the result for the random field Ising chain with same $\kappa$ and $\epsilon$ {\rm (}notably, $a=0$ and $b=1${\rm )}.
}
\label{fig:xi-hist}
\end{center}
\end{figure}

Up to errors $o(\epsilon)$, Theorem~\ref{theo-InvMeasure} implies that $\mu_s^\epsilon(dz)$ has no weight on $\mathbb{R}\setminus [-1,1]$, and that $\mu^\epsilon_+(dz)=\mu^\epsilon_-(dz)$ so that $\mu_s^\epsilon(dz)=\one[|z|\leq 1]\frac{1-z}{2}\,dz$, where $\one[|z|\leq 1]$ denotes the indicator function on $\{|z|\leq 1\}$. In particular, this implies that $\mu_\pm^\epsilon(dz)$ both converge weakly as $\epsilon\to 0$ to the triangular distribution $\frac{1}{2} \one[|z|\leq 1]\frac{1-z}{2}\,dz$. This triangular distribution is shown in the first plot of Figure~\ref{fig-Furstenberg}. The second plot then shows its transformation back to the Pr\"ufer phases. For the random hopping model, numerical data from simulations illustrating Theorem~\ref{theo-InvMeasure} is presented in Figure~\ref{fig:xi-hist}.

\vspace{.2cm}

Finally, let us briefly outline the second novel ingredient to the proof of Theorem~\ref{theo-Lyapunov}. As usual, the computation of the Lyapunov exponent is reduced to the evaluation of a Birkhoff sum of the function $\theta\mapsto\EM\,\log(\|T^\epsilon e_\theta\|)$. This can then be transformed to a function $z\mapsto f^\epsilon(z)$ in the $z$-picture. While this function is shown to be positive, it is too singular to be computed by Theorem~\ref{theo-InvMeasure}. However, one can add a suitable cocycle difference in order to produce an integrand that can be expanded in $1/\log(\epsilon^{-1})$ to complete the proof of Theorem~\ref{theo-Lyapunov}, as explained in Section~\ref{sec:gamma}.

\vspace{.2cm}

As promised right after the statement of Theorem~\ref{theo-Lyapunov}, let us now come to a different type of balanced hyperbolic critical points, namely those considered in the recent works~\cite{GG,Col} where the random matrices are still given by~\eqref{eq:expansion}, but with $b>|a|$ instead of $a>| b|$. As explained in Section~\ref{sec-Examples} this is relevant for the study of the classical random field Ising chain. Let us describe the drastic effects of this change on the dynamics. Given an i.i.d. sequence of random matrices $(T^\epsilon_n)_{n\geq 1}$ together with an initial condition $\theta_0\in \mathbb{R}$P$(1)$, one obtains in both cases a sequence $(\theta^\epsilon_n)_{n\geq 0}$ in $\mathbb{R}$P$(1)$. According to  {\it e.g.} eq. (15) in~\cite{DSS}, this sequence is given by the following two-step dynamics $\theta^\epsilon_{n-1}\mapsto \theta^\epsilon_{n'}\mapsto\theta^\epsilon_n$:
\begin{equation}
\label{eq-TwoStepDynPhases}
\theta^\epsilon_{n'}
\;=\;
\arccot\big(
\kappa_n^2(1+\epsilon c_n)^2\,\cot(\theta^\epsilon_{n-1})\big)
\;,
\qquad
\theta^\epsilon_{n}
\;=\;
\theta^\epsilon_{n'}
\,+\,\epsilon\big(a_n+b_n\cos(2\theta^\epsilon_{n'})\big)\,+\,\Oo(\epsilon^2)
\;.
\end{equation}
Now the first step of the dynamics has the two fixed points $\theta=0,\frac{\pi}{2}$ so that both intervals, $(0,\frac{\pi}{2})$ and $(\frac{\pi}{2},\pi)$, are invariant under the dynamics of this first step. For the second step of the dynamics, the (possibly random) values of $a$ and $b$ can lead to very distinct behaviors. For $a>|b|$, the above two intervals are {\it not} left invariant, but one can only pass the boundary points $\theta=0,\frac{\pi}{2}$ in the positive direction (for $a<-|b|$ in the negative direction) and this leads to the non-trivial rotation numbers analyzed in~\cite{DS,DSS}. On the other hand, for $b>|a|$ the second step of the dynamics leaves the interval $(0,\frac{\pi}{2})$ invariant (because the function $\theta\mapsto \cos(2\theta)$ is positive/negative near $\theta=0$/$\frac{\pi}{2}$ respectively), while the dynamics can leave the interval $(\frac{\pi}{2},\pi)$ (for the same reason) and will actually do so with probability $1$. Hence for $b>|a|$ there is no rotation and the dynamics is confined to $(0,\frac{\pi}{2})$. For this reason, the case $b>|a|$  will be referred to as a balanced hyperbolic critical point of {\it confined type}. Clearly, the Furstenberg measure is then supported on $[0,\frac{\pi}{2}]$ which corresponds to the whole fiber $\nu=-$ in~\eqref{eq-VariableTransforms} due to the sign in $x=-\cot(\theta)$. On the other hand, if $b<-|a|$, the invariant measure is supported on $[\frac{\pi}{2},\pi]$ corresponding to $\nu=+$. Let us next state the counterpart of Theorems~\ref{theo-Lyapunov} and~\ref{theo-InvMeasure}, focussing on the former case. The result was already obtained in~\cite{GG,Col} by other means and under different hypothesis. 

\vbox{
\begin{theorem}
\label{theo-InvMeasure2}
Consider a confined balanced hyperbolic critical point of confined type with $b>|a|$. Then for every closed interval $I\subset (-1,1)$, 
\begin{equation}
\label{eq-FurstenbergApprox2}
\mu^\epsilon_+(I)
\;=\;
0
\;,
\qquad
\mu^\epsilon_-(I)
\;=\;
\int_I\frac{1}{2}\,dz
\;+\;
\Oo\Big(\frac{\log(\log(\epsilon^{-1}))}{\log(\epsilon^{-1})}\Big)
\;.
\end{equation}
Moreover, the Lyapunov exponent satisfies~\eqref{eq-Lyap}.
\end{theorem}
}

Section~\ref{sec-Confined} describes the modifications of the arguments in Sections~\ref{sec:mu} and~\ref{sec:gamma} which are necessary for the proof of Theorem~\ref{theo-InvMeasure2}. Of course, there are balanced hyperbolic critical points which are of neither rotating nor confined type so that neither Theorems~\ref{theo-Lyapunov} and~\ref{theo-InvMeasure} nor Theorem~\ref{theo-InvMeasure2} apply. These critical points will be the object of a future investigation.

\section{The dynamics near a critical point}
\label{sec-Dyn}

This section begins with a review of the notations and prepartory statements, some of which are already contained in the prior works~\cite{DS,DSS}. In particular, Section~\ref{sec-DefinitionCritical} offers a broader perspective on critical points in one-dimensional random media and states the main technical assumptions, Section~\ref{sec-Examples} provides some examples, then Section~\ref{sec-DysonSchmidt} recalls and extends facts on the random dynamics of Dyson-Schmidt variables in the presence of hyperbolic critical points. These facts are transposed to the logarithmic Dyson-Schmidt variables in Section~\ref{sec-LogDysonSchmidt}, namely the form in which  they will be applied in Section~\ref{sec:mu}. Finally, Section~\ref{sec-RescaledLogarithmic} provides a perturbative formula for the dynamics of the rescaled logarithmic Dyson-Schmidt variables.

\subsection{Definition of critical points}
\label{sec-DefinitionCritical}

The random matrices in~\eqref{eq:expansion} are the product of one matrix that is close to the identity and a random diagonal matrix $D^0=\diag(\kappa,\kappa^{-1})$. Modifications of this structural property in the sense of the following definition appear in several one-dimensional random problems. 

\begin{definition}
\label{def-CriticalEnergy}
Let $(\Sigma,{\bf p})$ be a probability space and, for each $\sigma\in\Sigma$, suppose given a sufficiently smooth map $E\in\mathbb{R}\mapsto \mathcal{T}^E_\sigma\in\mbox{\rm SL}(2,\mathbb{R})$. Then, $E_c\in\mathbb{R}$ is a critical point {\rm (}or critical energy{\rm )} of the family  $(\mathcal{T}^E_\sigma)_{\sigma\in\Sigma}$ if the matrices $(\mathcal{T}^{E_c}_{\sigma})_{\sigma\in\Sigma}$ all commute. The critical point is then called 

\vspace{-.2cm}

\begin{itemize}

\item[{\rm (i)}] elliptic if $|\Tr(\mathcal{T}^{E_c}_{\sigma})|<2$ for all $\sigma\in\Sigma$;

\vspace{-.2cm}

\item[{\rm (ii)}] parabolic if $|\Tr(\mathcal{T}^{E_c}_{\sigma})|=2$ and $\mathcal{T}^{E_c}_{\sigma}$ is the same non-trivial Jordan block for all $\sigma\in\Sigma$;

\vspace{-.2cm}

\item[{\rm (iii)}] hyperbolic if $|\Tr(\mathcal{T}^{E_c}_{\sigma})|>2$ for a set of $\sigma\in\Sigma$ of positive measure.

\end{itemize}
\end{definition}

The two-site transfer matrix of the random hopping model given in~\eqref{eq-TransferHopping} hence has $E_c=0$ as a hyperbolic critical energy. More generally, if $E_c$ is a critical point of a family $E\mapsto \mathcal{T}^E_\sigma$, then one can find an invertible real $2\times 2$ matrix $M$ such that $M\mathcal{T}^{E_c}_\sigma M^{-1}$ is a rotation matrix, a standard Jordan block or a diagonal matrix in the elliptic, parabolic and hyperbolic cases respectively. In this representation one can then expand the so-called $M$-modified transfer matrices $T^\epsilon_\sigma=M\mathcal{T}^{E_c+\epsilon}_{\sigma} M^{-1}$ exactly as in~\eqref{eq:expansion} with zeroth order on the r.h.s. given either by random rotation matrices, a standard Jordan block or random diagonal matrices $D^0=\diag(\kappa,\kappa^{-1})$. If $T^\epsilon_\sigma$ is obtained as a product of a finite number of one-site transfer matrices (such as in~\eqref{eq-TransferHopping} where two one-site transfer matrices appear), one speaks of a random polymer model~\cite{JSS,DS}. Note that for the random hopping model the notation $T^\epsilon_\sigma=M\mathcal{T}^{E_c+\epsilon}_{\sigma} M^{-1}$ is consistent with~\eqref{eq:expansion} because $M$ is the identity and $E_c=0$. 

\vspace{.2cm}

Let us briefly give a few examples of critical energies. For a random polymer model in which $\Sigma$ only consists of two points, it generically happens that there exists an elliptic critical energy~\cite{DWP,JSS}. At such an elliptic critical energy $E_c$, it is then known that the Lyapunov exponent  behaves like $\gamma^{\epsilon}=C\epsilon^2+\Oo(\epsilon^3)$ with a computable constant $C>0$~\cite{JSS}. On the other hand, parabolic critical energies appear at the band edges of random Jacobi matrices (such as the one-dimensional Anderson model) and lead to a rich scaling behavior for the Lyapunov exponent in their vicinity~\cite{SaS}. Also the lower band edges of certain random Kronig-Penney models are parabolic critical energies and their Lyapunov exponent then behaves for $\epsilon>0$ like $\gamma^{-\epsilon}=C_- \epsilon+\Oo(\varepsilon^{\frac{3}{2}})$ and $\gamma^{\epsilon}=C_+\epsilon^{\frac{1}{2}}+\Oo(\epsilon)$ to the outside and inside of the spectrum respectively, again with computable constants $C_\pm>0$~\cite{DKS}. This paper is about hyperbolic critical points which are further distinguished in several cases.

\begin{definition}
\label{def-HypCriticalEnergy}
A hyperbolic critical point of $E\in\mathbb{R}\mapsto \mathcal{T}^E_\sigma\in\mbox{\rm SL}(2,\mathbb{R})$ is called

\vspace{-.2cm}

\begin{itemize}

\item[{\rm (i)}] unbalanced if $\EM(\log(\kappa_\sigma))\not=0$;

\vspace{-.2cm}

\item[{\rm (ii)}] balanced  if $\EM(\log(\kappa_\sigma))=0$;

\vspace{-.2cm}

\item[{\rm (iii)}] balanced  of rotating type if $\EM(\log(\kappa_\sigma))=0$ and $a_\sigma>|b_\sigma|$ almost surely;

\vspace{-.2cm}

\item[{\rm (iv)}] balanced of confined type if $\EM(\log(\kappa_\sigma))=0$ and $b_\sigma>|a_\sigma|$ almost surely. 

\end{itemize}
\end{definition}

In the unbalanced case the Lyapunov exponent is simply given by $\gamma^{\epsilon}=\EM(\log(\kappa_\sigma))+o(\epsilon)$~\cite{DS,GG}. Here the focus is on balanced hyperbolic critical points for which several examples will be given in Section~\ref{sec-Examples}. Let us note that one also has a balanced hyperbolic critical point of rotating type if $a_\sigma<-|b_\sigma|$ almost surely, and of confined type if $b_\sigma<-|a_\sigma|$ almost surely, but these cases reduce to the above after a conjugation of $T^\epsilon_\sigma$ with $J=\diag(1,-1)$.

\vspace{.2cm}

In the remainder of the paper, several quantitative bounds on the matrix entries of $T^\epsilon_\sigma$ as well as the rotating or confined type of the balanced hyperbolic critical point will be used. To state these bounds and fix the corresponding constants, let us rewrite~\eqref{eq:expansion} as
\begin{equation}
\label{eq:expansion2}
T^\epsilon_\sigma
\;=\;
J\,Q^\epsilon_\sigma \,D^\epsilon_\sigma \, J
\;,
\end{equation}
where $J=\diag(1,-1)$ as above and
\begin{align}
\label{eq-QFormula}
\begin{aligned}
D^\epsilon_\sigma
& 
\;=\;
\begin{pmatrix}
\kappa_\sigma(1+\epsilon c_\sigma) & 0 \\
0 & (\kappa_\sigma(1+\epsilon c_\sigma)
)^{-1}
\end{pmatrix}
\;,
\\
\qquad
Q^\epsilon_\sigma
& 
\;=\;
\mathbf{1}\,-\,\epsilon a_\sigma
\begin{pmatrix}
0 & -1 \\
1 & 0
\end{pmatrix}
\,-\,\epsilon b_\sigma
\begin{pmatrix}
0 & 1 \\
1 & 0
\end{pmatrix}
\,+\,
\epsilon^2 \,A^\epsilon_\sigma 
\;,
\end{aligned}
\end{align}
with $A^\epsilon_\sigma$ being a random $2\times 2$ matrix, bringing the term $\Oo(\epsilon^2)$ in~\eqref{eq:expansion} into an explicit form. This corresponds to (17) in~\cite{DSS}, except for the additional assumption $c = 0$ made there. The factor $J$ in~\eqref{eq:expansion2} is merely inserted for notational convenience because further down it brings the dynamics~\eqref{eq-DysonDyn} into the particularly simple form~\eqref{eq-TwoStepDyn}. The following technical assumptions are assumed to hold throughout this work:

\vspace{.2cm}

\noindent {\bf Main Hypothesis:} {\it The family $\epsilon\mapsto T^\epsilon_\sigma$ of random matrices have a balanced hyperbolic critical point at $\epsilon=0$ of the form~\eqref{eq:expansion} with random variables $\kappa_\sigma>0$, $a_\sigma$, $b_\sigma$, $c_\sigma$ and $A^\epsilon_\sigma$ all having distributions with compact support. The random variable $\log(\kappa_\sigma)$ is balanced, namely $\EM(\log(\kappa_\sigma))=0$ and its  distribution is supposed to be non-trivial in the sense that  ${\bf p}\big(\{\sigma\in\Sigma:\log(\kappa_\sigma) > 0\}\big)>0$. Furthermore,  with $\esssup$ and $\essinf$ w.r.t. ${\bf p}$, let us introduce the finite constants
$$
C_0 
\;:=\;
\esssup \,|\log(\kappa_\sigma)| \,\in\, (0,\infty)\,,
$$
and
$$
C_1 \;:= \;\essinf_{\sigma \in \Sigma}(a_{\sigma} - |b_{\sigma}|)\,,
\qquad
C'_1 \;:= \;\essinf_{\sigma \in \Sigma}(b_{\sigma}-|a_{\sigma}|)\,,
$$
as well as 
$$
C_2 \;:=\; \esssup_{\sigma \in \Sigma}(|a_{\sigma}| + |b_{\sigma}| +|c_{\sigma}|)\,,\quad
C_3\; :=\; \sup_{|\epsilon|\leq 1}\esssup_{\sigma \in \Sigma}\|A_{\sigma}^{\epsilon}\|\,.
$$
For a balanced hyperbolic critical point of rotating type, it is supposed that $C_1>0$, while for a confining type, it is supposed that $C'_1>0$.}

\vspace{.2cm}

\noindent {\bf Notations and conventions:} For sake of simplicity, we focus on $\epsilon>0$ throughout the paper. To improve readability from this point on, random variables like $\kappa_\sigma$ or $A_\sigma^\epsilon$ will often not contain the index $\sigma$ or even the index $\epsilon$, whenever confusion seems unlikely. We will also write $\kappa_n$ or $A_n^\epsilon$ for $\kappa_{\sigma_n}$ or $A_{\sigma_n}^\epsilon$ from now on. Moreover, we will denote sets like $\{\sigma\in\Sigma:\log(\kappa_\sigma) > 0\}$ simply by $\{\log(\kappa) > 0\}$ and then denote the indicator functions on such a set by $\mathbf{1}\big[\log(\kappa) > 0\big]$. It will also be useful to introduce the centered random variable 
\begin{equation}
\label{eq-ChiVar}
\ChiVar\; :=\; \frac{1}{C_0}\log(\kappa)
\;.
\end{equation}
According to the Main Hypothesis it satisfies $|\ChiVar| \leq 1$ ${\bf p}$-a.s. and the support of $\ChiVar$ contains a positive and a negative real number and furthermore either $-1$ or $1$. 

\subsection{Examples of hyperbolic critical points}
\label{sec-Examples}

A first example of a hyperbolic critical energy appears in the random hopping model already described in Section~\ref{sec-Intro}. If the distribution of the hopping elements $t_{n}$ on the even and odd sites is different, one is typically in the unbalanced case which is studied in more detail in~\cite{DS}. If the hopping terms are all i.i.d., then $\EM(\log(\kappa_n))=\EM(\log(t_{2n}))-\EM(\log(t_{2n-1}))=0$ and the hyperbolic critical energy is balanced. Moreover,~\eqref{eq-TransferHopping} shows that it is of rotating type. As already claimed above, Theorems~\ref{theo-Lyapunov} and~\ref{theo-InvMeasure} therefore apply.

\vspace{.2cm}

As already stressed in the prior work~\cite{DSS}, balanced hyperbolic critical energies always appear at topological phase transitions in chiral one-dimensional (and, more generally, one-channel) topological insulators. At these phase transitions, the critical energy is then $E_c=0$, the center of the spectrum. There are then structural arguments~\cite[Proposition~3]{DS} showing that for these hyperbolic critical energies the coefficients in the expansion~\eqref{eq:expansion} satisfy the deterministic inequality $a\geq \sqrt{b^2+c^2}$, implying that these critical energies are of rotating type. A concrete example is the disordered Su-Schrieffer-Heeger model~\cite{SSH,MSHP}, but also more general models in this class can be analyzed if one works with reduced transfer matrices~\cite{DSS}. As for the random hopping model, the inverse of the Lyapunov exponent defined by~\eqref{eq-LyapDef} is the localization length only up to a factor given by the average polymer length (see~\cite{JSS,DS,DSS} for details). 

\vspace{.2cm}

Next let us consider a random Dirac operator $H=\gamma_2 \imath \partial_x+\sum_{n\in\ZM}W_n\delta_n$ on $L^2(\mathbb{R},\CM^2)$ where $W_n=(W_n)^*$ are i.i.d. real $2\times 2$ matrices and $\gamma_1,\gamma_2,\gamma_3$ denote the standard Pauli matrices. The rigorous definition of the operator is implemented by boundary conditions $\lim_{\varepsilon\downarrow 0}\psi(n+\varepsilon)=e^{\imath\gamma_2 W_n}\lim_{\varepsilon\downarrow 0}\psi(n-\varepsilon)$ on sufficiently smooth $\psi\in L^2(\mathbb{R},\CM^2)$, see~\cite{SS} for details. Let us now implement the chiral symmetry $\gamma_3 H\gamma_3=-H$ and suppose that $\EM(W_n)=0$. This implies that $W_n$ is off-diagonal with a random entry $w_n$. The fundamental solution over $[n-1,n)$ at energy $\epsilon$ is then given by
$$
T^\epsilon_n
\;=\;
e^{-\imath\gamma_2\epsilon} \,e^{\imath\gamma_2 W_n} 
\;=\;
\left[\mathbf{1}\,+\,\epsilon
\begin{pmatrix}
0 & -1 \\
1 & 0
\end{pmatrix}
\,+\,
\Oo(\epsilon^2)
\right]
\begin{pmatrix}
e^{w_n} & 0 \\ 0 &e^{-w_n}
\end{pmatrix}
\;.
$$
This is hence of the form~\eqref{eq:expansion}, showing that the chiral random Dirac operator has a balanced hyperbolic critical energy at $E_c=0$, which is clearly of rotating type. Hence the Lyapunov exponent is again given by Theorem~\ref{theo-Lyapunov}.

\vspace{.2cm}

Finally, let us come to an example of a balanced hyperbolic critical point of confined type. For that purpose, let us consider the partition function of a classical Ising chain with spin coupling $J>0$ and random external magnetic field $(h_n)_{n\in\ZM}$ which, at inverse temperature $\beta=1$, volume $N$ and with periodic boundary conditions, is given by
\begin{align*}
Z_{\vphantom{\widehat{k}}N}(J)
& 
\;=\;
\sum_{\sigma\in\{-1,1\}^N}
\exp\left(-\sum_{1\leq n\leq N}J\sigma_n\sigma_{n+1}-\sum_{1\leq n\leq N}h_n\sigma_n
\right)
\;,
\end{align*}
where the outer sum runs over all spin configurations $\sigma$ and $\sigma_{N+1}=\sigma_1$, assuring periodic boundary conditions. One is then interested in the free energy density
$$
f(J)
\;=\;
\lim_{N\to\infty}\,\frac{1}{N}\,\log\big(Z_{\vphantom{\widehat{k}}N}(J)\big)
\;,
$$
and its behavior in the limit of strong coupling $J\to\infty$. As is well-known and easy to check ({\it e.g.}~\cite{DH}), the partition function can be rewritten as the trace of a product of random $2\times 2$ matrices:
\begin{align*}
Z_{\vphantom{\widehat{k}}N}(J)
& 
\;=\;
\Tr
\left(
\prod_{n=1}^N
\begin{pmatrix} 
e^{J} & e^{-J} \\ e^{-J} & e^{J}
\end{pmatrix}
\begin{pmatrix} 
e^{h_n} & 0 \\ 0 & e^{-h_n}
\end{pmatrix}
\right)
\\
&
\;=\;
\left(e^{2J} -e^{-2J}\right)^{\frac{N}{2}}
\Tr
\left(
\prod_{n=1}^N
(1-e^{-4J})^{-\frac{1}{2}}
\begin{pmatrix} 
1 & e^{-2J} \\ e^{-2J} & 1
\end{pmatrix}
\begin{pmatrix} 
e^{h_n} & 0 \\ 0 & e^{-h_n}
\end{pmatrix}
\right)
\;.
\end{align*}
In the second equality, the factor was taken out so that inside of the trace appears a product of random matrices with unit determinant given by
\begin{align*}
T^\epsilon_n
& 
\;=\;
(1-e^{-4J})^{-\frac{1}{2}}
\begin{pmatrix} 
1 & e^{-2J} \\ e^{-2J} & 1
\end{pmatrix}
\begin{pmatrix} 
e^{h_n} & 0 \\ 0 & e^{-h_n}
\end{pmatrix}
\;=\;
\left[
\one
\;+\;
\epsilon
\begin{pmatrix}
0 & 1 \\ 1 & 0
\end{pmatrix}
\;+\;
\Oo(\epsilon^2)
\right]
\begin{pmatrix}
\kappa_n & 0 \\ 0 & \tfrac{1}{\kappa_n}
\end{pmatrix}
\;,
\end{align*}
where we set $\epsilon=e^{-2J}$ and $\kappa_n=e^{h_n}$. Hence, one has a product of random matrices of the form~\eqref{eq:expansion} with $a=0$, $b=1$ and $c=0$. If $\gamma^\epsilon$ denotes the Lyapunov exponent associated to this random product, the free energy density is thus given by
$$
f(J)
\;=\;
\tfrac{1}{2}\log(e^{2J}-e^{-2J})\,+\,\gamma^\epsilon
\;,
\qquad
\epsilon\,=\,e^{-2J}
\;.
$$
While the first summand is $J(1+\Oo(\epsilon))$ and thus dominates the second, it is still of interest to compute the Lyapunov exponent. The small parameter is now $\epsilon=e^{-2J}$ and has a hyperbolic critical point. It is balanced if $\EM(h_n)=0$ and, moreover, of confined type because $b=1$ and $a=0$. Hence, Theorem~\ref{theo-InvMeasure2} can be applied to the random field Ising chain in this situation.

\subsection{Random dynamics in Dyson-Schmidt variables}
\label{sec-DysonSchmidt}

It is well-known ({\it e.g.} \cite{BL,PF,JSS,DSS}) that the action of invertible real $2\times 2$ matrices on $\mathbb{R}\mbox{\rm P}(1)$ that appears on the r.h.s. of~\eqref{eq-FurstenbergDef} is, under the stereographic projection~\eqref{eq-DysonSchmidt},  implemented by the standard M\"obius transformation $\binom{\alpha\;\beta}{\gamma\;\delta}\cdot x=\frac{\alpha x+\beta}{\gamma x+\delta}$ on the Dyson-Schmidt variables. Associated to an i.i.d. sequence $(T^\epsilon_n)_{n\in\NM}$ of matrices given by~\eqref{eq:expansion} with an initial condition $x_0\in\dot{\mathbb{R}}$, one hence obtains a random dynamical system $(x^\epsilon_n)_{n\geq 0}$ by
\begin{equation}
\label{eq-DysonDyn}
x^\epsilon_n\;=\;-T^\epsilon_n\cdot (-x^\epsilon_{n-1})
\;,
\end{equation}
in which again the overall sign in~\eqref{eq:expansion} is irrelevant. The minus signs in~\eqref{eq-DysonDyn} result from the sign in~\eqref{eq-DysonSchmidt} and maintain the orientation in the below. These sign changes are also implemented by a M\"obius transformation, namely $J\cdot x=-x$ for $J=\diag(1,-1)$ so that $x^\epsilon_n=JT^\epsilon_nJ\cdot x^\epsilon_{n-1}$. As $T^\epsilon_n$ is of the form~\eqref{eq:expansion2} and given by a product of matrices, the group action property of the M\"obius transformation shows
\begin{equation}
\label{eq-TwoStepDyn}
x^\epsilon_n\;=\;Q^\epsilon_n\cdot (D^\epsilon_n \cdot x^\epsilon_{n-1})
\;.
\end{equation}
This explains why it is advantageous to include the factor $J$ in~\eqref{eq:expansion2}. The random dynamics~\eqref{eq-TwoStepDyn} is precisely the two-step dynamics of~\eqref{eq-TwoStepDynPhases} in the $x$-picture, namely after the transformation~\eqref{eq-DysonSchmidt}. It is analyzed in great detail in the references~\cite{DS,DSS} and this section  reproduces and appends several facts that are relevant for the present work.  The formula~\eqref{eq-TwoStepDyn} shows that the dynamics is given by the alternation of a diagonal M\"obius action $x \mapsto D^\epsilon\cdot x=\kappa^2(1+\epsilon c)^2x$ followed by a perturbation $Q^\epsilon \cdot $ of the order of $\epsilon$. For all realizations, the action $D^\epsilon\cdot$ has two fixed points $0$ and $\infty$ (corresponding to $\theta=0$ and $\theta=\frac{\pi}{2}$ in $\mathbb{R}\mbox{\rm P}(1)$) and leaves the two intervals $(-\infty,0)$ and $(0,\infty)$ invariant.  Due to $a>0$, the $\epsilon$-dependent perturbation $Q^\epsilon \cdot$ only leads to passages through $0$ from left to right, and from $+\infty$ to $-\infty$ (for $\epsilon>0$), so that the random dynamics enters the intervals $(-\infty,0)$ and $(0,\infty)$ only from the left. The random times at which passages are completed are given by
\begin{align}
\label{eq-PassageTimes}
\{N\geq 1 \,:\, \sgn(x_{\vphantom{\widehat{k}}N-1}) \neq \sgn(x_{\vphantom{\widehat{k}}N})\} &= \{N\geq 1 \,:\, x_{\vphantom{\widehat{k}}N-1} \leq 0 < x_{\vphantom{\widehat{k}}N} \mbox{ or }   x_{\vphantom{\widehat{k}}N} \leq 0 < x_{\vphantom{\widehat{k}}N-1}\}
\,,
\end{align}
and its order statistics are denoted by $N_0 < N_1  < N_2 < \dots$. For (without loss of generality) $x_{N_0} > 0$, the first run through $(0,\infty)$ takes $N_1 - N_0$ steps, the first one through $(-\infty,0)$ takes $N_2 - N_1$ steps, and so on in an alternating manner so that during the $k$-th passage the sign $\nu$ in \eqref{eq-normalizedLogDysonSchmidt} is given by $\nu=(-1)^k$. Then  $(N_{k+1}-N_k)_{k\geq 0}$ are called the random passage times. They are neither independent nor identically distributed as they depend on the initial condition $x_{N_k}$ of the passage which in turn depends on the full history. In order to deal with this difficulty, Section~\ref{sec:mu} introduces a slower and a faster comparison process (similar to the constructions in~\cite{DSS}). Up to errors, this allows to decouple the passages so that renewal theory can be applied in the following. For the computation of the invariant measure (Theorem~\ref{theo-InvMeasure}) it will then be relevant to control the dynamics within each passage. As the passages are either through $(0,\infty)$ or $(-\infty,0)$, two cases have to be considered. It is, however, possible to reduce the analysis on the negative interval to that of the positive one by applying the orientation preserving bijection $x\in(-\infty,0)\mapsto -\frac{1}{x}=J'\cdot x\in(0,\infty)$ where $J'=\binom{0\,-1}{1\;\;0}$. Indeed, $J'^*D^\epsilon J'=-(D^\epsilon)^{-1}$ and $J'^*Q^\epsilon J'$ merely has a changed sign before $b$ in~\eqref{eq-QFormula} and the higher order term $A^\epsilon$ is conjugated by $J'$, but the sign before $a$ does not change. Hence the Main Hypothesis directly transposes (note that this does not hold in the unbalanced case dealt with in~\cite{DS,DSS} because $\EM\,\log(\kappa)$ and $\EM\,\log(\frac{1}{\kappa})$ then have a different sign). In the remainder of this paper, only the dynamics on $(0,\infty)$ in the $x$-picture will be analyzed. 

\vspace{.2cm}

The lemmata below provide quantitative deterministic statements on passages of the dynamics through $(0,\infty)$. The first lemma states a monotonicity property of the perturbation. It is stated and proved in~\cite[Lemma~4]{DSS}: 

\begin{lemma}
\label{lemma:Q-lower}
For all realizations and $\epsilon>0$, $x \in [0,\infty)$ and $Q^\epsilon \cdot x \geq 0$ imply $Q^\epsilon  \cdot x \geq x$.
\end{lemma}

The next lemma  will provide lower bounds on the dynamics. It will allow to construct the lower comparison process in Section~\ref{sec:slower} below. This latter process as well as all quantities associated to it will carry a hat. Let us  introduce the points
\begin{equation}
\label{eq-xHats}
\widehat{x}_- 
\;:=\; 
\tfrac{C_1\epsilon}{2}\,,
\qquad \widehat{x}_c 
\;:=\; 
\tfrac{C_1\epsilon}{2}(e^{-2C_0}+1)\,,
\qquad 
\widehat{x}_+ 
\;:=\; 
\tfrac{2e^{2C_0}}{C_1\epsilon}\,.
\end{equation}
\begin{figure}
\begin{center}
\begin{tikzpicture}[line join = round, line cap = round]
\coordinate (a) at (0.0,0.3);
\coordinate (b) at (0.0,-0.2);
\coordinate (bb) at (0.0,-0.8);
\coordinate (r) at (0.2,0.0);
\coordinate (l) at (-0.2,0.0);
\coordinate (l0) at (-6.5,0);
\vtick{(l0)};
\coordinate (l0a) at ($(l0) + (a)$);
\coordinate[label=below:{$0$}] (l0b) at ($(l0) + (b)$);
\coordinate (l1) at (-5.5,0);
\vtick{(l1)};
\coordinate (l1a) at ($(l1) + (a)$);
\coordinate (l1ar) at ($(l1) + (a) + (r)$);
\draw[->] (l0a) to[out=90, in=105, looseness=1.5] node[midway,above,inner sep=4pt] {\eqref{stat:slower-start}} (l1ar);
\coordinate[label=below:{$\widehat{x}_-$}] (l1b) at ($(l1) + (b)$);
\coordinate (l2) at (-4.5,0);
\vtick{(l2)};
\coordinate (l2ar) at ($(l2) + (a) + (r)$);
\draw[->] (l1a) to[out=90, in=105, looseness=1.5] node[midway,above,inner sep=4pt] {\eqref{stat:slower-between}} (l2ar);
\coordinate[label=below:{$\widehat{x}_c$}] (l2b) at ($(l2) + (b)$);
\coordinate (l3) at (-4,0);
\coordinate (l4) at (-3,0);
\coordinate (l5) at (-2.5,0);
\vtick{(l5)};
\coordinate (l5a) at ($(l5) + (a)$);
\coordinate[label=below:{$\widehat{x}_+$}] (l5b) at ($(l5) + (b)$);
\coordinate (lr) at (-1.5,0);
\coordinate[label=left:{$x$}] (lrr) at (-1,0);
\coordinate (lrra) at ($(lrr) + (a)$);
\draw[->] (l5a) to[out=90, in=120, looseness=1.5] node[midway,above,inner sep=4pt] {\eqref{stat:slower-end}} (lrra);
\draw [-,color=black,line width=0.3mm] (l0)--(l3);
\draw [-,color=black,dotted,line width=0.3mm] (l3)--(l4);
\draw [->,color=black,line width=0.3mm] (l4) -- (lr);
\coordinate (r0) at (1.5,0);
\tick{(r0)};
\coordinate (r0a) at ($(r0) + (a)$);
\coordinate[label=below:{$0$}] (r0b) at ($(r0) + (b)$);
\coordinate (r1) at (2.5,0);
\tick{(r1)};
\coordinate (r1a) at ($(r1) + (a)$);
\coordinate (r1al) at ($(r1) + (a) + (l)$);
\draw[->] (r0a) to[out=90, in=75, looseness=1.5] node[midway,above,inner sep=4pt] {\eqref{stat:faster-start}} (r1al);
\coordinate[label=below:{$\widetilde{x}_-$}] (r1b) at ($(r1) + (b)$);
\coordinate (r2) at (3.5,0);
\tick{(r2)};
\coordinate (r2al) at ($(r2) + (a) + (l)$);
\draw[->] (r1a) to[out=90, in=75, looseness=1.5] node[midway,above,inner sep=4pt] {\eqref{stat:faster-between}} (r2al);
\coordinate[label=below:{$\widetilde{x}_c$}] (r2b) at ($(r2) + (b)$);
\coordinate (r3) at (4,0);
\coordinate (r4) at (5,0);
\coordinate (r5) at (5.5,0);
\tick{(r5)};
\coordinate (r5a) at ($(r5) + (a)$);
\coordinate[label=below:{$\widetilde{x}_+$}] (r5b) at ($(r5) + (b)$);
\coordinate (rr) at (6.5,0);
\coordinate[label=left:{$x$}] (rrr) at (7.0,0);
\coordinate (rrra) at ($(rrr) + (a)$);
\draw[->] (r5a) to[out=90, in=120, looseness=1.5] node[midway,above,inner sep=4pt] {\eqref{stat:faster-end}} (rrra);
\draw [-,color=black,line width=0.3mm] (r0)--(r3);
\draw [-,color=black,dotted,line width=0.3mm] (r3)--(r4);
\draw [->,color=black,line width=0.3mm] (r4) -- (rr);
\end{tikzpicture}
\caption{\it In both pictures, the arrows illustrate properties of the dynamics in the Dyson-Schmidt coordinates on $(0,\infty)$ as stated in {\rm Lemmata~\ref{lemma:slower}} {\rm (}on the left{\rm )} and {\rm~\ref{lemma:faster}} {\rm (}on the right{\rm )}.}
\label{fig:slower}
\end{center}
\end{figure}
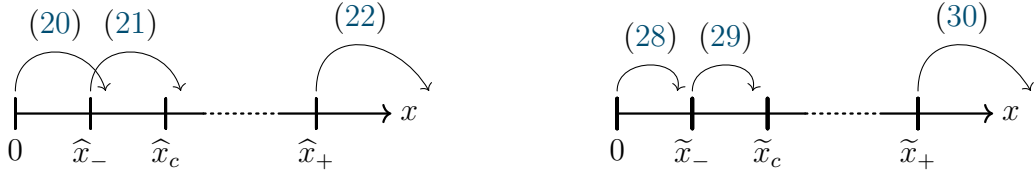
%

\begin{lemma}\label{lemma:slower}
For all realizations and $\epsilon>0$ small enough, one has
\begin{align}
\label{stat:slower-start}
\hspace{80pt}&x \in [0,\infty) &&\Longrightarrow && Q^\epsilon  \cdot (D^\epsilon \cdot x) \notin [0,\widehat{x}_-)\,,\hspace{80pt}\\\label{stat:slower-between}
\hspace{80pt}&x \in [\widehat{x}_-,\infty) &&\Longrightarrow &&Q^\epsilon  \cdot (D^\epsilon \cdot x) \notin [0,\widehat{x}_c)\,,\hspace{80pt}\\\label{stat:slower-end}
\hspace{80pt}& x \in [\widehat{x}_+,\infty) &&\Longrightarrow &&  Q^\epsilon  \cdot (D^\epsilon \cdot x) \in (-\infty,0) \,.\hspace{80pt}
\end{align}
\end{lemma}

Lemma~\ref{lemma:slower} is~\cite[Lemma~5]{DSS} where a proof is given (a minor modification is needed to account for the contribution of $c$ in $D^\epsilon$, which was set to $0$ in~\cite{DSS}). The claims of the lemma are graphically illustrated in Figure~\ref{fig:slower}. As stated above, these bounds are needed to construct a slower comparison process in Section~\ref{sec:mu}. The remainder of the section consists of bounds that are needed to also control a faster comparison process. For notational purposes, it will be useful to introduce the quantity
\begin{equation}
\label{eq-deltaDef}
\ParaEps 
\;:=\; 
\frac{1}{\log(\epsilon^{-1})}\,.
\end{equation}
Note that $\ParaEps>0$ for $\epsilon>0$ sufficiently small, and that $\ParaEps\to 0$ for $\epsilon \to 0$, even though $\epsilon\ll\delta^\alpha$ for any $\alpha\geq 1$. The next result shows that in the $x$-picture, up to some constant factor, the full action of $Q^\epsilon D^\epsilon$ can be bounded by that of $D^0$ on a large interval (recall that $D^0\cdot$ is just the multiplication by $\kappa^2$ in the $x$-picture).

\begin{lemma}
\label{lemma:Q-upper}
There exists a constant $C$ depending on $C_0$, $C_2$, $C_3$ such that after setting
\begin{equation}\label{eq:faster-x-lims}
\widetilde{x}_- 
\;:= \;
\frac{C\epsilon}{\delta^2}\,,\qquad
\widetilde{x}_+ 
\;:= \;
\frac{\delta^2}{C\epsilon}\,,
\end{equation}
for $x \in [\widetilde{x}_-,\widetilde{x}_+]$ it follows for $\epsilon > 0$ small enough and all realizations that
$$
e^{-2C_0\delta^2}(D^0 \cdot x) 
\;\leq\;
Q^\epsilon  \cdot (D^\epsilon \cdot x) 
\;\leq\; 
e^{2C_0\delta^2}(D^0 \cdot x)\;.
$$
\end{lemma}

\noindent{\bf Proof.} It will be shown that it is possible to take $C = \frac{8e^{3C_0}C_2}{C_0}$. As $D^\epsilon \cdot x \in [e^{-3C_0}x,e^{3C_0}x]$ and $D^0(D^{\epsilon})^{-1} \cdot x \in [e^{-C_0\delta^2}x,e^{C_0\delta^2}x]$ for all $x \in (0,\infty)$, it suffices to show that $x \in
[e^{-3C_0}\widetilde{x}_-,e^{3C_0}\widetilde{x}_+]$ implies
$e^{-C_0\delta^2}x \leq Q^\epsilon \cdot x \leq e^{C_0\delta^2}x$. Therefore take $x \in
[e^{-3C_0}\widetilde{x}_-,e^{3C_0}\widetilde{x}_+]$. Now $Q^\epsilon
\cdot x \leq e^{C_0\delta^2}x$ is (writing some contributions as error
terms)
$$
\frac{(1+\mathcal{O}(\epsilon^2))x + (a-b)\epsilon +
\mathcal{O}(\epsilon^2)}{1+\mathcal{O}(\epsilon^2)-(a+b+\mathcal{O}(\epsilon))\epsilon
x} \;\leq \;e^{C_0\delta^2}x\,,
$$
which is equivalent to the following inequality (again for all possible
realizations)
\begin{gather}\label{ineq:Q-upper}
(a+b+\mathcal{O}(\epsilon))\epsilon e^{C_0\delta^2}x^2 -
(e^{C_0\delta^2}-1 +
\mathcal{O}(\epsilon^2))x + (a-b)\epsilon + \mathcal{O}(\epsilon^2) \;\leq\;
0\,.
\end{gather}
The latter is indeed shown (estimating all random variables by the
Main Hypothesis) by
\begin{align}\label{ineq:Q-upper-bis}
\begin{aligned}
 &(a+b+\mathcal{O}(\epsilon))e^{C_0\delta^2}\epsilon x^2 -
(e^{C_0\delta^2}-1 +
\mathcal{O}(\epsilon^2))x + (a-b)\epsilon + \mathcal{O}(\epsilon^2)\\
&\qquad\leq\;
(C_2+\mathcal{O}(\epsilon))(1+\mathcal{O}(\delta^2))\epsilon x^2 -
(C_0\delta^2 + \mathcal{O}(\epsilon^2))x + C_2\epsilon +
\mathcal{O}(\epsilon^2)\\
&\qquad\leq \;2C_2\epsilon x^2 - \frac{C_0}{2}\delta^2 x + 2C_2\epsilon\\
&\qquad \leq\; 2C_2\epsilon x^2 - \left(\frac{C_0\delta^2}{4}+\frac{16C_2^2\epsilon^2}{C_0\delta^2}\right)x + 2C_2\epsilon
\\
& \qquad =\;
2C_2\epsilon(x-e^{-3C_0}\widetilde{x}_-)(x-e^{3C_0}\widetilde{x}_+)\\
&\qquad\leq\; 0
\end{aligned}
\end{align}
for $x \in [e^{-3C_0}\widetilde{x}_-,e^{3C_0}\widetilde{x}_+]$ and for
$\epsilon$ (and then also $\delta$) small enough.

\vspace{.1cm}

It remains to be shown that $x \in [e^{-3C_0}\widetilde{x}_-,e^{3C_0}\widetilde{x}_+]$ implies $Q^\epsilon \cdot x \geq e^{-C_0\delta^2}x$. The latter is equivalent to~\eqref{ineq:Q-upper} after replacing $\delta^2$ by $-\delta^2$ and inserting a global minus sign on the left hand side of this inequality. This modified statement is then once more shown (modifying the first line in the same way) by~\eqref{ineq:Q-upper-bis}.
\hfill $\Box$

\vspace{.2cm}

Now one can further introduce
\begin{equation}
\label{eq-xTildes}
\widetilde{x}_c 
\;:=\; 
e^{2C_0(1+\delta^2)}\widetilde{x}_-\,.
\end{equation}
Then the next result is~\cite[Lemma~10]{DSS} (the proof is identical, even though the definitions of $\widetilde{x}_\pm$ are different here). It is also illustrated in Figure~\ref{fig:slower}, see the right half there.

\begin{lemma}\label{lemma:faster}
For all realizations and $\epsilon>0$ small enough, one has
\begin{align}\label{stat:faster-start}
\hspace{80pt}&x \notin [0,\infty) &&\Longrightarrow &&Q^\epsilon  \cdot (D^\epsilon \cdot x) \notin [\widetilde{x}_-,\infty)\,,\hspace{80pt}\\\label{stat:faster-between}
\hspace{80pt}&x \notin [\widetilde{x}_-,\infty) &&\Longrightarrow &&Q^\epsilon  \cdot (D^\epsilon \cdot x) \notin [\widetilde{x}_c,\infty)\,,\hspace{80pt}\\
\label{stat:faster-end}
\hspace{80pt}&Q^\epsilon  \cdot (D^\epsilon \cdot x) \notin [0,\infty) &&\Longrightarrow &&x \notin [0,\widetilde{x}_+)\,.\hspace{80pt}
\end{align}
\end{lemma}

Recollecting all objects introduced above, one has
\begin{gather}\label{ineq:x}
0 \;<\; \widehat{x}_-\; <\; \widehat{x}_c \;<\; \widetilde{x}_-\; <\; \widetilde{x}_c \;< \;1 \;<\; \widetilde{x}_+ \;<\; \widehat{x}_+ \;<\; \infty\,.
\end{gather}
%

\subsection{Random dynamics of logarithmic Dyson-Schmidt variables}
\label{sec-LogDysonSchmidt}

This section merely spells out the results of Section~\ref{sec-DysonSchmidt} after the transformation~\eqref{eq-VariableTransforms}  to logarithmic Dyson-Schmidt variables. As in the last section, the focus will be merely on positive Dyson-Schmidt variables, so only the $+$ component of~\eqref{eq-VariableTransforms} which reads $x\in(0,\infty)\mapsto y=\frac{1}{2C_0}\log(x)\in\mathbb{R}$. Taking logarithms of the objects in~\eqref{ineq:x}, namely $\widehat{y}_- = \frac{1}{2C_0}\log(\widehat{x}_-)$, $\widehat{y}_c = \frac{1}{2C_0}\log(\widehat{x}_c)$, $\widetilde{y}_- = \frac{1}{2C_0}\log(\widetilde{x}_-) = -\frac{1}{2C_0}\log(\widetilde{x}_+) = -\widetilde{y}_+$, etc., yields the real constants
\begin{gather}
\label{ineq:y}
-\infty\;<\;\widehat{y}_- \;<\; \widehat{y}_c \;<\; \widetilde{y}_-\; <\; \widetilde{y}_c \;<\; 0 \;< \;\widetilde{y}_+ \;<\; \widehat{y}_+\;<\;\infty\,.
\end{gather}
All depend on $\epsilon$ (or, equivalently, on $\delta$). Their limit behavior is described in the next lemma.

\begin{lemma}\label{lemma:y-lims}
For $\epsilon$ small enough, it holds that
\begin{align*}
2C_0\ParaEps\widehat{y}_- & = -1 + \mathcal{O}(\ParaEps)\,,
& 
\quad
2C_0\ParaEps\widehat{y}_c & = -1 + \mathcal{O}(\ParaEps)\,,
&
\quad
2C_0\ParaEps\widehat{y}_+ & = 1 + \mathcal{O}(\ParaEps)\,,
\\
2C_0\ParaEps\widetilde{y}_- & = -1 + \mathcal{O}(\ParaEps\log(\ParaEps))\,,
&\quad
2C_0\ParaEps\widetilde{y}_c & = -1 + \mathcal{O}(\ParaEps\log(\ParaEps))\,,
&\quad
2C_0\ParaEps\widetilde{y}_+ & = 1 + \mathcal{O}(\ParaEps\log(\ParaEps))\,.
\end{align*}
\end{lemma}

\noindent{\bf Proof.} This is immediate from the definition, combined with~\eqref{eq-xHats},~\eqref{eq:faster-x-lims} and~\eqref{eq-xTildes}.
\hfill $\Box$

\vspace{.2cm}

In Section~\ref{sec:mu},  the dynamics will mainly be analyzed in the logarithmic Dyson-Schmidt variables because it takes a particularly simple form. Indeed, the M\"obius action $T^\epsilon \cdot$ takes the form 
\begin{equation}
\label{eq-TAst}
T^\epsilon \ast (y,\nu)
\;=\;
\Big(
\frac{1}{2C_0}\,\log\big(|T^\epsilon \cdot e^{2C_0 y}|\big)
,
\nu\, \sgn(y)\sgn\big(\log\big(|T^\epsilon \cdot e^{2C_0 y}|\big)\big)\Big)
\;.
\end{equation}
Somewhat abusing notations, we will also denote the first component simply by $T^\epsilon \ast y$. Recalling the notation~\eqref{eq-ChiVar}, the two-step dynamics~\eqref{eq-TwoStepDyn} in the $y$-picture therefore becomes
\begin{equation}
\label{eq-TwoStepDyn2}
y_n\;=\;
\ChiVar _n\;+\frac{1}{2C_0}\,\log\big((D^0_n)^{-1}\,Q^\epsilon_n\,D^{\epsilon}_n \cdot e^{2C_0 y_{n-1}}\big)
\;,
\qquad
\nu_n\,=\,\nu_{n-1}\sgn(y_n)\sgn(y_{n-1})
\;,
\end{equation}
which due to $Q^\epsilon_n=\one+\Oo(\epsilon)$ and $D^\epsilon_n = D^0_n + \mathcal{O}(\epsilon)$ leads to $y_n=y_{n-1}+\ChiVar_n+\Oo(\epsilon)$ as long as $|y_{n-1}|=o(\log(\epsilon))$. Up to errors, this is hence a standard random walk on the large interval $(\log(\epsilon),-\log(\epsilon))$ with a maximal step width that is conveniently normalized to $1$, and what happens at the boundaries is analyzed in the following two lemmata which will be used heavily in Section~\ref{sec:mu}. The proofs simply consist of transposing Lemmata~\ref{lemma:Q-lower} to~\ref{lemma:faster} to the logarithmic Dyson-Schmidt variables and are therefore not spelled out. 

\begin{lemma}\label{lemma:Q-lower-bis}
For each realization and $\epsilon$ small enough, $y < \widetilde{y}_+$ implies $Q^\epsilon  \ast y \geq y$.
\end{lemma}

\begin{lemma}\label{lemma:drift-bis}
For all realizations and $\epsilon>0$ small enough, it holds for all $y \in \mathbb{R}$ that
$$
y \in [\widetilde{y}_-,\widetilde{y}_+] \qquad\Longrightarrow\qquad |Q^\epsilon \ast (D^\epsilon \ast y) - D^0 \ast y| \,\leq\, \delta^2\,.
$$
\end{lemma}

\begin{lemma}\label{lemma:y}
For each realization, $y \in \mathbb{R}$ and $\epsilon>0$ small enough, it holds that
\begin{align}\label{stat:slower-faster-y}
\begin{aligned}
y &\in (-\infty,\widetilde{y}_+)  \qquad &&\Longrightarrow \qquad  &((Q^\epsilon D^\epsilon)\ast y) &\in (\widehat{y}_-,\infty)\,,\\
y &\in (\widehat{y}_-,\widetilde{y}_+)  \qquad &&\Longrightarrow \qquad  &((Q^\epsilon D^\epsilon)\ast y) &\in (\widehat{y}_c,\infty)\,,\\
((Q^\epsilon D^\epsilon)\ast y) &\in (\widetilde{y}_-,\infty)  \qquad &&\Longrightarrow \qquad &y &\in (-\infty,\widehat{y}_+)\,,\\
((Q^\epsilon D^\epsilon)\ast y) &\in (\widetilde{y}_c,\infty)  \qquad &&\Longrightarrow \qquad &y &\in (\widetilde{y}_-,\widehat{y}_+)\,.
\end{aligned}
\end{align}
\end{lemma}

\subsection{Rescaled logarithmic Dyson-Schmidt variables}
\label{sec-RescaledLogarithmic}

In this section, the transformation~\eqref{eq-normalizedLogDysonSchmidt} to rescaled logarithmic Dyson-Schmidt variables is briefly analyzed. Focussing again merely on positive Dyson-Schmidt variables, it will hence analyze the map $x\in(0,\infty)\mapsto z=\frac{1}{\log(\epsilon^{-1})}\log(x) =\delta\log(x)$. It only differs from~\eqref{eq-VariableTransforms} by an $\epsilon$-dependent factor. Hence all quantities in~\eqref{ineq:x}, or equivalently~\eqref{ineq:y}, transpose and define $\widehat{z}_-$, $\widetilde{z}_-$, etc. As the dynamics will mainly be controlled in the $y$-picture, they will not be used as frequently though. Therefore also the Lemmata~\ref{lemma:Q-lower} to~\ref{lemma:faster} will not be spelled out, but as the perturbation theory for the Lyapunov exponent will be carried out in the $z$-picture in Section~\ref{sec:gamma}, it will be necessary to implement the M\"obius dynamics. Of course, this is similar to~\eqref{eq-TAst}. The first component of $T^\epsilon \star (z,\nu)$ is
\begin{equation}
\label{eq:defxi-dyn}
T^\epsilon \star z
\;=\;
\frac{\log(T^\epsilon \cdot \epsilon^{-z})}{\log(\epsilon^{-1})}
\;=\;
\delta\; \log(T^\epsilon \cdot \epsilon^{-z})
\;,
\end{equation}
and then~\eqref{eq-DysonDyn} becomes
\begin{equation}
\label{eq-LogDysonDyn}
z^\epsilon_n\;=\;T^\epsilon_n\star z^\epsilon_{n-1}
\;,
\qquad
\nu_n\,=\,\nu_{n-1}\sgn(z_n)\sgn(z_{n-1})
\;.
\end{equation}
For the dynamics~\eqref{eq-TwoStepDyn}, one can now write
\begin{gather}\label{eq:xi-dyn}
T^\epsilon \star z
\;=\; 
z \,+\, \frac{\log(\kappa^2)}{\log(\epsilon^{-1})} \,+\, 
\frac{r^{\epsilon}(z)\epsilon^{1-|z|}}{\log(\epsilon^{-1})}\;,
\end{gather}
in which the remainder $r^{\epsilon}$ simply collects the corrections to the the lowest order random walk-like dynamics~\eqref{eq:xi-dyn}. One then has the following estimate on this remainder:

\begin{lemma}\label{lemma:r}
For $z \in [-Z,Z]$ with fixed $Z \in (0,1)$ and $\epsilon$ sufficiently small, it holds ${\bf p}$-a.s. that
\begin{gather}
\label{ineq:r}
|r^{\epsilon}(z)| \;\leq \;7\,C_2\exp(2C_0) \,+\, \mathcal{O}(\epsilon^{1-|z|})\;.
\end{gather}
\end{lemma}

\noindent\textbf{Proof.} In \eqref{eq:xi-dyn} the dynamics $D^\epsilon$ is split into two pieces $D^0$ and $D^\epsilon(D^0)^{-1}$ of which the latter is included in $r^\epsilon$. It is convenient to redistribute the contribution by writing
$$
T^\epsilon \star z
\;=\; 
z \,+\, \frac{\log(\kappa^2)}{\log(\epsilon^{-1})} \,+\,
\frac{\log((1+\epsilon c )^2)}{\log(\epsilon^{-1})}
\,+\,
\frac{\hat{r}^{\epsilon}(z)\epsilon^{1-|z|}}{\log(\epsilon^{-1})}\;,
$$
by definition of $\hat{r}^\epsilon(z)$. The extra summand can be bounded as
$$
\frac{|\log((1+\epsilon c )^2)|}{\log(\epsilon^{-1})}
\;\leq\;
\frac{2|c|\epsilon}{\log(\epsilon^{-1})}
\;\leq\;
\frac{2C_2\epsilon}{\log(\epsilon^{-1})}
\;\leq\;
\frac{2C_2\epsilon^{1-|z|}}{\log(\epsilon^{-1})}
\;.
$$
The contribution $\hat{r}^{\epsilon}(z)$ is then explicitly given by
\begin{gather}
\label{eq:r}
\hat{r}^{\epsilon}(z) 
\;=\; 
\epsilon^{|z|-1}\log\left[\frac{1+A^\epsilon_{1,1}\epsilon^2 + (a-b-A^\epsilon_{1,2}\epsilon)\kappa^{-2}\epsilon^{1+z}}{1+A^\epsilon_{2,2}\epsilon^2 - (a+b+A^\epsilon_{2,1}\epsilon)\kappa^2\epsilon^{1-z}}\right]\,,
\end{gather}
where $A^\epsilon_{i,j}$, $i,j=1,2$, denote the matrix entries of $A^\epsilon$ defined in~\eqref{eq:expansion2}, and~\eqref{eq:r} holds as long as the argument of the logarithm is an element of $(0,\infty)$. Note that for $z\in[-Z,Z]\subset (-1,1)$, the Main Hypothesis implies that argument of the logarithm is larger than $1$ for $\epsilon$ sufficiently small, so that then also $\hat{r}^{\epsilon}(z) >0$. Again using the Main Hypothesis, one has $\kappa^2\leq e^{2C_0}$ and concludes
\begin{align*}
\hat{r}^{\epsilon}(z) 
&
\;=\; 
\epsilon^{|z|-1}
\log\left[\frac{1 + (|a|+|b|)e^{2C_0}\epsilon^{1+z} + \mathcal{O}(\epsilon^{2-|z|})}{1 - ((|a|+|b|)e^{2C_0}\epsilon^{1-z} + \mathcal{O}(\epsilon^{2-|z|})}\right]
\\
&
\;\leq\; 
\epsilon^{|z|-1}
\log\left[\frac{1 + 2C_2e^{2C_0}\epsilon^{1-|z|} + \mathcal{O}(\epsilon^{2-|z|})}{1 -2C_2 e^{2C_0}\epsilon^{1-|z|} + \mathcal{O}(\epsilon^{2-|z|})}\right]
\\
&
\;=\; 
\epsilon^{|z|-1}\log\left[1 + 4C_2e^{2C_0}\epsilon^{1-|z|} + \mathcal{O}(\epsilon^{2-2|z|})\right]
\\
&
\;\leq \;
4\,C_2\,e^{2C_0} + \mathcal{O}(\epsilon^{1-|z|})\;,
\end{align*}
in which the requirement $z \in [-Z,Z]$ guarantees that the logarithm can be expanded for $\epsilon$ small enough (a bound which depends on $Z$, but not on the chosen $z$). Now combining this bound with $|{r}^{\epsilon}(z) |\leq 2C_2+\hat{r}^{\epsilon}(z)$ concludes the proof.
\hfill $\Box$

\section{Estimates on the Furstenberg measure}
\label{sec:mu}

This section will provide the proof of the following result, which is a slight generalization of Theorem~\ref{theo-InvMeasure} because it allows to consider intervals which depend on $\epsilon$. Recall the notation $\delta=\log(\epsilon^{-1})^{-1}$ from~\eqref{eq-deltaDef}.

\begin{theorem}
\label{theo:mu}
For $\nu\in\{-,+\}$ and $z^\epsilon \in (-1,1)$ satisfying $\limsup_{\epsilon \to 0} |z^\epsilon| < 1$, the Furstenberg measure $(\mu^\epsilon_+,\mu^\epsilon_-)$ in logarithmic Dyson-Schmidt variables satisfies 
$$
\mu^{\epsilon}_\nu([z^\epsilon,1]) 
\;=\; 
\frac{1}{2}\,\left(\frac{1-z^\epsilon}{2}\right)^2 
\,+\, 
\mathcal{O}(\delta\log(\delta))\,,
$$
and there exist constants $C_-, C_+ \in \mathbb{R}_0^+$ such that
$$
\mu^{\epsilon}_\nu\left((-\infty,-1-C_-\delta) \cup (1+C_+\delta,\infty) \cup \{\infty\}\right)
\; =\; 
\mathcal{O}(\delta^2)
\,.
$$
\end{theorem}

\noindent\textbf{Proof} of Theorem~\ref{theo-InvMeasure}:
This follows from Theorem~\ref{theo:mu} with $z^{\epsilon}$ independent of $\epsilon$.
\hfill $\square$

\begin{remark}
\label{rem-MarkovChain}
{\rm
Before delving into the proof of Theorem~\ref{theo:mu}, let us provide an elementary explanation why $\mu_s^\epsilon$ is approximated by the triangular distribution. In fact, roughly the random dynamics is given by a random walk on a finite interval with one side being a hard wall boundary and the other one a semipermeable barrier allowing transfer to the hard wall boundary on the other end. This can be modeled by a discrete Markov chain on a finite state space $\{1,2,\dots,N\}$, viewed as discrete approximation for the interval. The random walker makes step to left and right with probability $\frac{1}{2}$, is reflected at $1$ and when reaching $N+1$ it is moved to $1$. Hence the transition matrix is
$$
\frac{1}{2}\begin{pmatrix}
1 & 1 & & & & & 1\\
1 & 0 & 1 & & & &\\
& 1 & 0 & 1 & & &\\
\phantom{\ddots} & \phantom{\ddots} & \ddots & \ddots & \ddots & \phantom{\ddots} & \phantom{\ddots}\\
& & & 1 & 0 & 1 &\\
& & & & 1 & 0 & 1\\
& & & & & 1 & 0
\end{pmatrix}\,,
$$
in which all missing entries are equal to zero.  One can readily check that the transition matrix has $(N,N-1,\dots,1) \in \mathbb{R}^{N}$ as a (non-normalized column) eigenvector with (Perron-Frobenius) eigenvalue $1$. This eigenvector indeed is a discrete approximation of the triangular distribution in the first plot of  Figure~\ref{fig-Furstenberg}.
\hfill $\diamond$
}
\end{remark}

\subsection{Strategy of proof}

The Furstenberg measure $(\mu_+^\epsilon,\mu_-^\epsilon)$ in the $z$-picture is the invariant measure of the action $\star$, namely for all Borel functions $f:\dot{\mathbb{R}}\times\{-1,1\}\to \mathbb{R}$ the invariance equation 
$$
\sum_\nu\int \mu^{\epsilon}_\nu(dz) \,\EM\,f\big(T^\epsilon\star(z,\nu)\big) 
\;=\; 
\sum_\nu\int \mu^{\epsilon}_\nu(dz) \,f(z,\nu)
\,,
$$
holds. Iterating and averaging yields, for $N\geq 1$,
\begin{gather}\label{eq:mu-Birkhoff}
\sum_\nu\int \mu^{\epsilon}_\nu (dz) \,f(z,\nu)
\;=\;
\sum_\nu\int \mu^{\epsilon}_\nu(dz) \,
\mathbb{E} \frac{1}{N}\sum_{n=1}^{N} f(z_n,\nu_n)\,,
\end{gather}
where $(z_{n},\nu_n)= T^\epsilon_n \star (z_{n-1},\nu_{n-1})$ as in~\eqref{eq-LogDysonDyn} and the last integral is taken over the initial condition $(z,\nu)  = (z_0,\nu_0)$.  Due to~\eqref{eq:mu-Birkhoff}, it is sufficient to compute Birkhoff sums of orbits of the random dynamics~\eqref{eq-LogDysonDyn}. Hence the following three lemmata will directly lead to a proof of Theorem~\ref{theo:mu}.

\begin{lemma}
\label{lemma:mu-slower}
For $\nu\in\{-,+\}$ and $z^{\epsilon} \in (-1,1)$  obeying $\limsup\limits_{\epsilon \to 0} |z^{\epsilon}| < 1$, it holds that
$$
\lim_{N \to \infty}\mathbb{E}\,\frac{1}{N}\sum_{n=1}^N \mathbf{1}\big[z_n \in [z^{\epsilon},1],\nu_n=\nu\big] 
\;\geq\; 
\frac{1}{2}\,\left(\frac{1-z^{\epsilon}}{2}\right)^2 \,+\, \mathcal{O}(\delta)\,.
$$
\end{lemma}

\begin{lemma}\label{lemma:mu-faster}
For $\nu\in\{-,+\}$ and  $z^{\epsilon} \in (-1,1)$, obeying $\limsup\limits_{\epsilon \to 0} |z^{\epsilon}| < 1$, it holds that
$$
\lim_{N \to \infty}\mathbb{E}\,\frac{1}{N}\sum_{n=1}^N \mathbf{1}\big[z_n \in [z^{\epsilon},1],\nu_n=\nu\big] 
\;\leq\; 
\frac{1}{2}\,\left(\frac{1-z^{\epsilon}}{2}\right)^2 \,+\, \mathcal{O}(\delta\log(\delta))\,.
$$
\end{lemma}

\begin{lemma}\label{coro:xi-support}
It holds that
$$
\lim_{N \to \infty}\mathbb{E}\,\frac{1}{N}\sum_{n=1}^N \mathbf{1}\big[z_n \notin [\widehat{z}_-,\widehat{z}_+]\big] 
\;=\; 
\mathcal{O}(\delta^2)\,.
$$
\end{lemma}

\noindent\textbf{Proof}  of Theorem~\ref{theo:mu}.
The second statement follows by taking $f(z,\nu) = \mathbf{1}\big[z \notin [\widehat{z}_-,\widehat{z}_+]\big]$ in~\eqref{eq:mu-Birkhoff},  and Lemma~\ref{lemma:y-lims} to show the claim involving the constants $C_-$ and $C_+$. The first statement is shown analogously, this time by taking $f(z,\nu) = \mathbf{1}\big[z \in [z^{\epsilon},1],\nu'=\nu\big]$ and combining the bounds from Lemmata~\ref{lemma:mu-slower} and~\ref{lemma:mu-faster}.
\hfill $\square$

\vspace{.2cm}

Now let $z=z^\epsilon$ be such that $\limsup\limits_{\epsilon \to 0} |z^\epsilon| < 1$. Then $y=(2C_0\ParaEps)^{-1} z$ satisfies $y\in ( \widetilde{y}_c, \widetilde{y}_+ )$ for $\epsilon$ sufficiently small. Clearly, combining the statements of Lemmata~\ref{lemma:mu-slower} and~\ref{lemma:mu-faster} yields up to lower order corrections an equality instead of two inequalities, as in Lemma~\ref{coro:xi-support} and Theorem~\ref{theo:mu}. Nevertheless, the two results are stated separately in order to stress structural aspects of the proof. In both cases, to bound the Birkhoff sums it is desirable to control the quantity
\begin{align}\label{eq:S}
\begin{aligned}
S_{k}
\;:&=\; \#\big\{n \in \{N_k,N_k+1,\dots,N_{k+1}-1\} \,:\, z_n \in [z,1]\big\}
\\
&=\; \#\big\{n\geq 1 \,:\, z_n \in [z,1] \text{ during the } k\text{-th passage}\big\}
\,.
\end{aligned}
\end{align}
Note that $\nu_n=(-1)^k$ for $n\in[N_k,N_{k+1})$ as already explained after~\eqref{eq-PassageTimes}. This information can then be combined with the total number of steps during the $k$-th passage (which is analyzed in~\cite{DSS}) in order to estimate the proportion of steps of the process $(z_n)_{n \geq 0}$ that take a value in $[z,1]$. The proof of this claim will be given in Section~\ref{sec-ProofInvMeas} below. However, just as the full passage times $(N_{k+1}-N_k)_{k  \geq 0}$, the family of random variables $(S_{k})_{k  \geq 0}$ is neither identically distributed nor independent. Just as in~\cite{DSS}, this difficulty will be handled by introducing a faster and slower dynamical system for each passage. In order to take care of the additional dependence on the variable $z$, these comparison processes will be slightly more complicated than those introduced in ~\cite{DSS}.  Nonetheless, the construction of the processes given in Section~\ref{sec:faster} and Section~\ref{sec:slower} respectively will satisfy the following properties which justify the terminology as faster and slower comparison process. The construction of these processes and also the control of the Birkhoff sums will be carried out in $y$-picture, rather than in the $z$-picture.

\begin{lemma}\label{lemma:Y-faster}
There exist faster comparison processes $\{(\widetilde{y}_{k,n},\widetilde{o}_{k,n})_{n \geq 1}\}_{k \geq 0}$ on $\dot{\mathbb{R}} \times [0,1+\delta^2]$, obeying
\begin{gather}\label{ineq:Y-faster}
y_{\vphantom{\widehat{k}}N_k+n} \;\leq\; \widetilde{y}_{k,n}
\end{gather}
a.s. for all $k\geq 0$ and $n \in \{1,\dots,N_{k+1} - N_k-1\}$, as well as
\begin{gather}\label{eq:Y-faster}
\widetilde{y}_{k,N_{k+1} - N_k} \,=\, \infty\,.
\end{gather}
\end{lemma}

\begin{lemma}
\label{lemma:Y-slower}
There exist slower comparison processes $\{(\widehat{y}_{k,n},\widehat{o}_{k,n})_{n \geq 1}\}_{k  \geq 0}$ on $\dot{\mathbb{R}} \times [0,1]$, obeying
\begin{gather}\label{ineq:Y-slower}
\widehat{y}_{k,n} \;\leq \;y_{\vphantom{\widehat{k}}N_k+n}
\end{gather}
a.s. for all $k\geq 0$ and $n \in \{1,\dots,N_{k+1} - N_k-1\}$.
\end{lemma}

Further facts about these processes are, first of all, that they are indeed i.i.d. and, second of all, that they are given by a random walk with a positive and a negative respectively, both with a hard wall on the left and an absorbing boundary on the right, similarly as in Remark~\ref{rem-MarkovChain}. This allows to compute or at least bound quantities $\widehat{S}_k$ and $\widetilde{S}_k$ which are defined similarly as $S_k$ in \eqref{eq:S}. Combining these facts leads to a proof of the three Lemmata~\ref{lemma:mu-slower},~\ref{lemma:mu-faster} and~\ref{coro:xi-support} in Section~\ref{sec-ProofInvMeas}.

\subsection{Faster comparison processes}
\label{sec:faster}

\begin{figure}
\begin{center}
\begin{tikzpicture}[line join = round, line cap = round]
\coordinate (a) at (0.0,0.3);
\coordinate (b) at (0.0,-0.2);
\coordinate (ll) at (-2.75,0.0);
\coordinate (l0) at (-2.25,0.0);
\vtick{(l0)};
\coordinate[label=below:{$\phantom{\delta^2}\widetilde{y}_-\phantom{\delta^2}$}] (l0b) at ($(l0) + (b)$);
\coordinate (l1) at (-0.25,0.0);
\vtick{(l1)};
\coordinate (l1a) at ($(l1) + (a)$);
\coordinate[label=below:{$\phantom{\delta^2}\widetilde{y}_c\phantom{\delta^2}$}] (l1b) at ($(l1) + (b)$);
\coordinate (lr) at (0.25,0.0);
\coordinate (ml) at (1.5,0.0);
\coordinate (m1) at (2.0,0.0);
\coordinate (m1a) at ($(m1) + (a)$);
\coordinate (m1aa) at ($(m1a) + (a)$);
\vtick{(m1)};
\coordinate[label=below:{$y-1-\delta^2+\widetilde{o}$}] (m1b) at ($(m1) + (b)$);
\coordinate (m2) at (3.5,0.0);
\vtick{(m2)};
\coordinate[label=below:{$\downarrow$}] (m2b) at ($(m2) + (b)$);
\coordinate[label=below:{$\phantom{\delta^2}y\phantom{\delta^2}$}] (m2bbb) at ($(m2b) + (b) + (b)$);
\coordinate (m3) at (4.25,0.0);
\coordinate (m3a) at ($(m3) + (a)$);
\coordinate (m3aa) at ($(m3a) + (a)$);
\coordinate (m3aaa) at ($(m3aa) + (a)$);
\vtick{(m3)};
\coordinate[label=below:{$\phantom{\delta^2}y+\widetilde{o}\phantom{\delta^2}$}] (m3b) at ($(m3) + (b)$);
\coordinate (mr) at (4.75,0.0);
\coordinate (rl) at (6.0,0.0);
\coordinate (r1) at (6.5,0.0);
\coordinate (r1aa) at ($(r1) + (a) + (a)$);
\vtick{(r1)};
\coordinate[label=below:{$\widetilde{y}_+-1-\delta^2+\widetilde{o}$}] (r1b) at ($(r1) + (b)$);
\coordinate (r2) at (8.0,0.0);
\vtick{(r2)};
\coordinate[label=below:{$\downarrow$}] (r2b) at ($(r2) + (b)$);
\coordinate[label=below:{$\phantom{\delta^2}\phantom{{}_+}\widetilde{y}_+\phantom{\delta^2}$}] (r2bbb) at ($(r2b) + (b) + (b)$);
\coordinate (rr) at (8.5,0.0);
\draw [-,color=black,line width=0.3mm] (ll)--(lr);
\draw [-,color=black,dotted,line width=0.3mm] (lr)--(ml);
\draw [-,color=black,line width=0.3mm] (ml)--(mr);
\draw [-,color=black,dotted,line width=0.3mm] (mr)--(rl);
\draw [->,color=black,line width=0.3mm] (rl)--(rr);
\draw[->,dashed] (l1a) to[out=60, in=120, looseness=1.0] (m3aaa);
\draw[->] (m3aa) to[out=120, in=60, looseness=1.0] (m1aa);
\draw[->] (m3aa) to[out=60, in=150, looseness=1.0] (r1aa);
\draw[->,dash dot] (m1a) to[out=60, in=120, looseness=1.0] (m3a);
\end{tikzpicture}
\caption{\it Visual representation of the stages of the process $(\widetilde{y}_n, \widetilde{o}_n)_{n\geq 1}$. The steps taken when the second variable equals $1+\delta^2$ right after starting the process are indicated by the dashed arrow. This part of a passage ends at $y+\widetilde{o}$, in which $\widetilde{o}$ is the random overshoot over $y$. It is determined at the time on which the faster comparison process becomes larger or equal to $y$ for the first time. Afterwards, the process follows one of the solid arrows, either stopping the process (setting it to $\infty$) after becoming larger than $\widetilde{y}_+-1-\delta^2+\widetilde{o}$ or the faster comparison process becomes less than $y-1-\delta^2+\widetilde{o}$. In the latter case, the dash-dotted arrow leads to the next recalculation of the overshoot, after which the previous sentence applies again.}
\label{fig:mu-faster}
\end{center}
\end{figure}

This section deals with the faster comparison processes, of which the construction is closely related to the one given in Section 6 of~\cite{DSS} and will almost identically apply to the confined type in Section~\ref{sec-Confined}. For $k\geq 0$ indicating the $k$-th passage, a Markov process $(\widetilde{y}_{k,n},\widetilde{o}_{k,n})_{n \geq 1}$ on the space $\dot{\mathbb{R}} \times [0,1+\delta^2]$ is constructed by setting $\widetilde{y}_{k,1} = \widetilde{y}_c$, $\widetilde{o}_{k,1} = 1+\delta^2$ and for $n\geq 1$,
\begin{align}\label{eq:faster-Y-dyn}
\begin{aligned}
\widetilde{y}_{k,n+1} &:= \begin{cases}
\widetilde{y}_c\,, &\text{ if } \widetilde{y}_{k,n} \leq \widetilde{y}_-\,,\\
\infty\,, &\text{ if } \widetilde{y}_{k,n} \geq  \widetilde{y}_+-1-\delta^2+\widetilde{o}_{k,n}\,,\\
\widetilde{y}_{k,n} + (-1)^k\ChiVar_{\vphantom{\widetilde{k}}N_k+n} + \delta^2\,, &\text{ else}\,,
\end{cases}
\\
\widetilde{o}_{k,n+1} &:= \begin{cases}
\widetilde{y}_{k,n+1} - y\,, &\text{ if } \widetilde{y}_{k,n+1} \geq y \text{ and } \widetilde{o}_{k,n} = 1+\delta^2\,,\\
1+\delta^2\,, &\text{ if } \widetilde{y}_{k,n+1} \leq y-1-\delta^2+\widetilde{o}_{k,n} \text{ and } \widetilde{o}_{k,n} \neq 1+\delta^2\,,\\
\widetilde{o}_{k,n}\,, &\text{ else}\,.
\end{cases}
\end{aligned}
\end{align}
Let us note that the sign factor $(-1)^k$ in~\eqref{eq:faster-Y-dyn} reflects the fact that the dynamics on the $\nu=+$ and $\nu=-$ fibers differs in a manner described before Lemma~\ref{lemma:Q-lower} (concretely, $J'^*D^\epsilon J'=-(D^\epsilon)^{-1}$), which leads, in particular, to an alternating sign change of $\ChiVar_n=\log(\kappa_n)$.

\begin{remark}\label{rem:faster}
{\rm 
The first component of $(\widetilde{y}_{k,n},\widetilde{o}_{k,n})_{n\geq 1}$ is almost the same as the faster comparison process used in~\cite{DSS}, including a concrete choice for the drift (taken to be $\delta^2$ in the $y$-picture instead of the free parameter $\lambda$ in~\cite{DSS}). The only difference lies in the condition which sets $\widetilde{y}_{k,n+1}$ to $\infty$: in~\cite{DSS}, this happens (described in Dyson-Schmidt coordinates) if $\widetilde{y}_{k,n} \geq \widetilde{y}_+$, whereas (in logarithmic Dyson-Schmidt variables) in~\eqref{eq:faster-Y-dyn} this condition is modified to $\widetilde{y}_{k,n} \geq  \widetilde{y}_+-1-\delta^2+\widetilde{o}_{k,n}$. The latter is less restrictive, as $\widetilde{y}_+ \geq \widetilde{y}_+-1-\delta^2+\widetilde{o}_{k,n}$ by construction.
}
\hfill $\diamond$
\end{remark}

It is immediate that~\eqref{eq:faster-Y-dyn} defines a Markov process. The dynamics of its first component is visualized in Figure~\ref{fig:mu-faster}. Note that
$$
\liminf_{\delta \to 0} 2C_0\delta(y-\widetilde{y}_-) \;=\; 1+\liminf_{\epsilon \to 0} z > 0\;,
\qquad
\liminf_{\delta \to 0} 2C_0\delta(\widetilde{y}_+-y) \;=\; 1-\limsup_{\epsilon \to 0} z > 0\;,
$$
by Lemma~\ref{lemma:y-lims}, so both $y-\widetilde{y}_-$ and $\widetilde{y}_+-y$ are large and of the order to $\mathcal{O}(\delta^{-1})$, so that the process needs at least $\mathcal{O}(\delta^{-1})$ time steps before becoming constant and equal to $\infty$. By Lemma~\ref{lemma:Y-slower}, the first index $k$ labels the passage in which the constructed process $(\widetilde{y}_{k,n})_{n \geq 1}$ needs to be compared with the shifted dynamical system $(y_{\vphantom{\widehat{k}}N_k+n})_{n \geq 1}$.

\vspace{.2cm}

\noindent\textbf{Proof} of Lemma~\ref{lemma:Y-faster}.
Based on Lemmata~\ref{lemma:drift-bis} and~\ref{lemma:y}, the given properties can readily be verified ({\it cf.} statements (21) and (22) from ~\cite{DSS}, stopping even faster as indicated in Remark~\ref{rem:faster}).
\hfill $\square$

\vspace{.2cm}

Let us now come to the second component $\widetilde{o}_{k,n}$. It is called the overshoot, as it quantifies how far the first component of the process jumps over $y$ during the $k$-th passage. This statement will be made more precise when discussing the possible values for this variable. From the start of the process, the overshoot is equal to $1+\delta^2$. This changes whenever $\widetilde{y}_{k,n} \geq y$ for some $n\geq 1$. Then the overshoot $\widetilde{o}_{k,n}$ is set equal to the difference between $\widetilde{y}_{k,n}$ and $y$. Then the Main Hypothesis implies $\widetilde{o}_{k,n} < 1+\delta^2$. Afterwards, it remains constant. If the first component again becomes smaller than $y-1-\delta^2+\widetilde{o}_{k,n} < y$, it is set back to $1+\delta^2$ until the first component is again larger than or equal to $y$. As the aim is to estimate the number of occurrences of the first component of the process in $(y,\widetilde{y}_+)$, it is hence of interest to set
$$
\widetilde{s}_{k,n}
\;:=\;
\mathbf{1}\big[\widetilde{o}_{k,n} \neq 1+\delta^2\big]\,\mathbf{1}\big[\widetilde{y}_{k,n} \neq \infty\big]\;\in\;\{0,1\}
\;,
$$
and 
$$
\widetilde{S}_k \;:=\; \sum_{n=1}^{\infty} \widetilde{s}_{k,n}\,.
$$
Note that $(\widetilde{S}_{2k})_{k\geq 0}$ and $(\widetilde{S}_{2k+1})_{k\geq 0}$ are both families of i.i.d. random variables (possibly with different distributions due to the differences of passages as described in Section~\ref{sec-DysonSchmidt}) and so is $(\widetilde{S}_{2k}+\widetilde{S}_{2k+1})_{k\geq 0}$. In Section~\ref{sec-ProofInvMeas}, it will be shown that $\widetilde{S}_k$ is a good approximation for $S_k$. Therefore the focus is here on the control of $\mathbb{E}(\widetilde{S}_k)$. This is done for the $k$-th passage, but all faster comparison processes are independent because they all start at the same point $\widetilde{y}_c$. This allows to suppress the index $k$ in the following. Let us introduce three stopping times:
\begin{align*}
\widetilde{T}_{\shortrightarrow} &\;:= \;\inf\{n\geq 1 \,:\, \widetilde{y}_n \geq y\}\,,\\
\widetilde{S}_{\leftrightarrow} &\;:=\; \inf\{n\geq 1 \,:\, \widetilde{y}_{\widetilde{T}_{\scriptshortrightarrow}+n} \notin (y-1-\delta^2+\widetilde{o}_{\widetilde{T}_{\scriptshortrightarrow}},\widetilde{y}_+-1-\delta^2+\widetilde{o}_{\widetilde{T}_{\scriptshortrightarrow}})\}\,,\\
\widetilde{T} &\;:=\; \inf\{n\geq 1 \,:\, \widetilde{y}_n = \infty\}\,.
\end{align*}
The former $\widetilde{T}_{\shortrightarrow}$ is the time it takes for the process $(\widetilde{y}_n)_{n\geq 1}$ to pass $y$ for the first time; it coincides with the first $n\geq 1$ for which $\widetilde{s}_n = 1$. The variable $\widetilde{S}_{\leftrightarrow}$ quantifies the time it takes for $\widetilde{y}_n$ afterwards to leave the interval $(y-1-\delta^2+\widetilde{o}_{\widetilde{T}_{\scriptshortrightarrow}},\widetilde{y}_+-1-\delta^2+\widetilde{o}_{\widetilde{T}_{\scriptshortrightarrow}})$, equal to the time until $\widetilde{s}_n$ becomes $0$ again. The last random variable $\widetilde{T}$ equals the total time until the process $(\widetilde{y}_n)_{n\geq 1}$ becomes constant. Denoting the $\widetilde{T}$ for the $k$-th passage by $\widetilde{T}_k$, one once more has that $(\widetilde{T}_{2k}+\widetilde{T}_{2k+1})_{k\geq 0}$ is an i.i.d. family of random variables. As a consequence, the elements of this family are interarrival times with a corresponding renewal process $(\widetilde{P}_N)_{\vphantom{\widehat{k}}N\geq 1}$, which is for all $N\geq 1$ defined by
\begin{gather}\label{eq:faster-P}
\widetilde{P}_N 
\;:=\; 
\max\left\{K\geq 1 \,:\, \sum_{k=0}^K (\widetilde{T}_{2k}+\widetilde{T}_{2k+1}) \leq N\right\}\,.
\end{gather}
%

\begin{lemma}\label{lemma:expectation-faster}
The expectation values of $\widetilde{S}_{\leftrightarrow}$ is finite.
\end{lemma}

\noindent\textbf{Proof.} The conditions on $\ChiVar$ imply that $\widetilde{A} := \mathbb{P}[(-1)^k\ChiVar \geq 0] > 0$. When setting $\widetilde{B} := \lceil\frac{\widetilde{y}_+-y}{\delta^2}\rceil$, the random variable $\widetilde{N} := \inf\{n\geq 1 \,:\, \forall\, N \in \{1,2,\dots,\widetilde{B}\} \,:\, (-1)^k\ChiVar_{n\widetilde{B}+N} + \delta^2 \geq \delta^2\}$ is geometrically distributed with success probability $\widetilde{A}^{\widetilde{B}}$, so $\mathbb{E}[\widetilde{N}] < \infty$. As $\widetilde{S}_{\leftrightarrow} \leq \widetilde{N}\widetilde{B}$ a.s. by construction, $\mathbb{E}[\widetilde{S}_{\leftrightarrow}] \leq \mathbb{E}[\widetilde{N}\widetilde{B}] = \widetilde{B}\mathbb{E}[\widetilde{N}] < \infty$ follows.\hfill $\square$

\begin{lemma}\label{lemma:faster-T-S}
For the faster comparison process $(\widetilde{y}_n,\widetilde{o}_n)_{n\geq 1}$, one has
$$
\mathbb{E}[\widetilde{S}] 
\;=\;
\frac{\delta^{-2}}{\mathbb{E}[\log(\kappa)^2]}\left(\frac{1-z}{2}\right)^2\left[1 + \mathcal{O}(\delta\log(\delta))\right]\,.
$$
\end{lemma}

\noindent\textbf{Proof.} 
The first element of the proof is an identity suggested by the diagram
\begin{center}
\begin{tikzcd}
\widetilde{S}:\widetilde{S}_{\leftrightarrow} \arrow[rr, "\leq y-1-\delta^2+\widetilde{o}_{\widetilde{T}_{\scriptshortrightarrow}}", dashed] \arrow[rrd, "\geq \widetilde{y}_+-1-\delta^2+\widetilde{o}_{\widetilde{T}_{\scriptshortrightarrow}}"', dashed] &  & \widetilde{S} \\
&  & 0            
\end{tikzcd}
\end{center}
which graphically depicts the steps that are counted in order to arrive at the quantity $\widetilde{S}$. A first contribution to $\widetilde{S}$ is given by $\widetilde{S}_{\leftrightarrow}$. At this time $n=\widetilde{T}_{\shortrightarrow}+\widetilde{S}_{\leftrightarrow}$, either $\widetilde{y}_n \leq y-1-\delta^2+\widetilde{o}_n$ or $\widetilde{y}_n \geq \widetilde{y}_+-1-\delta^2+\widetilde{o}_n$ where actually $\widetilde{o}_n=\widetilde{o}_{\widetilde{T}_{\shortrightarrow}}$. In the latter case, no more contributions are added to $\widetilde{S}$. In the former case, the process is initialized again, resulting in another $\widetilde{S}$ steps.  Hence one expects that 
$$
\mathbb{E}[\widetilde{S}] 
\;=\; 
\mathbb{E}[\widetilde{S}_{\leftrightarrow}] + \mathbb{P}[\widetilde{y}_{\widetilde{T}_{\scriptshortrightarrow}+\widetilde{S}_{\leftrightarrow}} \leq y-1-\delta^2+\widetilde{o}_{\widetilde{T}_{\scriptshortrightarrow}}] \cdot \mathbb{E}[\widetilde{S}] + \mathbb{P}[\widetilde{y}_{\widetilde{T}_{\scriptshortrightarrow}+\widetilde{S}_{\leftrightarrow}} \geq \widetilde{y}_+-1-\delta^2+\widetilde{o}_{\widetilde{T}_{\scriptshortrightarrow}}] \cdot 0\,.
$$
This formula is formally verified by the calculation leading to~\eqref{eq-DiagrammFormula2} given below. For this, set
$$
\widetilde{S}^{(N)} \;:=\; \sum_{n\geq N+\widetilde{T}_{\scriptshortrightarrow}+1} \widetilde{s}_n\,,
$$
for $N\geq 1$. The strong Markov property then implies that
\begin{align*}
&\mathbb{E}\left[\widetilde{S}\mathbf{1}\big[\widetilde{y}_{\widetilde{T}_{\scriptshortrightarrow}+N} \leq y-1-\delta^2+\widetilde{o}_{\widetilde{T}_{\scriptshortrightarrow}}\big]\mathbf{1}\big[\widetilde{S}_{\leftrightarrow} = N\big]\right]\\
&\qquad= \;\mathbb{E}\left[(N + \widetilde{S}^{(N)})\mathbf{1}\big[\widetilde{y}_{\widetilde{T}_{\scriptshortrightarrow}+N} \leq y-1-\delta^2+\widetilde{o}_{\widetilde{T}_{\scriptshortrightarrow}}\big]\mathbf{1}\big[\widetilde{S}_{\leftrightarrow} = N\big]\right]\,,\\
&\mathbb{E}\left[\widetilde{S}\mathbf{1}\big[\widetilde{y}_{\widetilde{T}_{\scriptshortrightarrow}+\widetilde{S}_{\leftrightarrow}} \geq \widetilde{y}_+-1-\delta^2+\widetilde{o}_{\widetilde{T}_{\scriptshortrightarrow}}\big]\right]\, = \,\mathbb{E}\left[\widetilde{S}_{\leftrightarrow}\mathbf{1}\big[\widetilde{y}_{\widetilde{T}_{\scriptshortrightarrow}+\widetilde{S}_{\leftrightarrow}} \geq \widetilde{y}_+-1-\delta^2+\widetilde{o}_{\widetilde{T}_{\scriptshortrightarrow}}\big]\right]\,,\\
&\mathbb{E}\left[\widetilde{S}^{(N)}\mathbf{1}\big[\widetilde{y}_{\widetilde{T}_{\scriptshortrightarrow}+N} \leq y-1-\delta^2+\widetilde{o}_{\widetilde{T}_{\scriptshortrightarrow}}\big]\mathbf{1}\big[\widetilde{S}_{\leftrightarrow} = N\big]\right]
\\ 
&
\qquad
= \;\mathbb{E}[\widetilde{S}]
\,\mathbb{P}\big[\widetilde{y}_{\widetilde{T}_{\scriptshortrightarrow}+\widetilde{S}_{\leftrightarrow} } \leq y-1-\delta^2+\widetilde{o}_{\widetilde{T}_{\scriptshortrightarrow}}
\wedge \widetilde{S}_{\leftrightarrow}=N \big]
\,.
\end{align*}
Together with $\mathbb{P}\big[\widetilde{y}_{\widetilde{T}_{\scriptshortrightarrow}+\widetilde{S}_{\leftrightarrow}} \in (y-1-\delta^2+\widetilde{o}_{\widetilde{T}_{\scriptshortrightarrow}},\widetilde{y}_+-1-\delta^2+\widetilde{o}_{\widetilde{T}_{\scriptshortrightarrow}})\big] = 0$ and the fact that $\widetilde{S}_{\leftrightarrow} < \infty$ a.s. by Lemma~\ref{lemma:expectation-faster}, this implies
\begin{align*}
& 
\mathbb{E}[\widetilde{S}] 
\;=\; 
\mathbb{E}\left[\widetilde{S}\mathbf{1}\big[\widetilde{y}_{\widetilde{T}_{\scriptshortrightarrow}+\widetilde{S}_{\leftrightarrow}} \geq \widetilde{y}_+-1-\delta^2+\widetilde{o}_{\widetilde{T}_{\scriptshortrightarrow}}\big]\right] + 
\mathbb{E}\left[\widetilde{S}\mathbf{1}\big[\widetilde{y}_{\widetilde{T}_{\scriptshortrightarrow}+\widetilde{S}_{\leftrightarrow}} \leq  y-1-\delta^2+\widetilde{o}_{\widetilde{T}_{\scriptshortrightarrow}}\big]\right] 
\\
&
\;=\; 
\mathbb{E}\left[\widetilde{S}\mathbf{1}\big[\widetilde{y}_{\widetilde{T}_{\scriptshortrightarrow}+\widetilde{S}_{\leftrightarrow}} \geq \widetilde{y}_+-1-\delta^2+\widetilde{o}_{\widetilde{T}_{\scriptshortrightarrow}}\big]\right] + \sum_{\vphantom{\widehat{k}}N = 1}^{\infty} \mathbb{E}\left[\widetilde{S}\mathbf{1}\big[\widetilde{y}_{\widetilde{T}_{\scriptshortrightarrow}+N} \leq y-1-\delta^2+\widetilde{o}_{\widetilde{T}_{\scriptshortrightarrow}}\big]\mathbf{1}\big[\widetilde{S}_{\leftrightarrow} = N\big]\right]\\
&
\;=\; 
\mathbb{E}\left[\widetilde{S}_{\leftrightarrow}\mathbf{1}\big[\widetilde{y}_{\widetilde{T}_{\scriptshortrightarrow}+\widetilde{S}_{\leftrightarrow}} \geq \widetilde{y}_+-1-\delta^2+\widetilde{o}_{\widetilde{T}_{\scriptshortrightarrow}}\big]\right]\\
&
\qquad + \sum_{\vphantom{\widehat{k}}N = 1}^{\infty} \mathbb{E}\left[(N + \widetilde{S}^{(N)})\mathbf{1}\big[\widetilde{y}_{\widetilde{T}_{\scriptshortrightarrow}+N} \leq y-1-\delta^2+\widetilde{o}_{\widetilde{T}_{\scriptshortrightarrow}}\big]\mathbf{1}\big[\widetilde{S}_{\leftrightarrow} = N\big]\right]\\
&
\;=\; 
\mathbb{E}\left[\widetilde{S}_{\leftrightarrow}\mathbf{1}\big[\widetilde{y}_{\widetilde{T}_{\scriptshortrightarrow}+\widetilde{S}_{\leftrightarrow}} \geq \widetilde{y}_+-1-\delta^2+\widetilde{o}_{\widetilde{T}_{\scriptshortrightarrow}}\big]\right]
\\
&
\qquad + \mathbb{E}\left[\widetilde{S}_{\leftrightarrow}\mathbf{1}\big[\widetilde{y}_{\widetilde{T}_{\scriptshortrightarrow}+\widetilde{S}_{\leftrightarrow}} \leq y-1-\delta^2+\widetilde{o}_{\widetilde{T}_{\scriptshortrightarrow}}\big]\right] + \mathbb{E}[\widetilde{S}]\,\mathbb{P}\big[\widetilde{y}_{\widetilde{T}_{\scriptshortrightarrow}+\widetilde{S}_{\leftrightarrow}} \leq y-1-\delta^2+\widetilde{o}_{\widetilde{T}_{\scriptshortrightarrow}}\big]\\
&
\;=\; 
\mathbb{E}\big[\widetilde{S}_{\leftrightarrow}\big] + \mathbb{E}[\widetilde{S}]\,\mathbb{P}\big[\widetilde{y}_{\widetilde{T}_{\scriptshortrightarrow}+\widetilde{S}_{\leftrightarrow}} \leq y-1-\delta^2+\widetilde{o}_{\widetilde{T}_{\scriptshortrightarrow}}\big]\,,
\end{align*}
which is equivalent to
\begin{equation}
\label{eq-DiagrammFormula2}
\mathbb{E}[\widetilde{S}] 
\;=\; 
\frac{\mathbb{E}[\widetilde{S}_{\leftrightarrow}]}{\mathbb{P}[\widetilde{y}_{\widetilde{T}_{\scriptshortrightarrow}+\widetilde{S}_{\leftrightarrow}} \geq \widetilde{y}_+-1-\delta^2+\widetilde{o}_{\widetilde{T}_{\scriptshortrightarrow}}]}
\;.
\end{equation}
In order to prove an upper bound on $\mathbb{E}\big[\widetilde{S}_{\leftrightarrow}\big]$, the additional nonnegative constants
\begin{align*}
& 
\widetilde{\ell} 
\;:=\; 
\mathbb{E}\big[\widetilde{y}_{\widetilde{T}_{\scriptshortrightarrow}}\big] - \mathbb{E}\big[\widetilde{y}_{\widetilde{T}_{\scriptshortrightarrow}+\widetilde{S}_{\leftrightarrow}} \,\big|\, \widetilde{y}_{\widetilde{T}_{\scriptshortrightarrow}+\widetilde{S}_{\leftrightarrow}} \leq y-1-\delta^2+\widetilde{o}_{\widetilde{T}_{\scriptshortrightarrow}}\big]\,,
\\
& 
\widetilde{r} 
\;:= \;
\mathbb{E}\big[\widetilde{y}_{\widetilde{T}_{\scriptshortrightarrow}+\widetilde{S}_{\leftrightarrow}} \,\big|\, \widetilde{y}_{\widetilde{T}_{\scriptshortrightarrow}+\widetilde{S}_{\leftrightarrow}} \geq \widetilde{y}_+-1-\delta^2+\widetilde{o}_{\widetilde{T}_{\scriptshortrightarrow}}\big] - \mathbb{E}\big[\widetilde{y}_{\widetilde{T}_{\scriptshortrightarrow}}\big]
\,,
\end{align*}
will turn out to be useful. Now note that  $(\widetilde{y}_{\widetilde{T}_{\scriptshortrightarrow}+n} - n\delta^2)_{n\geq 1}$ is a martingale (to be precise, up to the stopping time $\widetilde{S}_{\leftrightarrow}$). Moreover, its increments are uniformly bounded for all $n < \widetilde{S}_{\leftrightarrow}$, as the Main Hypothesis implies $|\widetilde{y}_{\widetilde{T}_{\scriptshortrightarrow}+n+1}-(n+1)\delta^2-\widetilde{y}_{\widetilde{T}_{\scriptshortrightarrow}+n}+n\delta^2| \leq 1$. By the optional stopping theorem and Lemma~\ref{lemma:expectation-faster} it therefore follows that
\begin{align*}
\mathbb{E}\big[\widetilde{y}_{\widetilde{T}_{\scriptshortrightarrow}}\big]
&
\;=\; 
\mathbb{E}\big[\widetilde{y}_{\widetilde{T}_{\scriptshortrightarrow}+\widetilde{S}_{\leftrightarrow}}\big] - \delta^2\mathbb{E}[\widetilde{S}_{\leftrightarrow}]
\\
&
\;=\; 
(\mathbb{E}\big[\widetilde{y}_{\widetilde{T}_{\scriptshortrightarrow}}\big] - \widetilde{\ell})\mathbb{P}\big[\widetilde{y}_{\widetilde{T}_{\scriptshortrightarrow}+\widetilde{S}_{\leftrightarrow}} \leq y-1-\delta^2+\widetilde{o}_{\widetilde{T}_{\scriptshortrightarrow}}\big]\\
&\qquad+ (\widetilde{r} + \mathbb{E}\big[\widetilde{y}_{\widetilde{T}_{\scriptshortrightarrow}}\big])\mathbb{P}\big[\widetilde{y}_{\widetilde{T}_{\scriptshortrightarrow}+\widetilde{S}_{\leftrightarrow}} \geq \widetilde{y}_+-1-\delta^2+\widetilde{o}_{\widetilde{T}_{\scriptshortrightarrow}}\big] - \delta^2\mathbb{E}[\widetilde{S}_{\leftrightarrow}]\,.
\end{align*}
Inserting this in~\eqref{eq-DiagrammFormula2} yields
\begin{gather}\label{ineq:faster-S}
\mathbb{E}[\widetilde{S}] 
\;=\; 
\frac{\mathbb{E}\big[\widetilde{S}_{\leftrightarrow}\big]}{\mathbb{P}\big[\widetilde{y}_{\widetilde{T}_{\scriptshortrightarrow}+\widetilde{S}_{\leftrightarrow}} \geq \widetilde{y}_+-1-\delta^2+\widetilde{o}_{\widetilde{T}_{\scriptshortrightarrow}}\big]} 
\;=\; 
\frac{\widetilde{r}}{\delta^2} - \frac{\widetilde{\ell}}{\delta^2} \cdot \frac{\mathbb{P}\big[\widetilde{y}_{\widetilde{T}_{\scriptshortrightarrow}+\widetilde{S}_{\leftrightarrow}} \leq y-1-\delta^2+\widetilde{o}_{\widetilde{T}_{\scriptshortrightarrow}}\big]}{\mathbb{P}\big[\widetilde{y}_{\widetilde{T}_{\scriptshortrightarrow}+\widetilde{S}_{\leftrightarrow}} \geq \widetilde{y}_+-1-\delta^2+\widetilde{o}_{\widetilde{T}_{\scriptshortrightarrow}}\big]}\,.
\end{gather}
Now consider the map $\rho \in \mathbb{R} \mapsto \mathbb{E}[e^{-C_0\rho(\ChiVar + \delta^2)}] \in (0,\infty)$. It is differentiable at $\rho = 0$, with
$$
\left.\partial_{\rho}\mathbb{E}[e^{-C_0\rho(\ChiVar + \delta^2)}]\right|_{\rho = 0} 
\;=\; 
-C_0(\mathbb{E}[\ChiVar] + \delta^2) 
\;=\; 
-C_0\delta^2 
\;<\; 0\,.
$$
This implies that $\mathbb{E}[e^{-C_0\rho(\ChiVar + \delta^2)}] < 1$ for $\rho > 0$ sufficiently small. The given map is continuous, and the Main Hypothesis implies $\lim_{\rho \to \pm\infty} \mathbb{E}[e^{-C_0\rho(\ChiVar + \delta^2)}] = \infty$. Then the intermediate value theorem applies on $(0,\infty)$, yielding a solution of $\mathbb{E}[e^{-C_0\rho(\ChiVar + \delta^2)}] = 1$ for $\rho$ on $(0,\infty)$, which is denoted by $\tau$ (the strict convexity of the map implies that the solution is unique). All this yields the existence of some $\tau \in (0,\infty)$ obeying $\mathbb{E}[e^{-C_0\tau(\ChiVar + \delta^2)}] = 1$, from which it follows that $(e^{-C_0\tau(\widetilde{y}_{\widetilde{T}_{\scriptshortrightarrow}+n}-\widetilde{y}_{\widetilde{T}_{\scriptshortrightarrow}})})_{n\geq 1}$ is a martingale (once more, up to the stopping time $\widehat{S}_{\leftrightarrow}$). This time, the martingale increment $|e^{-C_0\tau(\widetilde{y}_{\widetilde{T}_{\scriptshortrightarrow}+n+1}-\widetilde{y}_{\widetilde{T}_{\scriptshortrightarrow}})} - e^{-C_0\tau(\widetilde{y}_{\widetilde{T}_{\scriptshortrightarrow}+n}-\widetilde{y}_{\widetilde{T}_{\scriptshortrightarrow}})}|\mathbf{1}\big[n+1 \leq \widetilde{S}_{\leftrightarrow}\big] = |e^{-C_0\tau(\ChiVar+\delta^2)}-1|e^{-C_0\tau(\widetilde{y}_{\widetilde{T}_{\scriptshortrightarrow}+n}-\widetilde{y}_{\widetilde{T}_{\scriptshortrightarrow}})}\mathbf{1}\big[n < \widetilde{S}_{\leftrightarrow}\big]$ is bounded by $(e^{C_0\tau(1+\delta^2)}-1)e^{C_0\tau(1+\delta^2)}$ by the Main Hypothesis and the definition of $\widetilde{S}_{\leftrightarrow}$. The optional stopping theorem then yields
\begin{align}
1 &
\;=\; 
\mathbb{E}\big[e^{-C_0\tau(\widetilde{y}_{\widetilde{T}_{\scriptshortrightarrow}+\widetilde{S}_{\leftrightarrow}}-\widetilde{y}_{\widetilde{T}_{\scriptshortrightarrow}})}\big]
\nonumber
\\
&
\;=\; 
\mathbb{E}\big[e^{-C_0\tau(\widetilde{y}_{\widetilde{T}_{\scriptshortrightarrow}+\widetilde{S}_{\leftrightarrow}}-\widetilde{y}_{\widetilde{T}_{\scriptshortrightarrow}})} \,\big|\, \widetilde{y}_{\widetilde{T}_{\scriptshortrightarrow}+\widetilde{S}_{\leftrightarrow}} \leq y-1-\delta^2+\widetilde{o}_{\widetilde{T}_{\scriptshortrightarrow}}\big]\mathbb{P}\big[\widetilde{y}_{\widetilde{T}_{\scriptshortrightarrow}+\widetilde{S}_{\leftrightarrow}} \leq y-1-\delta^2+\widetilde{o}_{\widetilde{T}_{\scriptshortrightarrow}}\big]
\label{eq:faster-prob}
\\
&
\quad\;\;+ \mathbb{E}\big[e^{-C_0\tau(\widetilde{y}_{\widetilde{T}_{\scriptshortrightarrow}+\widetilde{S}_{\leftrightarrow}}-\widetilde{y}_{\widetilde{T}_{\scriptshortrightarrow}})} \,\big|\, \widetilde{y}_{\widetilde{T}_{\scriptshortrightarrow}+\widetilde{S}_{\leftrightarrow}} \geq \widetilde{y}_+-1-\delta^2+\widetilde{o}_{\widetilde{T}_{\scriptshortrightarrow}}\big]\mathbb{P}\big[\widetilde{y}_{\widetilde{T}_{\scriptshortrightarrow}+\widetilde{S}_{\leftrightarrow}} \geq \widetilde{y}_+-1-\delta^2+\widetilde{o}_{\widetilde{T}_{\scriptshortrightarrow}}\big]\,,
\nonumber
\end{align}
which can be inserted into~\eqref{ineq:faster-S} to resolve the unknown probabilities there. Before doing so, let us study how $\tau$ depends on $\delta$. Its definition as a solution out of $(0,\infty)$ for $\rho$ can be rewritten to $\delta^2 = \frac{\log\left(\mathbb{E}[e^{-C_0\rho\ChiVar}]\right)}{C_0\rho}$. Upon setting this equal to $0$ for $\rho = 0$, this is an analytic function of $\rho \in \mathbb{R}$, and its derivative at $\rho = 0$ is $\frac{C_0\mathbb{E}[\ChiVar^2]}{2} \neq 0$. The Lagrange inversion theorem for analytic functions then yields $\tau = \frac{2\delta^2}{C_0\mathbb{E}[\ChiVar^2]} + \mathcal{O}(\delta^4)$. Therefore, one can expand the r.h.s. of~\eqref{eq:faster-prob} to obtain
\begin{align*}
1 &
\,=\, \left(1 - \frac{2\delta^2\widetilde{r}}{\mathbb{E}[\ChiVar^2]} + \frac{2\delta^4\mathbb{E}\big[(\widetilde{y}_{\widetilde{T}_{\scriptshortrightarrow}+\widetilde{S}_{\leftrightarrow}}-\widetilde{y}_{\widetilde{T}_{\scriptshortrightarrow}})^2 \,\big|\, \widetilde{y}_{\widetilde{T}_{\scriptshortrightarrow}+\widetilde{S}_{\leftrightarrow}} \geq \widetilde{y}_+-1-\delta^2+\widetilde{o}_{\widetilde{T}_{\scriptshortrightarrow}}\big]}{(\mathbb{E}[\ChiVar^2])^2} + \mathcal{O}(\delta^6\widetilde{y}_+^3)\right)\\
&\quad\;\cdot \mathbb{P}\big[\widetilde{y}_{\widetilde{T}_{\scriptshortrightarrow}+\widetilde{S}_{\leftrightarrow}} \geq \widetilde{y}_+-1-\delta^2+\widetilde{o}_{\widetilde{T}_{\scriptshortrightarrow}}\big] + \left(1 - \frac{2\delta^2(-\widetilde{\ell})}{\mathbb{E}[\ChiVar^2]} + \mathcal{O}(\delta^4)\right)\mathbb{P}\big[\widetilde{y}_{\widetilde{T}_{\scriptshortrightarrow}+\widetilde{S}_{\leftrightarrow}} \leq y-1-\delta^2+\widetilde{o}_{\widetilde{T}_{\scriptshortrightarrow}}\big]\,,
\end{align*}
which with $\widetilde{\ell} = \mathcal{O}(1)$ and $\widetilde{r} = \mathcal{O}(\widetilde{y}_+) = \mathcal{O}(\delta^{-1})$ from Lemma~\ref{lemma:y-lims} implies
\begin{align*}
&\frac{\mathbb{P}\big[\widetilde{y}_{\widetilde{T}_{\scriptshortrightarrow}+\widetilde{S}_{\leftrightarrow}} 
\;\leq \;
y-1-\delta^2+\widetilde{o}_{\widetilde{T}_{\scriptshortrightarrow}}\big]}{\mathbb{P}\big[\widetilde{y}_{\widetilde{T}_{\scriptshortrightarrow}+\widetilde{S}_{\leftrightarrow}} \geq \widetilde{y}_+-1-\delta^2+\widetilde{o}_{\widetilde{T}_{\scriptshortrightarrow}}\big]}\\
&\qquad=\; \frac{1}{\widetilde{\ell}} \cdot \left[\frac{\widetilde{r}}{1+\mathcal{O}(\delta^2)} - \frac{\delta^2\mathbb{E}\big[(\widetilde{y}_{\widetilde{T}_{\scriptshortrightarrow}+\widetilde{S}_{\leftrightarrow}}-\widetilde{y}_{\widetilde{T}_{\scriptshortrightarrow}})^2 \,\big|\, \widetilde{y}_{\widetilde{T}_{\scriptshortrightarrow}+\widetilde{S}_{\leftrightarrow}} \geq \widetilde{y}_+-1-\delta^2+\widetilde{o}_{\widetilde{T}_{\scriptshortrightarrow}}\big] + \mathcal{O}(\delta^4\widetilde{y}_+^3)}{\mathbb{E}[\ChiVar^2][1+\mathcal{O}(\delta^2)]}\right]\\
&\qquad=\; \frac{1}{\widetilde{\ell}} \cdot \left[\widetilde{r} + \mathcal{O}(\delta) - \frac{(2C_0\delta\widetilde{y}_+-2C_0\delta y + \mathcal{O}(\delta))^2}{4C_0^2\mathbb{E}[\ChiVar^2]}\right]\\
&\qquad= \;\frac{1}{\widetilde{\ell}} \cdot \left[\widetilde{r} - \frac{(1-z)^2}{4\mathbb{E}[\log(\kappa)^2]} + \mathcal{O}(\delta\log(\delta))\right]\,.
\end{align*}
Inserting this into~\eqref{ineq:faster-S} finishes the proof.\hfill $\square$

\subsection{Slower comparison processes}
\label{sec:slower}
\begin{figure}
\begin{center}
\begin{tikzpicture}[line join = round, line cap = round]
\coordinate (a) at (0.0,0.3);
\coordinate (b) at (0.0,-0.2);
\coordinate (ll) at (-2.75,0.0);
\coordinate (l0) at (-2.25,0.0);
\vtick{(l0)};
\coordinate[label=below:{$\phantom{\delta^2}\widehat{y}_-\phantom{\delta^2}$}] (l0b) at ($(l0) + (b)$);
\coordinate (l1) at (-0.25,0.0);
\vtick{(l1)};
\coordinate (l1a) at ($(l1) + (a)$);
\coordinate[label=below:{$\phantom{\delta^2}\widehat{y}_c\phantom{\delta^2}$}] (l1b) at ($(l1) + (b)$);
\coordinate (lr) at (0.25,0.0);
\coordinate (ml) at (1.5,0.0);
\coordinate (m1) at (2.0,0.0);
\coordinate (m1a) at ($(m1) + (a)$);
\coordinate (m1aa) at ($(m1a) + (a)$);
\vtick{(m1)};
\coordinate[label=below:{$\phantom{1}y+\widehat{o}\phantom{1}$}] (m1b) at ($(m1) + (b)$);
\coordinate (m2) at (3.0,0.0);
\vtick{(m2)};
\coordinate[label=below:{$\downarrow$}] (m2b) at ($(m2) + (b)$);
\coordinate[label=below:{$y+1$}] (m2bbb) at ($(m2b) + (b) + (b)$);
\coordinate (m3) at (4.25,0.0);
\coordinate (m3a) at ($(m3) + (a)$);
\coordinate (m3aa) at ($(m3a) + (a)$);
\coordinate (m3aaa) at ($(m3aa) + (a)$);
\vtick{(m3)};
\coordinate[label=below:{$y+1+\widehat{o}$}] (m3b) at ($(m3) + (b)$);
\coordinate (mr) at (4.75,0.0);
\coordinate (rl) at (6.0,0.0);
\coordinate (r1) at (6.5,0.0);
\vtick{(r1)};
\coordinate[label=below:{$\downarrow$}] (r1b) at ($(r1) + (b)$);
\coordinate[label=below:{$\phantom{1}\phantom{{}_+}\widehat{y}_+\phantom{1}$}] (r1bbb) at ($(r1b) + (b) + (b)$);
\coordinate (r2) at (7.5,0.0);
\coordinate (r2aa) at ($(r2) + (a) + (a)$);
\vtick{(r2)};
\coordinate[label=below:{$\widehat{y}_++\widehat{o}$}] (r2b) at ($(r2) + (b)$);
\coordinate (rr) at (8.0,0.0);
\draw [-,color=black,line width=0.3mm] (ll)--(lr);
\draw [-,color=black,dotted,line width=0.3mm] (lr)--(ml);
\draw [-,color=black,line width=0.3mm] (ml)--(mr);
\draw [-,color=black,dotted,line width=0.3mm] (mr)--(rl);
\draw [->,color=black,line width=0.3mm] (rl)--(rr);
\draw[->,dashed] (l1a) to[out=60, in=120, looseness=1.0] (m3aaa);
\draw[->] (m3aa) to[out=120, in=60, looseness=1.0] (m1aa);
\draw[->] (m3aa) to[out=60, in=150, looseness=1.0] (r2aa);
\draw[->,dash dot] (m1a) to[out=60, in=120, looseness=1.0] (m3a);
\end{tikzpicture}
\caption{\it Visual representation of the stages of the process $(\widehat{y}_n, \widehat{o}_n)_{n\geq 1}$. The steps taken when the second variable equals $1-\delta^2$ right after starting the process are indicated by the dashed arrow. This part of a passage ends at $y+1+\widehat{o}$, in which $\widehat{o}$ is the random overshoot over $y+1$. It is determined at the time on which the faster comparison process becomes larger or equal to $y+1$ for the first time. Afterwards, the process follows one of the solid arrows, either stopping the process (setting it to $\infty$) after becoming larger than $\widehat{y}_++\widehat{o}$ or the faster comparison process becomes less than $y+\widehat{o}$. In the latter case, the dash-dotted arrow leads to a new determination of the overshoot, after which the previous sentence applies again.}
\label{fig:mu-slower}
\end{center}
\end{figure}
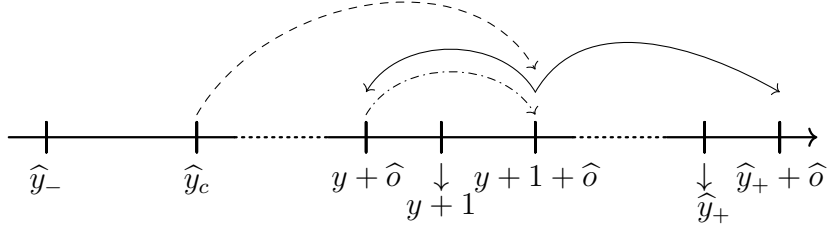
Proceeding like in the previous section, a slower comparison process will be constructed, showing Lemma~\ref{lemma:Y-slower} as a counterpart to Lemma~\ref{lemma:Y-faster}. For $k\geq 0$, define a Markov process $(\widehat{y}_{k,n},\widehat{o}_{k,n})_{n\geq 1}$ on the space $\dot{\mathbb{R}} \times [0,1-\delta^2]$ by setting $\widehat{y}_{k,1} = \widehat{y}_-$, $\widehat{o}_{k,1} = 1-\delta^2$ and for $n\geq 1$,
\begin{align}\label{eq:slower-Y-dyn}
\begin{aligned}
\widehat{y}_{k,n+1} &\;:=\; \begin{cases}
\widehat{y}_c\,, &\text{ if } \widehat{y}_{k,n} \leq \widehat{y}_-\,,\\
\infty\,, &\text{ if } \widehat{y}_{k,n} \geq  \widehat{y}_++\widehat{o}_{k,n}\,,\\
\widehat{y}_{k,n} + (-1)^k\ChiVar_{\vphantom{\widetilde{k}}N_k+n} - \delta^2\,, &\text{ else}\,,
\end{cases}
\\
\widehat{o}_{k,n+1} &\;:=\; \begin{cases}
\widehat{y}_{k,n+1} - y-1\,, &\text{ if } \widehat{y}_{k,n+1} \geq y+1 \text{ and } \widehat{o}_{k,n} = 1-\delta^2\,,\\
1-\delta^2\,, &\text{ if } \widehat{y}_{k,n+1} \leq y+\widehat{o}_{k,n} \text{ and } \widehat{o}_{k,n} \neq 1-\delta^2\,,\\
\widehat{o}_{k,n}\,, &\text{ else}\,.
\end{cases}
\end{aligned}
\end{align}
%

\begin{remark}\label{rem:slower}
{\rm The first component of this slower comparison process is quite different from the slower comparison process used in~\cite{DSS}, as it now includes a negative drift ($-\delta^2$ in the $y$-picture). On the other hand, it resembles the faster comparison process defined in~\eqref{eq:faster-Y-dyn}. Concretely, besides the sign change of the drift, the starting point changed from $\widetilde{y}_c$ to $\widehat{y}_-$ or $\widehat{y}_c$ after one step (which is of minor importance) and the conditions $\widetilde{y}_{k,n} \leq \widetilde{y}_-$ and $\widetilde{y}_{k,n} \geq  \widetilde{y}_+-1-\delta^2+\widetilde{o}_{k,n}$ are replaced by $\widehat{y}_{k,n} \leq \widehat{y}_-$ and $\widehat{y}_{k,n} \geq  \widehat{y}_++\widehat{o}_{k,n}$ . In a nutshell, the drift is reversed, $\widehat{y}_-$ takes the role of $\widetilde{y}_-$ and $\widehat{y}_+$ that of $\widetilde{y}_+$ (up to contributions that are small with respect to $\delta^{-1}$, see Lemma~\ref{lemma:y-lims}).
}
\hfill $\diamond$
\end{remark}

The dynamics of the first component defined in~\eqref{eq:slower-Y-dyn} is visualized in Figure~\ref{fig:mu-slower}.

\vspace{.2cm}

\noindent\textbf{Proof} of Lemma~\ref{lemma:Y-slower}. Analogous to the proof of Lemma~\ref{lemma:Y-faster} ({\it cf.} statement (21) from ~\cite{DSS}).\hfill $\square$

\vspace{.2cm}

The second component of each process is again an overshoot, this time over $y+1$ instead of $y$. It is bounded by $1-\delta^2$ and is set back to this value if the first component again becomes smaller than $y+\widehat{o}_{k,n}$. Up to the precise details of its definition, its interpretation is the same as for the faster comparison process. The same holds for the following quantities. Again set
\begin{align*}
\widehat{s}_n &\;:=\; \mathbf{1}[\widehat{o}_n \neq 1+\delta^2]\mathbf{1}[\widehat{y}_n \neq \infty]\,,\\
\widehat{S} &\;:=\; \sum_{n=1}^{\infty} \widehat{s}_n\,,\\
\widehat{T}_{\shortrightarrow} &\;:=\; \inf\{n\geq 1 \,:\, \widehat{y}_n \geq y+1\}\,,\\
\widehat{S}_{\leftrightarrow} &\;:=\; \inf\{n\geq 1 \,:\, \widehat{y}_{\widehat{T}_{\scriptshortrightarrow}+n} \notin (y+\widehat{o}_{n},\widehat{y}_++\widehat{o}_n)\}\,,\\
\widehat{T} &\;:=\; \inf\{n\geq 1 \,:\, \widehat{y}_n = \infty\}\,,\\
\widehat{\ell} &\;:=\; \mathbb{E}\big[\widehat{y}_{\widehat{T}_{\scriptshortrightarrow}}\big] - \mathbb{E}\big[\widehat{y}_{\widehat{T}_{\scriptshortrightarrow}+\widehat{S}_{\leftrightarrow}} \,\big|\, \widehat{y}_{\widehat{T}_{\scriptshortrightarrow}+\widehat{S}_{\leftrightarrow}} \leq y+\widehat{o}_{\widehat{T}_{\scriptshortrightarrow}}\big]\,,\\
\widehat{r} &\;:=\; \mathbb{E}\big[\widehat{y}_{\widehat{T}_{\scriptshortrightarrow}+\widehat{S}_{\leftrightarrow}} \,\big|\, \widehat{y}_{\widehat{T}_{\scriptshortrightarrow}+\widehat{S}_{\leftrightarrow}} \geq \widehat{y}_++\widehat{o}_{\widehat{T}_{\scriptshortrightarrow}}\big] - \mathbb{E}\big[\widehat{y}_{\widehat{T}_{\scriptshortrightarrow}}\big]
\,.
\end{align*}
All of these notions are defined for a single passage, that is, an index $k\geq 0$ is suppressed. If it again labels the passage for which the objects introduced above are defined, it follows that $(\widehat{T}_{2k})_{k\geq 0}$, $(\widehat{T}_{2k+1})_{k\geq 0}$, $(\widehat{S}_{2k})_{k\geq 0}$ and $(\widehat{S}_{2k+1})_{k\geq 0}$ are families of i.i.d. random variables. As a consequence, the elements of the family of random variables $(\widehat{T}_{2k}+\widehat{T}_{2k+1})_{k\geq 0}$ are interarrival times with a corresponding renewal process $(\widehat{P}_N)_{\vphantom{\widehat{k}}N\geq 1}$, which is for all $N\geq 1$ defined by
\begin{gather}\label{eq:slower-P}
\widehat{P}_N \;:=\; \max\left\{K\geq 1 \,:\, \sum_{k=0}^K (\widehat{T}_{2k}+\widehat{T}_{2k+1}) \leq N\right\}\,.
\end{gather}

\begin{lemma}\label{lemma:expectation-slower}
The expectation values of $\widehat{S}_{\leftrightarrow}$ is finite.
\end{lemma}

\noindent\textbf{Proof.}
The proof is identical to that of Lemma~\ref{lemma:expectation-faster} after replacing $\ChiVar$ by $-\ChiVar$.
\hfill $\square$

\begin{lemma}\label{lemma:slower-T-S}
For the slower comparison process $(\widehat{y}_n,\widehat{o}_n)_{n\geq 1}$, one has
$$
\mathbb{E}[\widehat{S}] 
\;=\;
\frac{\delta^{-2}}{\mathbb{E}[\log(\kappa)^2]}\left(\frac{1-z}{2}\right)^2\left[1 + \mathcal{O}(\delta)\right]\,.
$$
\end{lemma}

\noindent\textbf{Proof.}
Exactly as in the proof of Lemma~\ref{lemma:faster-T-S}, it should be clear that the diagram given by
\begin{center}
\begin{tikzcd}
{\widehat{S}: \widehat{S}_{\leftrightarrow}} \arrow[rr, dashed, "\leq y+\widehat{o}_{\widehat{T}_{\scriptshortrightarrow}}"] \arrow[rrd, dashed, "\geq \widehat{y}_++\widehat{o}_{\widehat{T}_{\scriptshortrightarrow}}"'] &  & \widehat{S}  \\&  & 0               
\end{tikzcd}
\end{center}
gives a correct description of the contributions to $\mathbb{E}[\widetilde{S}]$, and consequently the calculation leading to~\eqref{eq-DiagrammFormula2} holds upon replacing all hats with tildes, swapping $y-1-\delta^2+\widetilde{o}_{\widetilde{T}_{\scriptshortrightarrow}}$ for $y+\widehat{o}_{\widehat{T}_{\scriptshortrightarrow}}$ and $\widetilde{y}_+-1-\delta^2+\widetilde{o}_{\widetilde{T}_{\scriptshortrightarrow}}$ for $\widehat{y}_++\widehat{o}_{\widehat{T}_{\scriptshortrightarrow}}$. Finally, reversing the drift (writing $-\delta^2$ for $\delta^2$), the derivation of~\eqref{ineq:faster-S} now yields
\begin{gather}\label{ineq:slower-S}
\mathbb{E}[\widehat{S}] 
\;=\; 
\frac{\mathbb{E}\big[\widehat{S}_{\leftrightarrow}\big]}{\mathbb{P}\big[\widehat{y}_{\widehat{T}_{\scriptshortrightarrow}+\widehat{S}_{\leftrightarrow}} \geq \widehat{y}_++\widehat{o}_{\widehat{T}_{\scriptshortrightarrow}}\big]} 
\;=\; 
\frac{\widehat{\ell}}{\delta^2} \cdot \frac{\mathbb{P}\big[\widehat{y}_{\widehat{T}_{\scriptshortrightarrow}+\widehat{S}_{\leftrightarrow}} \leq y+\widehat{o}_{\widehat{T}_{\scriptshortrightarrow}}\big]}{\mathbb{P}\big[\widehat{y}_{\widehat{T}_{\scriptshortrightarrow}+\widehat{S}_{\leftrightarrow}} \geq \widehat{y}_++\widehat{o}_{\widehat{T}_{\scriptshortrightarrow}}\big]} - \frac{\widehat{r}}{\delta^2}\,.
\end{gather}
Applying the same modifications (implying that $\tau = -\frac{2\delta^2}{C_0\mathbb{E}[\ChiVar^2]} + \mathcal{O}(\delta^4)$ will also be negative in this case) changes the result of the final calculation in the proof of Lemma~\ref{lemma:faster-T-S} to
$$
\frac{\mathbb{P}\big[\widehat{y}_{\widehat{T}_{\scriptshortrightarrow}+\widehat{S}_{\leftrightarrow}} \leq y+\widehat{o}_{\widehat{T}_{\scriptshortrightarrow}}\big]}{\mathbb{P}\big[\widehat{y}_{\widehat{T}_{\scriptshortrightarrow}+\widehat{S}_{\leftrightarrow}} \geq \widehat{y}_++\widehat{o}_{\widehat{T}_{\scriptshortrightarrow}}\big]} 
\;=\; 
\frac{1}{\widehat{\ell}} \cdot \left[\widehat{r} + \frac{(1-z)^2}{4\mathbb{E}[\log(\kappa)^2]} + \mathcal{O}(\delta)\right]\,.
$$
The sign change before the second term between brackets results from the reversal of the drift, and the error bound slightly improves due to the different error bounds in Lemma~\ref{lemma:y-lims} for variables with hats and with tildes. Again, Inserting this into~\eqref{ineq:slower-S} finishes the proof.
\hfill $\square$

\subsection{Proofs of the results on the invariant measure}
\label{sec-ProofInvMeas}

From their definitions as passage times and Lemmata~\ref{lemma:Y-slower} and~\ref{lemma:Y-faster}, it is apparent that
\begin{gather}\label{ineq:T}
\widetilde{T}_k 
\;\leq\; 
N_{k+1} - N_k 
\;\leq\; 
\widehat{T}_k
\end{gather}
holds (for all realizations) for all $k\geq 0$. With the definitions~\eqref{eq:slower-P} and~\eqref{eq:faster-P}, it follows for all $N\geq 1$ that
\begin{gather}\label{ineq:P}
\widehat{P}_N 
\;\leq\; P_N \,:=\, \max\left\{K\geq 1 \,:\, \sum_{k=0}^K ((N_{2k+1} - N_{2k})+(N_{2k+2} - N_{2k+1})) \leq N\right\} 
\;\leq \;
\widetilde{P}_N\,.
\end{gather}
As already indicated before, the passage times of the actual process do not need to be identically distributed or independent. The same holds for the expression in the middle of~\eqref{ineq:P}, that is, any two random variables of the family $(P_N)_{\vphantom{\widehat{k}}N\geq 1}$ can be correlated or differently distributed. The passage times of the comparison processes can be controlled, as indicated by the following result.

\begin{lemma}\label{lemma:balanced}
Assume that $\mathbb{E}[\log(\kappa)] = 0$. The comparison processes $\{(\widehat{y}_{k,n},\widehat{o}_{k,n})_{n\geq 1}\}_{k\geq 0}$ and $\{(\widetilde{y}_{k,n},\widetilde{o}_{k,n})_{n\geq 1}\}_{k\geq 0}$ defined in~\eqref{eq:slower-Y-dyn} and~\eqref{eq:faster-Y-dyn} obey for all $k\geq 0$
$$
\frac{1}{\mathbb{E}[\widehat{T}_k]} 
\;=\; 
\delta^2\mathbb{E}[\log(\kappa)^2][1 + \mathcal{O}(\delta)]\,,\qquad
\frac{1}{\mathbb{E}[\widetilde{T}_k]} 
\;=\; \delta^2\mathbb{E}[\log(\kappa)^2][1 + \mathcal{O}(\delta\log(\delta))]\,.
$$
\end{lemma}

\noindent\textbf{Proof.} Remark~\ref{rem:faster} states that the first components of the faster processes are almost identical to the corresponding ones (with the same notation, though provided in the $x$-picture) given in~\cite{DSS}. The only difference is present in the conditions that set the first component of the present processes to $\infty$. This means that the notation $\widetilde{T}$ is not precisely the same in~\cite{DSS} and the present work. After a close inspection of~\eqref{eq:faster-Y-dyn}, one can note that $\widetilde{y}_+-1-\delta^2 \leq \widetilde{y}_+-1-\delta^2+\widetilde{o}_{k,n} \leq \widetilde{y}_+$. This implies that the current faster processes are even faster than those in~\cite{DSS}. However, it can be verified that replacing $\widetilde{y}_+$ by $\widetilde{y}_+-1-\delta^2$ in~\cite{DSS} would not change the results of Proposition 3 there. In conclusion, Proposition 3 of~\cite{DSS} still holds for the notation used in this work. The proof of this result given in~\cite{DSS} can be modified for the case in which the sign of the additional drift is reversed, which moreover does not change the result. It then applies for the slower comparison processes used in this work (once more inserting the adapted starting point and stop condition as described in Remark~\ref{rem:slower}), which finishes the proof. A careful treatment yields the correct error bounds.
\hfill $\square$

\vspace{.2cm}

Now all preparations for the proofs of Lemmata,~\ref{lemma:mu-slower},~\ref{lemma:mu-faster} and~\ref{coro:xi-support} are available.

\vspace{.2cm}

\noindent\textbf{Proof} of Lemma~\ref{lemma:mu-slower}. A crucial observation is provided by
\begin{gather}\label{ineq:S-slower}
\widehat{S}_{k}\; \leq\; S_{k} \,+\, \left[\widehat{T}_k - (N_{k+1} - N_k)\right]
\end{gather}
for all $k\geq 0$. This inequality can be explained as follows: if $\widehat{s}_{k,n} = 1$, hence $\widehat{y}_{k,n} \geq y$, then~\eqref{ineq:Y-slower} from Lemma~\ref{lemma:Y-slower} implies that either $y_{\vphantom{\widehat{k}}N_k+n} \geq y$ (which is counted by $S_{k}$) or $y_{\vphantom{\widehat{k}}N_k+n}$ already left the $k$-th passage (estimated from above by the other contribution). Recall that the renewal process $\widehat{P}_N$ as defined in~\eqref{eq:slower-P} counts the number of completed pairs passages of the slower processes $\{(\widehat{y}_{k,n},\widehat{o}_{k,n})_{n\geq 1}\}_{k\geq 0}$ up to the time $N\geq 1$. The following calculation is heavily inspired by the derivation on pages 418-420 of~\cite{GS}. Now~\eqref{eq:S},~\eqref{eq:slower-P},~\eqref{ineq:P} and~\eqref{ineq:S-slower} prove the first three estimates below; then a summand is added and a nonnegative $\tau \in \mathbb{R}^+$ appears; the last equality follows from an alternative form of Wald's identity (Exercise 3.6. of~\cite{Eth}) with the inner sum yielding i.i.d. random variables (see~\cite{DSS}) for the stopping time (see page 418 of~\cite{GS}) $\widehat{P}_N + 1$. Recalling that the passages corresponding to odd $k$ are in $(-\infty,0)$ (see the discussion after \eqref{eq-PassageTimes}), one finds
\begin{align*}
\mathbb{E}\sum_{n=1}^N \mathbf{1} \big[z_n \in [z,1],\nu_n=-\big] \;&\geq\; \mathbb{E} \sum_{k=0}^{P_N} S_{2k+1}\\
&\geq \;\mathbb{E} \sum_{k=0}^{\widehat{P}_N} S_{2k+1}\\
&\geq\; \mathbb{E} \sum_{k=0}^{\widehat{P}_N} \left[\widehat{S}_{2k+1} - (\widehat{T}_{2k+1} - \widetilde{T}_{2k+1})\right]\\
&\geq \;\mathbb{E} \sum_{k=0}^{\widehat{P}_N+1} 
\left[\min\{\widehat{S}_{2k+1}, \tau\} - (\widehat{T}_{2k+1} - \widetilde{T}_{2k+1})\right] - \tau\\
&=\;\mathbb{E}[\widehat{P}_N + 1]\mathbb{E}\left[\min \{\widehat{S}_{1}, \tau\} -\widehat{T}_1 + \widetilde{T}_1\right] - \tau
\\
&\geq \;\mathbb{E}[\widehat{P}_N]\mathbb{E}\left[\min\{\widehat{S}_{1}, \tau\} - \widehat{T}_1 + \widetilde{T}_1\right] - \mathbb{E}[\widehat{T}_1] - \tau\,.
\end{align*}
Applying the elementary renewal theorem first and then taking $\tau \to \infty$ yield
\begin{align*}
\lim_{N \to \infty}\mathbb{E}\,\frac{1}{N}\sum_{n=1}^N \mathbf{1}\big[z_n \in [z,1],\nu_n=-\big] 
&
\;\geq \;
\frac{\mathbb{E}\left[\min\{\widehat{S}_{1}, \tau\} - \widehat{T}_1+\widetilde{T}_1\right]}{\mathbb{E}[\widehat{T}_1+\widehat{T}_2]}
\\
&\rightarrow \;\frac{\mathbb{E}[\widehat{S}_{1}]}{\mathbb{E}[\widehat{T}_1+\widehat{T}_2]}\, -\,\frac{\mathbb{E}[\widehat{T}_1-\widetilde{T}_1]}{\mathbb{E}[\widehat{T}_1+\widehat{T}_2]}\,.
\end{align*}
The same holds for $\nu=+$.  Inserting the results from Lemmata~\ref{lemma:slower-T-S} and~\ref{lemma:balanced} finishes the proof.\hfill $\square$

\vspace{.2cm}

\noindent\textbf{Proof} of Lemma~\ref{lemma:mu-faster}.
Though the steps of this proof are similar to that of Lemma~\ref{lemma:mu-slower}, the details that should be adapted are numerous enough to provide a complete proof here. A crucial observation is provided by
\begin{gather}\label{ineq:S-faster}
\widetilde{S}_{k} \;\geq\; S_{k} - \left[(N_{k+1} - N_k) - \widetilde{T}_k\right]
\end{gather}
for all $k\geq 0$. This inequality can be explained as follows: if the actual dynamics obeys $y_{\vphantom{\widehat{k}}N_k+n} \geq y$ for some $n \in \{1,\dots,N_{k+1} - N_k-1\}$, then~\eqref{ineq:Y-faster} from Lemma~\ref{lemma:Y-faster} implies that either $\widetilde{y}_{k,n} \geq y$ or $\widetilde{y}_{k,n} = \infty$. If then $\widetilde{y}_{k,n} \neq \infty$, this is counted by $\widetilde{S}_{k}$. Whenever the other case holds, this number of steps is bounded from above by the subtracted contribution. Completely similar to the proof of Lemma~\ref{lemma:mu-slower} (without the need for $\tau >0$ here), one finds as above for $\nu=-$
\begin{align*}
\lim_{N \to \infty}\mathbb{E}\frac{1}{N}\sum_{n=1}^N \mathbf{1} \big[z_n \in [z,1],\nu_n=-\big] 
\;&\leq \;\limsup_{N \to \infty}\mathbb{E}\frac{1}{N}  \sum_{k=0}^{P_N+1} S_{2k+1}
\\
&\leq \;\limsup_{N \to \infty}\mathbb{E}\frac{1}{N} \sum_{k=0}^{\widetilde{P}_N+1}S_{2k+1}\\
&\leq \;\lim_{N \to \infty}\mathbb{E}\frac{1}{N} \sum_{k=0}^{\widetilde{P}_N+1}\left[\widetilde{S}_{2k+1} + (\widehat{T}_{2k+1} - \widetilde{T}_{2k+1})\right]\\
&=\;\lim_{N \to \infty}\frac{\mathbb{E}[\widetilde{P}_N + 1]}{N}\,\mathbb{E}\left[\widetilde{S}_{1}+\widehat{T}_1 -\widetilde{T}_1\right]\\
&\leq \;\lim_{N \to \infty}\frac{\mathbb{E}[\widetilde{P}_N]}{N}\;\mathbb{E}\left[\widetilde{S}_{1}+\widehat{T}_1 - \widetilde{T}_1\right] \,+\, \lim_{N \to \infty}\frac{\mathbb{E}
[{\widehat{T}_1}]}{N}\\
&= \;\frac{\mathbb{E}[\widetilde{S}_{1}]}{\mathbb{E}[\widetilde{T}_1+\widetilde{T}_2]} 
\,+\, 
\frac{\mathbb{E}[\widehat{T}_1-\widetilde{T}_1]}{\mathbb{E}[\widetilde{T}_1+\widetilde{T}_2]} 
\,.
\end{align*}
Once more, the same estimate for $\nu=+$ combined with the results Lemmata~\ref{lemma:faster-T-S} and~\ref{lemma:balanced} then finish the proof. \hfill $\square$

\vspace{.2cm}

\noindent\textbf{Proof} of Lemma~\ref{coro:xi-support}.
From the first and the third line of~\eqref{stat:slower-faster-y} in Lemma~\ref{lemma:y}, it follows that the dynamics in the $y$-picture stays at most three steps outside the interval $(\widehat{y}_-,\widehat{y}_+)$ during a single passage (in almost all cases, the intermediate step at $\infty$ is superfluous, but any fixed finite number yields the same result). That is, possibly one at $\infty$, at most one in $\mathbb{R} \setminus (\widehat{y}_-,\infty)$ (by~\eqref{stat:slower-faster-y}, the first implication), and at most one in $\mathbb{R} \setminus (-\infty,\widehat{y}_+)$ (by~\eqref{stat:slower-faster-y}, the contrapositive of the third implication). Recall from the definition~\eqref{ineq:P} that $P_N$ counts the number of completed pairs of passages up to the time $N\geq 1$. Then the foregoing analysis,~\eqref{ineq:P}, an alternative form of Wald\^as
identity (Exercise 3.6. of~\cite{Eth}), the elementary renewal theorem and Lemma~\ref{lemma:balanced} imply
\begin{align*}
\lim_{N \to \infty}\mathbb{E}\frac{1}{N}\sum_{n=1}^N \mathbf{1}\big[y_n \notin [\widehat{y}_-,\widehat{y}_+]\big] 
& \;\leq \;\lim_{N \to \infty}\mathbb{E}\frac{1}{N}\sum_{k=0}^{P_N+1} 3
\;\leq\; \lim_{N \to \infty}\mathbb{E}\frac{1}{N}\sum_{k=0}^{\widetilde{P}_N+1} 3\\
&\;= \;3\lim_{N \to \infty}\frac{\mathbb{E}[\widetilde{P}_N+1]}{N}
\;=\; \frac{3}{\mathbb{E}[\widetilde{T}_1 + \widetilde{T}_2]}
\;=\; \mathcal{O}(\delta^2)\,,
\end{align*}
completing the proof. 
\hfill $\square$

\section{Perturbation theory for the Lyapunov exponent}
\label{sec:gamma}

This section proves Theorem~\ref{theo-Lyapunov} on the behavior of the Lyapunov exponent of families of random matrices of the form~\eqref{eq:expansion} satisfying the Main Hypothesis, namely on the Lyapunov exponent near a balanced hyperbolic critical point of rotating type (that is, obeying $C_1>0$).  The analysis of the Lyapunov exponent starts out with the well-known fact~\cite{BL} that the Lyapunov exponent~\eqref{eq-LyapDef} can be calculated by Furstenberg's formula
\begin{equation}
\label{eq-FurstenbergFormula}
\gamma^\epsilon
\;=\;
\int \mu^\epsilon(d\theta)\;
\EM\,\log\big(\|T^\epsilon e_\theta\|\big)
\;.
\end{equation}
In order to apply Theorem~\ref{theo:mu} it is necessary to make the change of variables to the $z$-picture (rescaled Dyson-Schmidt variables). This is done in the next lemma.

\begin{lemma}
\label{lem-Gamma-z}
The Lyapunov exponent satisfies
\begin{equation}
\label{eq-FurstenbergFormula2}
\gamma^\epsilon
\;=\;
\frac{1}{2}\,\int \mu_s^\epsilon(d z)\;f^\epsilon(z)
\,+\,\Oo(\epsilon)
\;,
\end{equation}
where $f^\epsilon:\dot{\mathbb{R}}\to \mathbb{R}$ is given by $f^\epsilon(\infty)=0$ and
\begin{equation}
\label{eq:f-Birkhoff}
f^\epsilon(z)
\;=\;
\mathbb{E}\,\log\left[\frac{\epsilon^{-z-\frac{\log(\kappa^2)}{\log(\epsilon^{-1})}} + \epsilon^{z+\frac{\log(\kappa^2)}{\log(\epsilon^{-1})}}}{\epsilon^{-z} + \epsilon^{z}}\right]
\;.
\end{equation}
\end{lemma}

\noindent {\bf Proof.} Due to~\eqref{eq-FurstenbergFormula} and the push-forward measure $\mu^\epsilon_\pm$, one has to compute $\EM\,\log\big(\|T^\epsilon e_\theta\|\big)$ with $\theta=\arccot(-\epsilon^z)$ up to corrections of order $\Oo(\epsilon)$. Here $\EM$ denotes merely the average over $\sigma\mapsto T_\sigma^\epsilon$ w.r.t. ${\bf p}$. Replacing~\eqref{eq:expansion}, one finds with uniform error bounds
\begin{align*}
2\,\log(\|T^\epsilon e_{\theta}\|) 
&
\;=\; \log\left[\left\|\left[\one + \mathcal{O}(\epsilon)\right]\begin{pmatrix}\kappa\cos(\theta)\\\kappa^{-1}\sin(\theta)\end{pmatrix}\right\|^2\right]\\
&
\;=\; 
\log(\kappa^2\cos(\theta)^2 + \kappa^{-2}\sin(\theta)^2) + \mathcal{O}(\epsilon)\\
&
\;=\; 
\log\left[\frac{\kappa^2\cos(\theta)^2 + \kappa^{-2}\sin(\theta)^2}{\cos(\theta)^2 + \sin(\theta)^2} \cdot \frac{(\sin(\theta)\cos(\theta))^{-1}}{(\sin(\theta)\cos(\theta))^{-1}}\right] + \mathcal{O}(\epsilon)\\
&
\;=\; 
\log\left[\frac{\kappa^2 \cot(\theta) + \kappa^{-2}\cot(\theta)^{-1}}{\cot(\theta) + \cot(\theta)^{-1}}\right] + \mathcal{O}(\epsilon)
\\
&
\;=\; 
\log\left[\frac{\kappa^2\epsilon^{-z} + \kappa^{-2}\epsilon^{z}}{\epsilon^{-z} + \epsilon^{z}}\right] + \mathcal{O}(\epsilon)
\,.
\end{align*}
This function depends on $z$, but is independent of $\nu$ and therefore immediately implies the claim. The value $z=\infty$ is chosen to assure continuity of $f^\epsilon$.
\hfill $\Box$

\vspace{.2cm}

\begin{figure}[h]
\begin{center}
\includegraphics[width=0.4\textwidth]{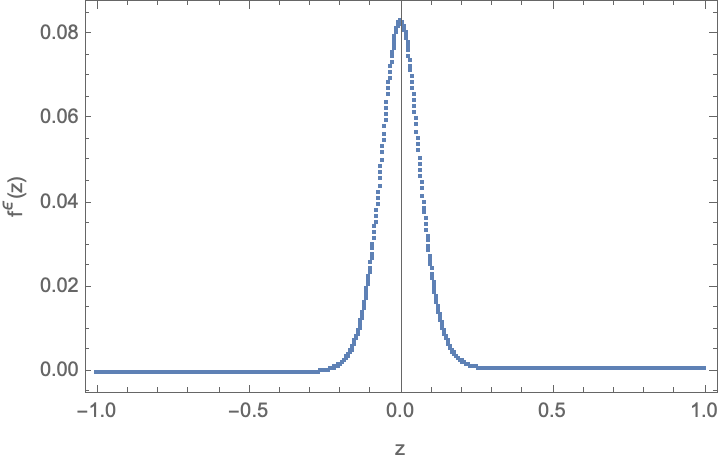}
\hspace{.3cm}
\includegraphics[width=0.4\textwidth]{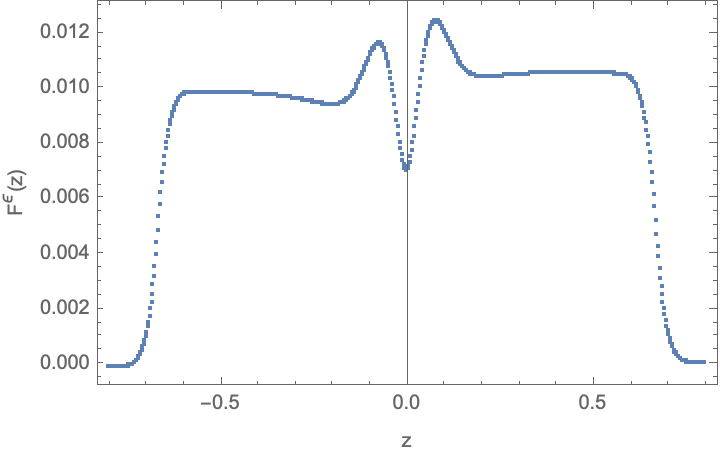}
\caption{\it Plot of the functions $f^{\epsilon}$ and $F^\epsilon$ for the random hopping model with $\epsilon = 10^{-6}$ and  $t - 1.1$ being uniformly distributed on $[-0.4,0.4]$. The expectations in $f^{\epsilon}$ and $F^{\epsilon}$ with $Z = \frac{2}{3}$  were computed as  averages over $10^6$ samples. Note that $f^{\epsilon}$ is symmetric here because the distribution of $\log(\kappa)$ for the random hopping model is symmetric as well. Also note that $F^\epsilon$ is approximately constant on $I_\gamma$ and vanishing on $I_\infty$, illustrating {\rm Lemma~\ref{lemma:cocycle}}. 
}
\label{fig:f}
\end{center}
\end{figure}

The next aim is to analyze the function $f^\epsilon$. The analytical results are given in Lemma~\ref{lemma:gamma-xi}  (in a non-optimal form that is to be applied in the proof of Lemma~\ref{lemma:cocycle}) and a numerical plot of the function for the case of a random hopping model is given in Figure~\ref{fig:f}. The crucial facts for the following are that its value at  $z = 0$ is independent of $\epsilon$ and, in particular, always macroscopic, and furthermore that the function is positive with a rapid decay to $0$ outside of an interval of size $\Oo(\delta)$ around $0$. 

\begin{lemma}
\label{lemma:gamma-xi}
For all $z \in \mathbb{R}$, the following statements hold:
\begin{align}
\label{stat:f-Birkhoff-bounds}
& f^{\epsilon}(z) \,\in\, (0,2C_0)\,,
\\
\label{eq:f-Birkhoff-zero}
& f^{\epsilon}(0) \,=\, \mathbb{E}\log\left[\tfrac{1}{2}(\kappa^2 + \kappa^{-2})\right]\,,
\\
\label{stat:f-Birkhoff-derivative}
& |z| > 2C_0\delta\quad \Longrightarrow\quad \operatorname{sgn}(z)\partial_{z}f^{\epsilon}(z) < 0\,,
\\
\label{stat:f-Birkhoff-decay}
& |z| \geq \delta\log(\delta^{-1})\quad \Longrightarrow\quad f^{\epsilon}(z) = \mathcal{O}\left(\epsilon^{2|z|}\right)\,.
\end{align}
\end{lemma}

%

\noindent\textbf{Proof.}
The upper bound in~\eqref{stat:f-Birkhoff-bounds} follows from the Main Hypothesis, using ${\bf p}[\kappa = 1] < 1$ for the strictness of the first inequality:
$$
f^{\epsilon}(z) 
\;<\; 
\mathbb{E}\log\left[\frac{\epsilon^{-z-\frac{2|\log(\kappa)|}{\log(\epsilon^{-1})}} + \epsilon^{z-\frac{2|\log(\kappa)|}{\log(\epsilon^{-1})}}}{\epsilon^{-z} + \epsilon^{z}}\right] 
\;\leq \;
\mathbb{E}\log\left[\frac{\epsilon^{-z-\frac{2C_0}{\log(\epsilon^{-1})}} + \epsilon^{z-\frac{2C_0}{\log(\epsilon^{-1})}}}{\epsilon^{-z} + \epsilon^{z}}\right] 
\;=\; 
2C_0\,.
$$
The strict positivity given in~\eqref{stat:f-Birkhoff-bounds} remains to be shown. By $\mathbb{E}(\ChiVar) = 0$, for $z \in \mathbb{R}$ it holds that
\begin{gather}
\label{eq:f-Birkhoff-bis}
\begin{aligned}
f^{\epsilon}(z) 
& \;=\; 
\mathbb{E}\log\left(1 + \epsilon^{2z}\exp(-4C_0\ChiVar)\right) - \log(1 + \epsilon^{2z}) 
\\
& \;=\; 
\mathbb{E}\log\left(1 + \epsilon^{-2z}\exp(4C_0\ChiVar)\right) - \log(1 + \epsilon^{-2z})\,.
\end{aligned}
\end{gather}
Note that 
$$
\frac{d^2}{d\ChiVar^2}\left[\log\left(1 + \epsilon^{2z}\exp(-4C_0\ChiVar)\right)\right] 
\;=\; 
\frac{d}{d\ChiVar}\left[\frac{4C_0}{\epsilon^{-2z}\exp(4C_0\ChiVar) + 1}\right] 
\;=\; 
\frac{16\,C_0^2\epsilon^{-2z}\exp(4C_0\ChiVar)}{(\epsilon^{-2z}\exp(4C_0\ChiVar) + 1)^2} \;>\; 0
$$ 
for all $\ChiVar \in [-1,1]$ and $z \in \mathbb{R}$. The application of a strict version of Jensen's inequality then implies 
$$
f^{\epsilon}(z) 
\;> \;
\log\left(1 + \epsilon^{2z}\exp(-4C_0\mathbb{E}(\ChiVar))\right) - \log(1 + \epsilon^{2z}) 
\;=\; 0
\;.
$$ 
Next,~\eqref{eq:f-Birkhoff-zero} is immediate. For the proof of~\eqref{stat:f-Birkhoff-derivative}, let us start from
\begin{gather}\label{eq:f-Birkhoff-derived}
\partial_{z} f^{\epsilon}(z) 
\;=\;
\log(\epsilon^{-1})\mathbb{E}\left[\frac{1}{\epsilon^{2z}\exp(-4C_0\ChiVar) + 1} - \frac{1}{\epsilon^{2z} + 1}\right]
\;,
\end{gather}
which follows from the second rewriting in~\eqref{eq:f-Birkhoff-bis}. Note that 
$$
\frac{d^2}{d\ChiVar^2}\left[\frac{1}{\epsilon^{2z}\exp(-4C_0\ChiVar) + 1}\right] 
\;=\; 
\frac{16C_0^2(\epsilon^{z}\exp(-2C_0\ChiVar) - \epsilon^{-z}\exp(2C_0\ChiVar))}{(\epsilon^{z}\exp(-2C_0\ChiVar) + \epsilon^{-z}\exp(2C_0\ChiVar))^3}
\;.
$$
As the denominator is always strictly positive and the numerator is proportional (with some positive constant) to $\sinh\left(-z\log(\epsilon^{-1})-2C_0\ChiVar\right)$, the full expression has the same sign as $-z\log(\epsilon^{-1})-2C_0\ChiVar$. As a consequence, a strict version of Jensen's inequality can be applied if $-z\log(\epsilon^{-1}) + 2C_0 < 0$ or $-z\log(\epsilon^{-1}) - 2C_0 > 0$. Inserting~\eqref{eq:f-Birkhoff-derived}, one concludes
\begin{align*}
 &z > 2C_0\delta &&\Longrightarrow &&\partial_{z} f^{\epsilon}(z) < \log(\epsilon^{-1})\left[\frac{1}{\epsilon^{2z}\exp(-4C_0\mathbb{E}(\ChiVar)) + 1} - \frac{1}{\epsilon^{2z} + 1}\right] = 0\,,\\
&z < -2C_0\delta &&\Longrightarrow &&\partial_{z} f^{\epsilon}(z) > \log(\epsilon^{-1})\left[\frac{1}{\epsilon^{2z}\exp(-4C_0\mathbb{E}(\ChiVar)) + 1} - \frac{1}{\epsilon^{2z} + 1}\right] = 0\,,
\end{align*}
which yields~\eqref{stat:f-Birkhoff-derivative}. For the proof of~\eqref{stat:f-Birkhoff-decay} let now $z \in \mathbb{R}$ (possibly depending on $\epsilon$) fullfill $\liminf\limits_{\epsilon \to 0} |z|\log(\epsilon^{-1}) = \infty$, then $\lim\limits_{\epsilon \to 0} \epsilon^{2|z|} = 0$ follows. A Taylor expansion of the first (for $z > 0$) or the second (for $z < 0$) equality of~\eqref{eq:f-Birkhoff-bis} implies~\eqref{stat:f-Birkhoff-decay}. 
\hfill $\square$

\vspace{.2cm}

In the following, only the first contribution to Lyapunov exponent $\gamma^\epsilon$ in~\eqref{eq-FurstenbergFormula2} will be considered because the error term is of lower order than the error in Theorem~\ref{theo-Lyapunov}. For this purpose, one may attempt to apply the information on the invariant measure $\mu^\epsilon_\pm(dz)$ as provided in Theorem~\ref{theo-InvMeasure}. However, by Lemma~\ref{lemma:gamma-xi} the integrand $f^\epsilon$ is of considerable (macroscopic) size only on an interval of size $\delta\log(\delta^{-1})$ and for intervals of such small size Theorem~\ref{theo-InvMeasure} gives little information because the error term is precisely of the same order. One remedy is to modify the function $f^\epsilon$ by a cocycle for the random action $T^\epsilon\star$ to
$$
F^\epsilon(z)
\;=\;
f^\epsilon(z)\,+g^\epsilon(z)\,-\,\EM\,g^\epsilon(T^\epsilon\star z)
\;,
$$
where $g^\epsilon:\dot{\mathbb{R}}\to\mathbb{R}$ is a bounded function which should be chosen such that the integral of $F^\epsilon$ can be accessed by Theorem~\ref{theo-InvMeasure}. Indeed, the invariance property of the invariant measure shows that the two last summands in $F^\epsilon$ cancel upon integration so that
\begin{equation}
\label{eq-FurstenbergFormula3}
\gamma^\epsilon
\;=\;
\int \mu_s^\epsilon(d z)\;F^\epsilon(z)
\,+\,\Oo(\epsilon)
\;,
\end{equation}
In order to find a suitable function $g^{\epsilon}$, it is helpful to note that $f^{\epsilon}$ is almost a cocycle itself, as can be seen by writing out~\eqref{eq:f-Birkhoff} and~\eqref{eq:xi-dyn} in the following manner:
$$
f^{\epsilon}(z) 
\,=\,
\mathbb{E}\log\left[\epsilon^{-z-\frac{\log(\kappa^2)}{\log(\epsilon^{-1})}} + \epsilon^{z+\frac{\log(\kappa^2)}{\log(\epsilon^{-1})}}\right] 
\,-\, 
\mathbb{E}\log\left[\epsilon^{-z} + \epsilon^{z}\right]\,, 
\quad 
T^\epsilon \star z \,=\, z + \frac{\log(\kappa^2)}{\log(\epsilon^{-1})} + \frac{r^{\epsilon}(z)\epsilon^{1-|z|}}{\log(\epsilon^{-1})}\,.
$$
It therefore makes sense to propose that $g^{\epsilon}(z) = \log\left[\epsilon^{-h(z)} + \epsilon^{h(z)}\right]$ for some other function $h:\dot{\mathbb{R}} \to \mathbb{R}$ which is chosen to be approximately linear near $z=0$ and constant outside some interval $[-Z,Z]$ for $Z\in(0,1)$. The linearity at $z=0$ then assures that $F^\epsilon$ is much smaller than $f^\epsilon$ near $z=0$ while the constancy outside $[-Z,Z]$ implies that $F^\epsilon$ inherits the decay property~\eqref{stat:f-Birkhoff-decay} from $f^\epsilon$. The following lemma implements this heuristics and shows that $h$ can simply be chosen to be a piecewise polynomial map (independent of $\epsilon$) and that, moreover, the function $F^\epsilon$ then turns out to be essentially constant on the interval $[-Z,Z]$.  A numerical plot of the outcome $F^\epsilon$ for this choice is given in Figure~\ref{fig:f}.

\begin{lemma}
\label{lemma:cocycle}
Let $Z \in (0,1)$ be a constant. Define the functions $g^{\epsilon}:\dot{\mathbb{R}} \to \mathbb{R}$ and $h:\mathbb{R} \to \mathbb{R}$ by $g^{\epsilon}(\infty) = \log\left[\epsilon^{-\frac{Z}{2}} + \epsilon^{\frac{Z}{2}}\right]$ and, for $z \in \mathbb{R}$,
\begin{gather}\label{eq:g-h}
g^{\epsilon}(z) 
\;:=\; 
\log\left[\epsilon^{-h(z)} + \epsilon^{h(z)}\right]\,,
\qquad 
h(z) 
\;:=\;  
\begin{cases}\sgn(z)\,\frac{Z}{2}\,, & \text{ if } |z| >Z\,,\\
z-\sgn(z)\,\frac{z^2}{2Z}\,, & \text{ if } |z| \leq Z\,.
\end{cases}
\end{gather}
These functions are continuous and bounded. Introducing the disjoint decomposition of $\dot{\mathbb{R}}$ into the four intervals
\begin{align*}
I_0 &\,:=\, \big\{z \in \mathbb{R} \,:\, |z| \in \big[0,\delta\log(\delta^{-1})\big]\big\}\,,
\\ 
I_{\gamma} &\,:=\, 
\big\{z \in \mathbb{R} \,:\, |z| \in \big(
\delta\log(\delta^{-1}),Z-\delta(2C_0+2C_2\exp(2C_0)\epsilon^{1-Z})
\big)\big\}
\,,
\\
I_{Z} &\,:=\, 
\big\{z \in \mathbb{R} \,:\, |z| \in \big[Z-\delta(2C_0+2C_2\exp(2C_0)\epsilon^{1-Z},
Z + \delta( 2C_0+2C_2\exp(2C_0)\epsilon^{1-Z})\big]\big\}\,,
\\
I_{\infty} &\,:=\, 
\big\{z \in \mathbb{R} \,:\, |z| \in \big(Z + \delta( 2C_0+2C_2\exp(2C_0)\epsilon^{1-Z}), \infty\big)\big\} \cup \{\infty\}\,,
\end{align*}
the following bounds hold:
\begin{align}
\label{stat:cocycle-semel}
&z \in I_0 &&\Longrightarrow &&
F^{\epsilon}(z) 
\; =\; 
\mathcal{O}\big(\delta\log(\delta)^2\big)
\,,
\\
\label{stat:cocycle-bis}
&z \in I_{\gamma} &&\Longrightarrow &&
F^{\epsilon}(z) \; = \;
\frac{2\delta}{Z}\,\mathbb{E}\left(\log(\kappa)^2\right) 
+ \mathcal{O}(\delta^2)
\,,
\\
\label{stat:cocycle-ter}
&z \in I_{Z} &&\Longrightarrow &&
F^{\epsilon}(z)\; =\; 
\mathcal{O}(\delta)
\,,
\\
\label{stat:cocycle-quater}
&z \in I_{\infty} &&\Longrightarrow &&
F^{\epsilon}(z) \; =\; \mathcal{O}(\epsilon^{2Z})\,.
\end{align}
\end{lemma}

\noindent\textbf{Proof.} The continuity and boundedness of $h$ are immediate. The same properties then follow for $g^{\epsilon}$, for which indeed also $\lim\limits_{z \to \pm \infty} g^{\epsilon}(z) = g^{\epsilon}(Z) = g^{\epsilon}(-Z) = g^{\epsilon}(\infty)$.

\vspace{.1cm}

When showing~\eqref{stat:cocycle-semel},~\eqref{stat:cocycle-bis} and~\eqref{stat:cocycle-ter}, one needs to take several contributions of $T^\epsilon\star z$ into account. Consequently, by~\eqref{eq:xi-dyn}, the quantity $\frac{r^{\epsilon}(z)\epsilon^{1-|z|}}{\log(\epsilon^{-1})}$ will show up. For $z \in I_0 \cup I_{\gamma} \cup I_{Z}$,~\eqref{ineq:r} implies $\frac{r^{\epsilon}(z)\epsilon^{1-|z|}}{\log(\epsilon^{-1})} = \mathcal{O}\left(\frac{\epsilon^{1-Z}}{\log(\epsilon^{-1})}\right)=\Oo(\delta\epsilon^{1-Z})$. Since the latter is substantially smaller than all error estimates in the conclusions of~\eqref{stat:cocycle-semel},~\eqref{stat:cocycle-bis} and~\eqref{stat:cocycle-ter}, the given term will be neglected throughout the rest of this argument.

\vspace{.1cm}

Let $z \in I_0$. It then follows that $\epsilon^{\pm\frac{z|z|}{2Z}} = 1 + \mathcal{O}\left(z^2\log(\epsilon^{-1})\right) = 1 +\Oo(\delta\log(\delta)^2)$. By~\eqref{ineq:r}, the two last terms of~\eqref{eq:xi-dyn} can also be estimated by $\mathcal{O}(\delta\log(\delta^{-1}))$, so $T^\epsilon \star z = \mathcal{O}(\delta\log(\delta^{-1}))$, from which also $\epsilon^{\pm\frac{(T^\epsilon \star z)|T^\epsilon \star z|}{2Z}} = 1 + \mathcal{O}(\delta\log(\delta)^2)$ follows. Therefore,
\begin{align*}
\epsilon^{-h(z)} + \epsilon^{h(z)} &= \epsilon^{-z+\frac{z|z|}{2Z}} + \epsilon^{z-\frac{z|z|}{2Z}}\\
 &= \epsilon^{-z}\left[1 + \mathcal{O}(\delta\log(\delta)^2)\right] + \epsilon^{z}\left[1 + \mathcal{O}(\delta\log(\delta)^2)\right]
 \\
 &= \left[\epsilon^{-z} + \epsilon^{z}\right]\left[1 + \frac{\epsilon^{-z}}{\epsilon^{-z} + \epsilon^{z}}\mathcal{O}(\delta\log(\delta)^2) + \frac{\epsilon^{z}}{\epsilon^{-z} + \epsilon^{z}}\mathcal{O}(\delta\log(\delta)^2)\right]\\
 &= \left[\epsilon^{-z} + \epsilon^{z}\right]\left[1 + \mathcal{O}(\delta\log(\delta)^2) + \mathcal{O}(\delta\log(\delta)^2)\right]\\
 &= \left[\epsilon^{-z} + \epsilon^{z}\right]\left[1 + \mathcal{O}(\delta\log(\delta)^2)\right]\,,
\end{align*}
and similarly $\epsilon^{-h(T^\epsilon \star z)} + \epsilon^{h(T^\epsilon \star z)} = \left[\epsilon^{-z-\delta\log(\kappa^2)} + \epsilon^{z+\delta\log(\kappa^2)}\right]\left[1 + \mathcal{O}(\delta\log(\delta)^2)\right]$. Then~\eqref{stat:cocycle-semel} is shown by
\begin{align*}
F^{\epsilon}(z)  
& 
\;=\; 
\mathbb{E}\log\left[\frac{\epsilon^{-z-\delta\log(\kappa^2)} + \epsilon^{z+\delta\log(\kappa^2)}}{\epsilon^{-z} + \epsilon^{z}} \cdot \frac{\epsilon^{-h(z)} + \epsilon^{h(z)}}{\epsilon^{-h(T^\epsilon \star z)} + \epsilon^{h(T^\epsilon \star z)}}\right]
\\
&
\;=\; 
\mathbb{E}\log\left[\frac{\epsilon^{-z-\delta\log(\kappa^2)} + \epsilon^{z+\delta\log(\kappa^2)}}{\epsilon^{-z} + \epsilon^{z}} \cdot \frac{\epsilon^{-z} + \epsilon^{z}}{\epsilon^{-z-\delta\log(\kappa^2)} + \epsilon^{z+\delta\log(\kappa^2)}} \cdot \frac{1 + \mathcal{O}(\delta\log(\delta)^2)}{1 + \mathcal{O}(\delta\log(\delta)^2)}\right]
\\
& 
\;=\; 
\mathcal{O}(\delta\log(\delta)^2)\,.
\end{align*}
Let $z \in I_{\gamma}$. Note that $\sgn(z) = \sgn(z + \delta\log(\kappa^2)) = \sgn(h(z)) = \sgn(h(T^\epsilon \star z))$, which implies
\begin{align}\label{eq:cocycle-bis}
\begin{aligned}
F^{\epsilon}(z)  
&
\;=\; 
\mathbb{E}\log\left[\frac{1 + \epsilon^{2\left|z+\delta\log(\kappa^2)\right|}}{1 + \epsilon^{2|z|}} \cdot \frac{1 + \epsilon^{2|h(z)|}}{1 + \epsilon^{2|h(T^\epsilon \star z)|}}\right]\\
&
\qquad+ \log(\epsilon^{-1})\,\mathbb{E}\left[\left|z+\delta\log(\kappa^2)\right| - |z| + |h(z)| - |h(T^\epsilon \star z)|\right]\\
&
\;=\; 
\mathbb{E}\log\left[\frac{1 + \epsilon^{2\left|z+\delta\log(\kappa^2)\right|}}{1 + \epsilon^{2|z|}} \cdot \frac{1 + \epsilon^{2|h(z)|}}{1 + \epsilon^{2|h(T^\epsilon \star z)|}}\right]\\
&
\qquad+ \log(\epsilon^{-1})\sgn(z)\,\mathbb{E}\left[h(z) + \delta\log(\kappa^2) - h(T^\epsilon \star z)\right]\,.
\end{aligned}
\end{align}
Then, using the fact that $\epsilon^{2|z|} = \left[\exp(|z|\log(\epsilon^{-1}))\right]^{-2} = \mathcal{O}(\delta^2)$, as well as analogously $\epsilon^{2\left|z + \delta\log(\kappa^2)\right|} = \mathcal{O}(\delta^2)$, $\epsilon^{2|h(z)|} = \mathcal{O}(\delta^2)$ and $\epsilon^{2|h(T^\epsilon \star z)|} = \mathcal{O}(\delta^2)$,
\begin{align}
\label{eq:cocycle-ter}
\mathbb{E}\log & \left[\frac{1 + \epsilon^{2\left|z+\delta\log(\kappa^2)\right|}}{1 + \epsilon^{2|z|}} \cdot \frac{1 + \epsilon^{2|h(z)|}}{1 + \epsilon^{2|h(T^\epsilon \star z)|}}\right] 
\;=\; 
\mathbb{E}\log\left[\frac{1 + \mathcal{O}(\delta^2)}{1 + \mathcal{O}(\delta^2)} \cdot \frac{1 + \mathcal{O}(\delta^2)}{1 + \mathcal{O}(\delta^2)}\right]
\;= \;
\mathcal{O}(\delta^2)
\end{align}
follows. Using $\mathbb{E}[\log(\kappa^2)] = 0$ in the first (together with the standard error bound on $r^{\epsilon}$) and the last step, the second term on the r.h.s. of~\eqref{eq:cocycle-bis} is equal to
\begin{align}\label{eq:cocycle-quater}
\begin{aligned}
\log(\epsilon^{-1})\sgn(z)
\,
\mathbb{E}&\left[h(z) + \delta\log(\kappa^2) - h(T^\epsilon \star z)\right]\\
 &
\; =\; 
\log(\epsilon^{-1})\sgn(z)\,
\mathbb{E}\left[-\frac{z|z|}{2Z} + \frac{(T^\epsilon \star z)|T^\epsilon \star z|}{2Z} + \mathcal{O}\left(\delta \epsilon^{1-Z}\right)\right]\\
 &
\;=\; 
\frac{\log(\epsilon^{-1})}{2Z}\,\mathbb{E}\left[-z^2 + \left(z + \delta\log(\kappa^2)\right)^2 + \mathcal{O}\left(\delta\epsilon^{1-Z}\right)\right]\\
 &
\;=\; 
\frac{\delta}{2Z}\,
\mathbb{E}\left(\left[\log(\kappa^2)\right]^2\right) + \mathcal{O}\left(\epsilon^{1-Z}\right)\,.
\end{aligned}
\end{align}
Combining~\eqref{eq:cocycle-bis},~\eqref{eq:cocycle-ter} and~\eqref{eq:cocycle-quater} yields~\eqref{stat:cocycle-bis}.

\vspace{.1cm}

Now let $z \in I_{Z}$, so in particular $|z| - Z = \mathcal{O}(\delta)$, which will be sufficient to show~\eqref{stat:cocycle-ter}. Note 
$$
z - \frac{z|z|}{2Z} 
\,=\, 
\frac{\sgn(z)}{2Z}\left[Z^2 - (|z| - Z)^2\right]
\,=\, 
\frac{Z}{2}\sgn(z) + \mathcal{O}(\delta^2)
\,.
$$
It follows from the definition of $h$ in~\eqref{eq:g-h} that the foregoing is then equal to both $h(z)$ and $h(T^\epsilon \star z)$. As a consequence, in full analogy to the proof of~\eqref{stat:cocycle-semel},~\eqref{stat:cocycle-ter} indeed follows from
\begin{align*}
g^{\epsilon}(z) - \mathbb{E}[g^{\epsilon}(T^\epsilon \star z)] 
&
\;=\; 
\mathbb{E}\log\left[\frac{\epsilon^{-h(z)} + \epsilon^{h(z)}}{\epsilon^{-h(T^\epsilon \star z)} + \epsilon^{h(T^\epsilon \star z)}}\right]
\\
&
\;=\; 
\mathbb{E}\log\left[\frac{\epsilon^{-\frac{Z}{2}\sgn(z) + \mathcal{O}(\delta^2)} + \epsilon^{\frac{Z}{2}\sgn(z) + \mathcal{O}(\delta^2)}}{\epsilon^{-\frac{Z}{2}\sgn(z) + \mathcal{O}(\delta^2)} + \epsilon^{\frac{Z}{2}\sgn(z) + \mathcal{O}(\delta^2)}}\right]
\\
&
\;=\; 
\mathbb{E}\log\left[\frac{\epsilon^{-\frac{Z}{2}\sgn(z)} + \epsilon^{\frac{Z}{2}\sgn(z)}}{\epsilon^{-\frac{Z}{2}\sgn(z)} + \epsilon^{\frac{Z}{2}\sgn(z)}} \cdot \frac{1 + \mathcal{O}(\delta)}{1 + \mathcal{O}(\delta)}\right]
\\
&
\;=\; 
\mathcal{O}(\delta)\,,
\end{align*}
as by~\eqref{stat:f-Birkhoff-decay} from Lemma~\ref{lemma:gamma-xi} it holds that $f^{\epsilon}(z) = \mathcal{O}(\epsilon^{2Z})$.

\vspace{.1cm}

As to~\eqref{stat:cocycle-quater}, it follows from the latter and~\eqref{ineq:r} implying $|h(z)| = \frac{Z}{2} = |h(T^\epsilon \star z)|$ for $z \in I_{\infty}$ except $z = \infty$, for which it can readily be verified as well.
\hfill $\square$

\vspace{.2cm}

\noindent\textbf{Proof of Theorem~\ref{theo-Lyapunov}.} The starting point is to apply Theorem~\ref{theo:mu} to the four subsets of $\mathbb{R}$ appearing in Lemma~\ref{lemma:cocycle}.  As the subsets are all symmetric around zero and can all be constructed as differences from sets of the form $[-C,C]$ or $(-C,C)$ for some $C>0$, a practical observation is that, for $C \in (0,1)$, 
\begin{equation}
\label{eq-MeasureSymInterval}
\mu^\epsilon_s([-C,C])
\; =\; 
\left(\frac{1+C}{2}\right)^2 - \left(\frac{1-C}{2}\right)^2 + \Oo\big(\delta\log(\delta^{-1})\big) 
\;=\; 
C + \Oo\big(\delta\log(\delta^{-1})\big)\,.
\end{equation}
From this one deduces
\begin{align}\label{eq:intervals}
\begin{aligned}
& \mu^\epsilon_s(I_0) \;=\; \Oo\big(\delta\log(\delta^{-1})\big)\,, \qquad &&
\mu^\epsilon_s(I_{\gamma}) \;=\; Z + \Oo\big(\delta\log(\delta^{-1})\big)\,,\\
& \mu^\epsilon_s(I_{Z}) \;= \;\Oo\big(\delta\log(\delta^{-1})\big)\,,\qquad &&
\mu^\epsilon_s(I_{\infty}) = 1-Z + \Oo\big(\delta\log(\delta^{-1})\big)\,.
\end{aligned}
\end{align}
Now let us use~\eqref{eq-FurstenbergFormula3} for the Lyapunov exponent with $F^\epsilon$ as given in Lemma~\ref{lemma:cocycle} so that the bounds~\eqref{stat:cocycle-semel},~\eqref{stat:cocycle-bis},~\eqref{stat:cocycle-ter} and~\eqref{stat:cocycle-quater} hold. As the integrand $F^\epsilon$ is constant up to error terms on all four subsets, one deduces
\begin{align*}
2\,\gamma^\epsilon 
\;=\; &
\mu^\epsilon_s(I_0)\cdot \mathcal{O}(\delta\log(\delta)^2)
\,+\,
\mu^\epsilon_s(I_{\gamma})\cdot 
\left(
\frac{2\delta}{Z}\,\mathbb{E}\left(\log(\kappa)^2\right) 
+ \mathcal{O}(\delta^2)
\right)
\\ & 
\,+\,
\mu^\epsilon_s(I_{Z})\cdot \mathcal{O}(\delta)
\,+\,
\mu^\epsilon_s(I_{\infty}) \cdot \Oo(\epsilon^{2Z})
\,+\,\Oo(\epsilon)
\\
\;=\; &
\Oo\big(\delta\log(\delta^{-1})\big) \cdot \mathcal{O}(\delta\log(\delta)^2)
 + \left(Z + \Oo\big(\delta\log(\delta^{-1})\big)\right) \cdot \left(\frac{2\delta}{Z}\,\mathbb{E}\left(\log(\kappa)^2\right) + \mathcal{O}(\delta^2)\right)
\\
&
\,+\, \mathcal{O}\big(\delta\log(\delta^{-1})\big) \cdot \mathcal{O}(\delta) + \left(1-Z + \mathcal{O}(\delta)\right) \cdot \mathcal{O}(\epsilon^{2Z}) + \mathcal{O}(\epsilon)
\\
\;=\; & 
2\,\delta\,\mathbb{E}\left(\log(\kappa)^2\right) 
\,+\, \mathcal{O}\left(\delta^2\log(\delta^{-1})^3\right)\,.
\end{align*}
This finishes the proof. \hfill $\square$
%

\section{Balanced hyperbolic critical points of confined type}
\label{sec-Confined}

This section provides the proof of Theorem~\ref{theo-InvMeasure2}. It is based on a modification of the arguments in Sections~\ref{sec-Dyn} and~\ref{sec:mu} which result from the essentially different properties of the dynamics, as already described in Section~\ref{sec-Intro}. In particular, there will again be passages in logarithmic Dyson-Schmidt variables (including times at which they are completed) and two families of comparison processes (and their corresponding passage times). We will denote these objects by the same symbols $N$, $(\widehat{y},\widehat{o})$, $(\widetilde{y},\widetilde{o})$, $\widehat{T}$ and $\widetilde{T}$, even though their definitions {\it differ} from the ones in Section~\ref{sec:mu}. However, in definitions and estimates, the constant  $C_1$ has to be replaced. However, instead of replacing with $C'_1$ given in the Main Hypothesis, we will rather use 
$$
C''_1 \;:=\; \essinf(-b-|a|)
\;,
$$
which is still supposed to be strictly positive. Indeed, for notational convenience, we will rather deal with the case $b < -|a|$ (corresponding to $C''_1>0$) instead of  $b > |a|$ (corresponding to $C'_1>0$). As explained after~\eqref{eq-TwoStepDynPhases}, the two cases are swapped by a conjugation sending $(a,b)$ to $(-a,-b)$, so this is no limitation. In the following, the constant $C''_1$ will then be contained implicitly in several quantities.  For example, $\widehat{x}_+$ is redefined as $\frac{2e^{2C_0}}{C''_1\epsilon}$ in this section.

\vspace{.2cm}

As argued in Section~\ref{sec-Intro}, the hypothesis $C''_1>0$ implies that the dynamics in the Pr\"ufer variables $\theta^\epsilon_n$ are now confined to $(-\frac{\pi}{2},0)$, leading once more to Dyson-Schmidt variables $x_n^\epsilon$ in $(0,\infty)$. According to \eqref{eq-VariableTransforms}, the logarithmic Dyson-Schmidt variables are then $y^\epsilon_n = \frac{1}{2C_0}\log(x^\epsilon_n)$. For the critical points of rotating type, the dynamics possessed an approximate symmetry encoded by the action of $J'=\binom{0\,-1}{1\;\;0}$, mapping $x \in \dot{\mathbb{R}}$ to $-x^{-1}$. As explained before Lemma~\ref{lemma:Q-lower}, it is therefore sufficient to only treat the case of $x \in (0,\infty)$, corresponding to half of the passages. A similar reduction is possible for critical points of confined type, based on the action of $J''=\binom{0\,1}{1\,0}$, mapping $x$ to $x^{-1}$. This map is orientation-reversing. It satisfies $J''D^\epsilon J''=(D^\epsilon)^{-1}$, and $J''Q^\epsilon J''$ is almost $Q^\epsilon$ as given in \eqref{eq-QFormula} as  merely the sign before $a$ changed and the higher order term is rather $J''A^\epsilon J''$ so that, in particular,  the Main Hypothesis can still be applied. Therefore, the action of the dynamics on $y \in \mathbb{R}$ in the $y$-picture is up to the orientation and errors equal to those of $J'' \ast y = -y$.

\vspace{.2cm}

This property is illustrated by the following local monotonicity statement, which in the confined type replaces Lemma~\ref{lemma:Q-lower-bis}.

\begin{lemma}\label{lemma:Q}
There exists a constant $C'$ such that for all realizations and $\epsilon>0$ small enough, $y \in (-\infty,-C')$ implies $Q^\epsilon \ast y \geq y$, whereas $y \in (C',\infty)$ implies $Q^\epsilon \ast y \leq y$.
\end{lemma}

\noindent {\bf Proof.} Put $C'=\frac{1}{2C_0}\log(\frac{8C_2}{C''_1})$. As the change of variables \eqref{eq-VariableTransforms} is an order-preserving bijection from $(0,\infty)$ to $\mathbb{R}$, the statement can be rephrased as follows: $x \in (0,\frac{C''_1}{8C_2})$ implies $Q^\epsilon \cdot x \geq x$, whereas $x \in (\frac{8C_2}{C''_1},\infty)$ implies $Q^\epsilon \cdot x \leq x$. Similar to~\eqref{ineq:Q-upper} in the proof of Lemma~\ref{lemma:Q-upper} (replacing $\delta^2$ by $0$, leaving out a global factor of $\epsilon$), $Q \cdot x \leq x$ is equivalent to
$$
(a+b+\mathcal{O}(\epsilon)) x^2  + \mathcal{O}(\epsilon)x + a-b + \mathcal{O}(\epsilon) 
\;\leq\;
0\,,
$$
and reversing this inequality is equivalent to $Q^\epsilon \cdot x \geq x$. By the Main Hypothesis, the l.h.s. of the above inequality becomes positive for $x \to 0$ and negative for $x \to \infty$. For $x \in (0,\infty)$, this expression only vanishes if $x$ takes the value
$$
\frac{a-b + \mathcal{O}(\epsilon)}{\mathcal{O}(\epsilon) + \sqrt{b^2 - a^2 + \mathcal{O}(\epsilon)}} 
\;=\; 
\frac{\mathcal{O}(\epsilon) + \sqrt{b^2 - a^2 + \mathcal{O}(\epsilon)}}{-a-b + \mathcal{O}(\epsilon)}
\;.
$$
Observing that the term on the left is bounded from below by $\frac{C''_1}{8C_2}$ and the one on the right is bounded from above by $\frac{8C_2}{C''_1}$ finishes the proof.
\hfill $\Box$

\vspace{.2cm}

From the statement and proof of Lemma~\ref{lemma:Q}, it becomes clear that $Q^\epsilon \cdot$ is a well-defined map on $(0,\infty)$ - which is false for the rotating type - or equivalently, $Q^\epsilon \ast$ properly acts on $\mathbb{R}$ (the copy for which $\nu = +$). One can verify that both $Q^\epsilon \ast$ and $D^\epsilon \ast$ preserve the order on $\mathbb{R}$, that is, for all $y,y' \in \mathbb{R}$ one has
\begin{gather}\label{stat:order}
y\,<\,y' \quad\Longrightarrow\quad ((Q^\epsilon D^\epsilon)\ast y) \,<\, ((Q^\epsilon D^\epsilon)\ast y')\,.
\end{gather}
This directly follows from the injectivity of all real M\"obius transforms not mapping to infinity (such as $Q^\epsilon\cdot$ and $D^\epsilon\cdot$). The role of the constants of Section~\ref{sec-LogDysonSchmidt} in the current context is clarified in the following two lemmata, analogous to Lemmata~\ref{lemma:y} and~\ref{lemma:Q-upper}.

\begin{lemma}\label{lemma:Ising}
For each realization, $y \in \mathbb{R}$ and $\epsilon>0$ small enough, it holds that
\begin{align}\label{stat:slower-start-right}
\hspace{80pt}&y \in (-\infty,\infty) &&\Longrightarrow && ((Q^\epsilon D^\epsilon)\ast y) \in (\widehat{y}_-,\infty)\,,\hspace{80pt}\\\label{stat:slower-start-left}
\hspace{80pt}&y \in (-\infty,\infty) &&\Longrightarrow && ((Q^\epsilon D^\epsilon)\ast y) \in (-\infty,\widehat{y}_+)\,,\hspace{80pt}\\\label{stat:slower-between-left}
\hspace{80pt}&y \in (\widehat{y}_-,\infty) &&\Longrightarrow && ((Q^\epsilon D^\epsilon)\ast y) \in (\widehat{y}_c,\infty)\,,\hspace{80pt}\\\label{stat:faster-between-left}
\hspace{80pt}&y \notin (\widetilde{y}_-,\infty) &&\Longrightarrow && ((Q^\epsilon D^\epsilon)\ast y) \notin (\widetilde{y}_c,\infty)\,.\hspace{80pt}
\end{align}
\end{lemma}

\noindent{\bf Proof.} Note that~\eqref{stat:slower-start-right} is the same statement as the first line of~\eqref{stat:slower-faster-y} from Lemma~\ref{lemma:y}, a rephrasing of~\eqref{stat:slower-start} of Lemma~\ref{lemma:slower}, identical to (25) of~\cite[Lemma~5]{DSS} (up to the fact that the assumption $y < \widehat{y}_+$ is no longer needed to guarantee that $(Q^\epsilon D^\epsilon)\ast y$ is well-defined). The proof is identical to the one given in~\cite{DSS} after replacing $C_1$ by $C''_1$ and using the order-preserving property by~\eqref{stat:order}. In full analogy,~\eqref{stat:slower-between-left} can be shown as (26) of~\cite[Lemma~5]{DSS}, and~\eqref{stat:faster-between-left} as (34) of~\cite[Lemma~10]{DSS} (replacing the role of~\cite[Lemma~9]{DSS} by the current Lemma~\ref{lemma:drift-bis}). Finally,~\eqref{stat:slower-start-left} follows by applying the symmetry given before Lemma~\ref{lemma:Q} to~\eqref{stat:slower-start-right}, including a factor of $e^{2C_0} > 1$ (in the definition of $\widehat{x}_+ = e^{2C_0}\widehat{x}_-^{-1}$) to account for additional error contributions.
\hfill $\Box$

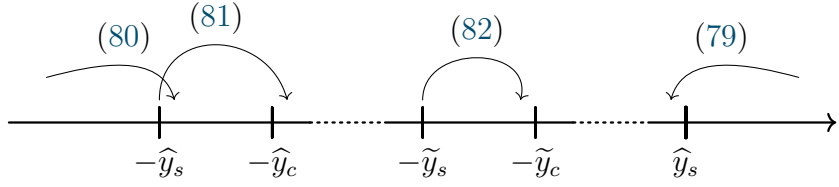
\begin{figure}
\begin{center}
\begin{tikzpicture}[line join = round, line cap = round]
\coordinate (a) at (0.0,0.3);
\coordinate (b) at (0.0,-0.2);
\coordinate (bb) at (0.0,-0.8);
\coordinate (r) at (0.2,0.0);
\coordinate (l) at (-0.2,0.0);
\coordinate (ll) at (-0.5,0);
\coordinate (l0) at (0,0);
\coordinate (l0a) at ($(l0) + (a) + (a)$);
\coordinate (l1) at (1.5,0);
\vtick{(l1)};
\coordinate (l1a) at ($(l1) + (a)$);
\coordinate (l1ar) at ($(l1) + (a) + (r)$);
\draw[->] (l0a) to[out=15, in=105, looseness=1.0] node[midway,above,inner sep=4pt] {\eqref{stat:slower-start-left}} (l1ar);
\coordinate[label=below:{$-\widehat{y}_s$}] (l1b) at ($(l1) + (b)$);
\coordinate (l2) at (3,0);
\vtick{(l2)};
\coordinate (l2ar) at ($(l2) + (a) + (r)$);
\draw[->] (l1a) to[out=90, in=105, looseness=1.5] node[midway,above,inner sep=4pt] {\eqref{stat:slower-between-left}} (l2ar);
\coordinate[label=below:{$-\widehat{y}_c$}] (l2b) at ($(l2) + (b)$);
\coordinate (lr) at (3.5,0);
\coordinate (ml) at (4.5,0);
\coordinate (m1) at (5,0);
\vtick{(m1)};
\coordinate (m1a) at ($(m1) + (a)$);
\coordinate[label=below:{$-\widetilde{y}_s$}] (m1b) at ($(m1) + (b)$);
\coordinate (m2) at (6.5,0);
\vtick{(m2)};
\coordinate[label=below:{$-\widetilde{y}_c$}] (m2b) at ($(m2) + (b)$);
\coordinate (m2al) at ($(m2) + (a) + (l)$);
\draw[->] (m1a) to[out=90, in=75, looseness=1.5] node[midway,above,inner sep=4pt] {\eqref{stat:faster-between-left}} (m2al);
\coordinate (mr) at (7,0);
\coordinate (rl) at (8,0);
\coordinate (r1) at (8.5,0);
\tick{(r1)};
\coordinate (r1al) at ($(r1) + (a) + (l)$);
\coordinate[label=below:{$\widehat{y}_s$}] (r1b) at ($(r1) + (b)$);
\coordinate (r0) at (10,0);
\coordinate (r0a) at ($(r0) + (a) + (a)$);
\draw[->] (r0a) to[out=165, in=75, looseness=1.0] node[midway,above,inner sep=4pt] {\eqref{stat:slower-start-right}} (r1al);
\coordinate (rr) at (10.5,0);
\draw [-,color=black,line width=0.3mm] (ll)--(lr);
\draw [-,color=black,dotted,line width=0.3mm] (lr)--(ml);
\draw [-,color=black,line width=0.3mm] (ml)--(mr);
\draw [-,color=black,dotted,line width=0.3mm] (mr)--(rl);
\draw [->,color=black,line width=0.3mm] (rl) -- (rr);
\end{tikzpicture}
\caption{\it Illustration of the dynamics in the logarithmic Dyson-Schmidt variables on $\mathbb{R}$ as stated in {\rm Lemma~\ref{lemma:Ising}}.}
\label{fig:confined}
\end{center}
\end{figure}

\vspace{.2cm}

As~\eqref{stat:slower-start-right} and~\eqref{stat:slower-start-left} state that the dynamics is essentially confined, the natural notion of a ``passage'' as for the rotating type is no longer provided. Lemma~\ref{lemma:drift-bis}, however, indicates a regime in which the dynamics can be controlled. All points outside of the given interval are either rather ``on the left'' or ``on the right'' of $\mathbb{R}$. It is therefore sensible to define the following random times. First set $N_0:=\min\{n\geq 1\,:\,y_n\notin[\widetilde{y}_-,\widetilde{y}_+]\}$ and then iteratively for $k\geq 0$
$$
N_{k+1}
\;:=\;
\min
\big\{n\geq 1\,:\,n>N_k\,,\,y_n\notin[\widetilde{y}_-,\widetilde{y}_+]\,, \sgn(y_n)\neq\sgn(y_{\vphantom{\widehat{k}}N_k})
\big\}
\,.
$$
Without loss of generality, suppose that $y_{\vphantom{\widehat{k}}N_0}<\widetilde{y}_-$ so that the first passage goes from left to right. It is now again possible to introduce slower and faster comparison processes.  
The crucial properties of these processes are given in the following lemma, which corresponds to Lemmata~~\ref{lemma:Y-slower} and~\ref{lemma:Y-faster} (let us stress again, however, that the symbols $N_k$, $(\widehat{y}_{k,n},\widehat{o}_{k,n})$, $(\widetilde{y}_{k,n},\widetilde{o}_{k,n})$, $\widehat{T}_k$ and $\widetilde{T}_k$ have a different meaning here than in earlier sections).

\begin{lemma}
\label{lemma:ConfinedComparisson}
There exist slower and faster comparison processes $\{(\widehat{y}_{k,n},\widehat{o}_{k,n})_{n\geq 1}\}_{k\geq 0}$ on $\dot{\mathbb{R}} \times [0,1]$ and $\{(\widetilde{y}_{k,n},\widetilde{o}_{k,n})_{n\geq 1}\}_{k\geq 0}$ on $\dot{\mathbb{R}} \times [0,1+\delta^2]$, obeying
\begin{gather}
\label{ineq:Confined}
\widehat{y}_{k,n} \;\leq \;(-1)^k\,y_{\vphantom{\widehat{k}}N_k+n}
\;\leq\; \widetilde{y}_{k,n}
\end{gather}
a.s. for all $k\geq 0$ and $n \in \{1,\dots,N_{k+1} - N_k-1\}$, as well as 
\begin{gather}\label{eq:ConfinedEnd}
\widetilde{y}_{k,N_{k+1} - N_k} \,=\, \infty\,.
\end{gather}
\end{lemma}

Let us explain the origin of the sign $(-1)^k$. We made the choice that the comparison processes always move from left to right, whereas the actual dynamics bounces between the left and right boundaries. These orientation changes are compensated by the sign $(-1)^k$. Accordingly, the interval on which the dynamics is bounded from below and from above by the slower and faster comparison processes respectively is no longer the full space $\mathbb{R}$ (as for the rotating type in the $y$-picture) but $(-\infty,\widetilde{y}_+)$ and $(\widetilde{y}_-,\infty)$ in an alternating way per passage. This corresponds to the definition of a passage, $\widetilde{y}_+=-\widetilde{y}_-$ and~\eqref{ineq:Confined}.

\vspace{.2cm}

By the foregoing observations, it will be sufficient for the current purposes to define the faster comparison processes $(\widetilde{y}_{k,n},\widetilde{o}_{k,n})$ exactly as given in~\eqref{eq:faster-Y-dyn}, with some symbols having a different meaning in the present context.

\vspace{.2cm}

When it comes to the slower comparison processes, only a single minor modification will be needed. That is, the role of $\widehat{y}_+$, involved in the stopping condition for the slower comparison processes defined earlier, will now be played by $\widetilde{y}_+$, due to the change in the definition of a passage. For the convenience of the reader, let us display the formal definition once more: for $k\geq 0$, define a Markov process $(\widehat{y}_{k,n},\widehat{o}_{k,n})_{n\geq 1}$ on the space $\dot{\mathbb{R}} \times [0,1-\delta^2]$ by setting $\widehat{y}_{k,1} = \widehat{y}_-$, $\widehat{o}_{k,1} = 1-\delta^2$ and for $n\geq 1$,
\begin{align}\label{eq:slower-Y-dyn-confined}
\begin{aligned}
\widehat{y}_{k,n+1} &:= \begin{cases}
\widehat{y}_c\,, &\text{ if } \widehat{y}_{k,n} \leq \widehat{y}_-\,,\\
\infty\,, &\text{ if } \widehat{y}_{k,n} \geq  \widetilde{y}_++\widehat{o}_{k,n}\,,\\
\widehat{y}_{k,n} + (-1)^k\ChiVar_{\vphantom{\widetilde{k}}N_k+n} - \delta^2\,, &\text{ else}\,,
\end{cases}
\\
\widehat{o}_{k,n+1} &:= \begin{cases}
\widehat{y}_{k,n+1} - y-1\,, &\text{ if } \widehat{y}_{k,n+1} \geq y+1 \text{ and } \widehat{o}_{k,n} = 1-\delta^2\,,\\
1-\delta^2\,, &\text{ if } \widehat{y}_{k,n+1} \leq y+\widehat{o}_{k,n} \text{ and } \widehat{o}_{k,n} \neq 1-\delta^2\,,\\
\widehat{o}_{k,n}\,, &\text{ else}\,.
\end{cases}
\end{aligned}
\end{align}
The proof that these definitions actually provide processes that satisfy all the properties stated in Lemma~\ref{lemma:ConfinedComparisson} is not spelled out in detail here because, as due to Lemmata~\ref{lemma:drift-bis},~\ref{lemma:Q} and~\ref{lemma:Ising}, it is merely a modification of the arguments in Sections~\ref{sec:faster} and \ref{sec:slower}. 

\vspace{.2cm}

\noindent {\bf Proof} of Theorem~\ref{theo-InvMeasure2}.
For the proof of the statement on the invariant measure, the arguments in Lemma~\ref{lemma:mu-slower} and Lemma~\ref{lemma:mu-faster} will be applied once for all even $k$ (passages from left to right) and odd $k$ (passages from right to left) separately. One can apply Lemma~\ref{lemma:faster-T-S} directly, but Lemmata~\ref{lemma:slower-T-S} and~\ref{lemma:balanced} are modified because the slower comparison processes of~\eqref{eq:slower-Y-dyn-confined} now are already stopped around $\widetilde{y}_+$ which simply leads to a weaker error bound of order $\Oo(\delta\log(\delta))$ for all presented results (just as in Lemma~\ref{lemma:faster-T-S}). One hence concludes that passages with even and odd $k$ are both approximately distributed by a triangular law (the one in Lemmata~\ref{lemma:mu-slower},~\ref{lemma:mu-faster} and~\ref{coro:xi-support}, the first two now having $\Oo(\delta\log(\delta))$ error bounds). Due to the orientation change, the distribution of the odd $k$ has to be flipped and then the addition of the two triangles leads to a uniform distribution on $[-1,1]$. Finally, the formula~\eqref{eq-Lyap} for the Lyapunov exponent still holds, simply because~\eqref{eq-MeasureSymInterval} still holds for the uniform distribution (note that it actually holds for all probability measures with a trapezoid distribution on $[-1,1]$).
\hfill $\Box$

\vspace{.2cm}

\noindent {\bf Comment:} While this work was being refereed, the follow-up \cite{CGGH} to the works \cite{GG,Col} became available as a preprint. 

\vspace{.2cm}

\noindent {\bf Declarations:} This work was supported by the DFG grant SCHU 1358/8-1 and the Chilean grant FONDECYT 1230949. Data sharing is not applicable to this article as no datasets were generated or analyzed during the current study. The authors have no relevant financial or non-financial interests to disclose.


\end{document}